\documentclass[english]{revtex4-2}
\usepackage[T1]{fontenc}
\usepackage[utf8]{inputenc}
\usepackage{color}
\usepackage{array}
\usepackage{float}
\usepackage{multirow}
\usepackage{amsmath}
\usepackage{graphicx}

\makeatletter

\providecommand{\tabularnewline}{\\}

\usepackage{nomencl}
\usepackage{etoolbox}
\usepackage{ifthen}

\renewcommand{\nomgroup}[1]{%
\ifthenelse{\equal{#1}{A}}{\item[\textbf{Roman Symbols}]}{%
\ifthenelse{\equal{#1}{G}}{\item[\textbf{Greek Symbols}]}{%
\ifthenelse{\equal{#1}{C}}{\item[\textbf{Abbreviations}]}{%
\ifthenelse{\equal{#1}{S}}{\item[\textbf{Subscripts and superscripts}]}{%
\ifthenelse{\equal{#1}{Z}}{\item[\textbf{Mathematical Symbols}]}
{}
}
}
}
}
}

\usepackage{color} 
\definecolor{myred}{rgb}{1,0,0.3}
\definecolor{myblue}{rgb}{0,0,1}
\definecolor{mywhite}{rgb}{1,1,1}
\definecolor{myblack}{rgb}{0,0,0}

\@ifundefined{showcaptionsetup}{}{%
 \PassOptionsToPackage{caption=false}{subfig}}
\usepackage{subfig}
\makeatother

\usepackage{babel}
\begin{document}
\title{On stable wrapper-based parameter selection method for efficient ANN-based
data-driven modeling of turbulent flows}
\author{Hyeongeun Yun, Yongcheol Choi, Youngjae Kim, Seongwon Kang}
\email{Corresponding author, skang@sogang.ac.kr}

\address{Department of Mechanical Engineering, Sogang University, Seoul 04107,
Korea}
\begin{abstract}
To model complex turbulent flow and heat transfer phenomena, this
study aims to analyze and develop a reduced modeling approach based
on artificial neural network (ANN) and wrapper methods. This approach
has an advantage over other methods such as the correlation-based
filter method in terms of removing redundant or irrelevant parameters
even under non-linearity among them. As a downside, the overfitting
and randomness of ANN training may produce inconsistent subsets over
selection trials especially in a higher physical dimension. This study
analyzes a few existing ANN-based wrapper methods and develops a revised
one based on the gradient-based subset selection indices to minimize
the loss in the total derivative or the directional consistency at
each elimination step. To examine parameter reduction performance
and consistency-over-trials, we apply these methods to a manufactured
subset selection problem, modeling of the bubble size in a turbulent
bubbly flow, and modeling of the spatially varying turbulent Prandtl
number in a duct flow. It is found that the gradient-based subset
selection to minimize the total derivative loss results in improved
consistency-over-trials compared to the other ANN-based wrapper methods,
while removing unnecessary parameters successfully. For the reduced
turbulent Prandtl number model, the gradient-based subset selection
improves the prediction in the validation case over the other methods.
Also, the reduced parameter subsets show a slight increase in the
training speed compared to the others.
\end{abstract}
\maketitle

\section{Introduction}

While the turbulent flows can be predicted accurately using the Navier-Stokes
equation via direct numerical simulation (DNS), a very high computational
cost of DNS is quite demanding for most engineering applications.
For efficient alternatives such as the Reynolds-averaged Navier-Stokes
(RANS), it is often difficult to construct an accurate closure model
with a theoretical approach due to the wide range of length and time
scales and strong non-linearity among physical parameters \citep{Duraisamy2019,brunton2020machine,Vinuesa2022}.
Several multi-physics problems such as the two-phase flows and reacting
flows have similar issues in closure modeling. In order to relax this
issue, there have been various efforts to use an artificial neural
network (ANN) for modeling complex non-linear phenomena. ANN is highly
flexible and effective in approximating complex physical relationships
arising from strong non-linearity and high dimensions without a predefined
model and prior knowledge of the underlying physics. While an ANN-based
model may suffer from a lack of self-regulation from competing physical
effects, successful outcome may be achieved with sufficient provision
of the database.

Among numerous ANN-based approaches for turbulence modeling in the
literature, a few of them are introduced briefly here. \citet{milano2002neural}
improved prediction of near-wall flow fields through utilization of
data-driven modeling of the wall pressure and shear stress, while
\citet{bin2022progressive} improved prediction of the near-wall eddy
viscosity using progressive machine learning for various types of
flow fields. Rather than improving the prediction of specific flow
fields, several studies utilized ANN models to improve the accuracy
of the turbulence modeling techniques itself. \citet{Sarghini2003}
proposed an efficient subgrid-scale model for LES using ANNs trained
with an LES database. \citet{subel2021data} presented data-driven
subgrid modeling of the Burgers turbulence for low generalization
error via time shifting and transfer learning. RANS modeling error
were reduced by \citet{Ling2016} through ANN-modeling of Reynolds
stress. Finally, \citet{Tracey2015} and \citet{volpiani2021machine}
improved the performance of the Spalart-Allmaras model by using ANN-trained
source term.

Despite the advantages associated with ANN-based modeling, the high-dimensional
nature of turbulence problems make constructing a well-trained ANN-model
especially difficult. ANN-based modeling becomes more efficient by
reducing the dimensionality of the input space. \citet{pmlr-v4-janecek08a,Ladha2011}
argued that increased dimensionality of the input space can make the
training data locally sparse and lead to undesirable errors amplified
by inter-parameter dependency, ultimately reducing the accuracy of
the trained model. From a computational viewpoint, \citet{Marsland:2009:MLA:1571643}
and \citet{pmlr-v4-janecek08a} argued that the memory consumption
can be very high for a training a model with high dimensionality of
the input space. Also, the trained model may show large errors due
to overfitting \citep{James2013}.

Therefore, ANN-based modeling of turbulence problems can benefit significantly
from systematic dimensionality reduction via primary parameter identification.
Consequently, there have been several studies on dimensionality reduction
techniques for flow modeling. A few studies \citep{Zhang2015,Duraisamy2015,Parish2016}
identified the modeling parameters using a sequential forward selection
approach with ANN and Gaussian process for turbulent modeling. \citet{Sun2019}
identified dominant parameters for the eddy viscosity using the optimal
brain surgeon, decision tree, and ReliefF algorithm. \citet{Moghaddam2018}
extracted new parameters using the convolutional ANN (CNN) from a
parameter set for the Reynolds stress. In \citet{Isaac2014}, a chemistry
model for a reacting flow was simplified by reducing the number of
transport equations using the principal component analysis (PCA).

In the above literature, dimensionality reduction techniques for data-driven
modeling can be classified into two groups, i.e. parameter (or feature)
selection and extraction. In parameter selection, a parameter subset
is selected from original parameter set in a database. The parameter
selection can be further classified into three sub-categories named
as the filter, wrapper, and embedded methods (\citep{Ladha2011,Khalid2014},
etc.). A filter method evaluates importance of the parameters based
on pre-determined discriminating criteria. The Pearson correlation
coefficient and information gain are often used for this purpose.
A wrapper method selects the parameters based on the accuracy of a
trained model. The sequential forward selection (SFS), sequential
backward elimination (SBE), and genetic algorithm belong to this category.
An embedded method selects the modeling parameters simultaneously
with training a model. Parameter selection using L-1 regularization
is an example \citep{liu2009multi,zhao2010efficient}. On the other
hand, the parameter extraction recombines and generates a new parameter
set optimized for modeling. PCA is a well-known example. It is based
on linear decomposition and may have a limitation for non-linear physics.
To relax this limitation, \citet{kramer1991nonlinear} suggested nonlinear
PCA based on ANN, which was extended and applied to turbulent combustion
\citep{mirgolbabaei2014nonlinear}. \citet{callaham2021learning}
employed the Gaussian mixture models and sparse PCA for feature extraction
in turbulent boundary layer.

For complex physical problems, different approaches are better for
different purposes. In some cases, parameter extraction may not be
ideal, because the number of involved physical variables is not reduced,
although the dimensionality of a model is reduced \citep{Guyon2003}.
It may be difficult to understand physical meaning of recombined modeling
parameters \citep{pmlr-v4-janecek08a}. A filter method may have an
issue for a large number of physical parameters with overlapping effects.
Because of limited consideration of inter-parameter dependency, redundant
parameters may not be removed well \citep{Biswas2016,Chandrashekar2014}.
Notably, there have been very limited efforts in the literature on
a comparative analysis of dimensionality reduction techniques for
modeling the flow problems.

The wrapper method has a relatively high computational cost compared
to the filter method; however, the method removes the redundant parameters
relatively well even when non-linear relationships exist among variety
of the input parameters. When the wrapper methods (e.g., SFS and SBE)
are combined with ANN, they show improved capability in handling non-linear
relationships but issue of inconsistency over selection trials may
arise due to the overfitting and random nature of ANN training. Also,
the ANN-based wrapper (ANN-wrapper) method may experience increased
generalization errors.

In this background, the first objective of the present study is to
analyze and compare different ANN-wrapper methods for efficient modeling
of turbulent flow problems. The consistency-over-trials (CoT) of different
ANN-wrapper methods are analyzed, which was not examined in the previous
studies. As the second objective, a novel ANN-wrapper method is proposed
to improve the CoT. The following method relies on a new gradient-based
subset selection indices which are elaborated upon in Sec.~\ref{sec:Descriptions-of-parameter selection}.
Finally, the performance and characteristics of the newly proposed
and existing ANN-wrapper methods are examined. The characteristics
of different methods are first examined using a manufactured subset
selection problem in Sec.~\ref{sec: Analysis-of-parameter}, followed
by actual turbulent flow applications in Secs. \ref{sec: Applications-bubbly}
and \ref{sec:Application-Prt}.

\section{Data-driven modeling for flow problems and parameter selection techniques\label{sec:Descriptions-of-parameter selection}}

In this section, a data-driven modeling approach for CFD using a reduced
parameter subset is introduced. A few ANN-wrapper methods for parameter
selection are explained and discussed. Then, a novel method with improved
consistency over selection trials is presented.

\subsection{An approach of data-driven modeling and simulation}

\begin{figure}[H]
\centering{}\includegraphics[scale=0.45]{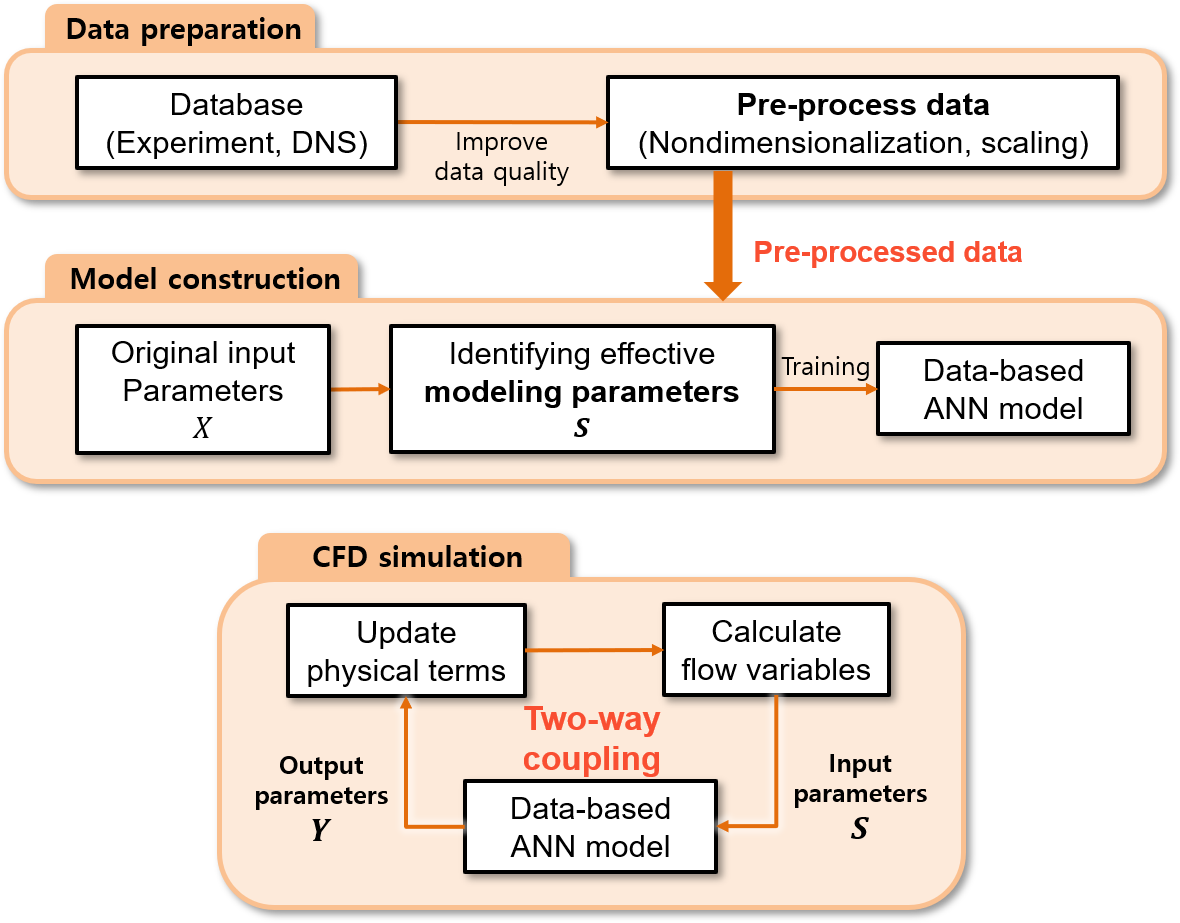}\caption{A process diagram of data-driven flow modeling and simulation using
ANN and parameter selection.\label{fig:Structure of data based simulation using RANS-1}}
\end{figure}
We consider parameter selection techniques for data-driven modeling
of flow problems. Fig.~\ref{fig:Structure of data based simulation using RANS-1}
shows a procedure to construct a reduced data-driven ANN model coupled
to a flow simulation. This approach is similar to the forward data-driven
modeling - one of five categories for data-driven thermal fluid models
by \citet{Chang2019}, but the present approach includes a parameter
selection (reduction) process for efficient modeling. Each step in
Fig.~\ref{fig:Structure of data based simulation using RANS-1} is
described below:
\begin{enumerate}
\item Data preparation step
\begin{enumerate}
\item To collect data from experiments or numerical simulations
\item To pre-process the database (e.g. non-dimensionalization, scaling)
\end{enumerate}
\item Model construction step
\begin{enumerate}
\item To determine a parameter subset ($S$) with a reduction technique
\item To train ANN using the pre-processed database and a parameter subset
\end{enumerate}
\item Data-driven simulation step
\begin{enumerate}
\item To perform flow simulation using a physics solver coupled with the
constructed ANN
\end{enumerate}
\end{enumerate}

\subsection{Details of ANN-based modeling}

\begin{figure}[H]
\begin{centering}
\includegraphics[scale=0.45]{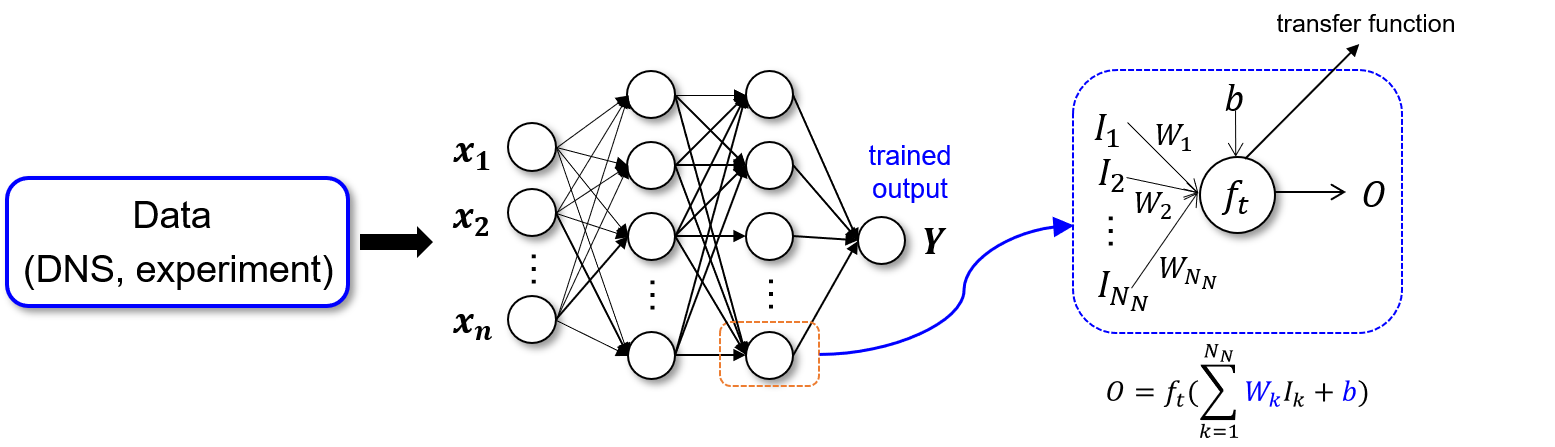}
\par\end{centering}
\centering{}\caption{A structure of a fully-connected multi-layer ANN. The symbols $W_{k}$,
$b$ and $f_{t}$ denote the weight, bias and activation function
in a neuron operation, respectively. $N_{N}$ denotes the number of
neurons per hidden layer.\label{fig:Structure of Artificial neural networks}}
\end{figure}
\begin{table}[H]
\begin{centering}
\begin{tabular}{|c|c|}
\hline 
ANN library & Google TensorFlow\tabularnewline
\hline 
Activation function & Hyperbolic tangent\tabularnewline
\hline 
Learning rate & 0.0001\tabularnewline
\hline 
Learning algorithm & Adam \citet{Kingma2014}\tabularnewline
\hline 
Loss function & Mean square error\tabularnewline
\hline 
\end{tabular}
\par\end{centering}
\caption{Hyper-parameters and conditions for training an ANN.\label{tab:Hyperparameter and conditions}}
\end{table}

In this study, a fully-connected multi-layer ANN is employed as shown
in Fig.~\ref{fig:Structure of Artificial neural networks}. Based
on ANN, a model function $f$ can be constructed as
\begin{equation}
Y=f\left(X\right)=f\left(x_{1},x_{2},\cdots,x_{N_{x}}\right),\label{eq: base functional}
\end{equation}
where $Y$ and $X$ denote the output and input parameter sets, respectively.
$N_{x}$ is the number (dimension) of the input parameter set. In
the present study, the problems with a single output parameter are
considered, i.e. $Y=\left(y_{1}\right)$ and $X=\left(x_{1},x_{2},\cdots,x_{N_{x}}\right)$.
The subscript for $y$ will be omitted conveniently hereafter. After
a parameter selection, a reduced model function $f_{M_{x}}$ using
a parameter subset will be derived as 
\[
Y=f_{M_{x}}\left(S_{M_{x}}\right)=f_{M_{x}}\left(x_{s1},x_{s2},\cdots,x_{M_{x}}\right),
\]
where $S_{M_{x}}$ is a selected parameter subset with the size $M_{x}$
($<N_{x}$).

For a single neuron operation, $O$ and $I_{k}$ represent the output
and input values, respectively. A few hyper-parameters and conditions
for training ANNs are listed in Table~\ref{tab:Hyperparameter and conditions}.
Considering the characteristics of the hyperbolic tangent activation
function, the database for training is normalized to a range $\left[x_{min}^{*},x_{max}^{*}\right]=\left[-0.9,0.9\right]$
using
\begin{equation}
x^{*}=x_{min}^{*}+\frac{x-x_{min}}{x_{max}-x_{min}}\left(x_{max}^{*}-x_{min}^{*}\right),\label{eq: normalization}
\end{equation}
where the asterisk represents a normalized parameter. The mean square
error (MSE) is used as the loss function and written as
\begin{equation}
\mathrm{MSE}=\overline{\left(Y^{*,ANN}-Y^{*,DB}\right)^{2}},\label{eq:Definition of RMSE}
\end{equation}
where the overbar denotes averaging over the samples in the database.
The superscripts $ANN$ and $DB$ denote the ANN-predicted and the
database values, respectively.

\subsection{Performance evaluation metrics of reduced model}

In the existing wrapper methods (introduced in the next section),
a component in the parameter subset is selected from the original
(full) parameter set based on a criterion. They often employ either
MSE or root mean square error (RMSE) for the purpose. Thus, we can
consider the RMSE as a performance evaluation metric (PEM) of a reduced
model, i.e.,
\begin{equation}
\mathrm{PEM-RMSE}=\sqrt{\mathrm{MSE}}.\label{eq: PEM-RMSE}
\end{equation}
Note that the existing ANN-wrapper methods can be advantageous by
using the (nearly) identical indices for the model training, selection
criterion, and PEM.

In order to achieve a revised ANN-wrapper method with improved consistency-over-trials
(CoT), separate indices for the selection criterion and PEM of a reduced
model may be necessary. For the latter, we consider two different
PEMs for CoT. The first is the average of $N$-highest possibilities
among all selected parameters from repeated parameter selections,
i.e.,
\begin{equation}
\mathrm{PEM-CoT1}=\frac{1}{N}\sum_{i=1}^{N}D\left(P_{N}\left(x_{s}\right)\right),\label{eq: metric of stability 1}
\end{equation}
\begin{equation}
P_{N}\left(x_{s}\right)=\frac{\mathrm{Number\:of\:times\:selecting\:}x_{s}}{\mathrm{Number\:of\:repeated\:\mathit{S_{N}}\:selections}},\label{eq: metric of stability 1 PN}
\end{equation}
where $x_{s}$ denotes a (each) parameter selected at least once,
$P_{N}\left(x_{s}\right)$ is the possibility of selecting $x_{s}$
over repeated selections, and $D$ is the sort function in the descending
order. A higher PEM-CoT1 implies that a selected parameter subset
has relatively more consistent components over repeated selections.
The second PEM for CoT is the highest possibility of selecting a specific
subset, i.e.,
\begin{equation}
\mathrm{PEM-CoT2}=\max\left(\frac{\mathrm{Number\:of\:times\:selecting\:a\:}S}{\mathrm{Number\:of\:repeated\:\mathit{S_{N}}\:selections}}\right),\label{eq:metric of stability 2}
\end{equation}
where $S$ denotes a specific parameter subset. PEM-CoT2 is more strict
than PEM-CoT1, thus relatively smaller. Both of them are computed
after repeated parameter selections, thus can not be used as the selection
criterion.

\subsection{ANN-based wrapper methods for parameter selection\label{subsec: filter and wrapper}}

\subsubsection{Sequential forward selection (SFS)}

\begin{figure}[H]
\begin{centering}
\includegraphics[scale=0.48]{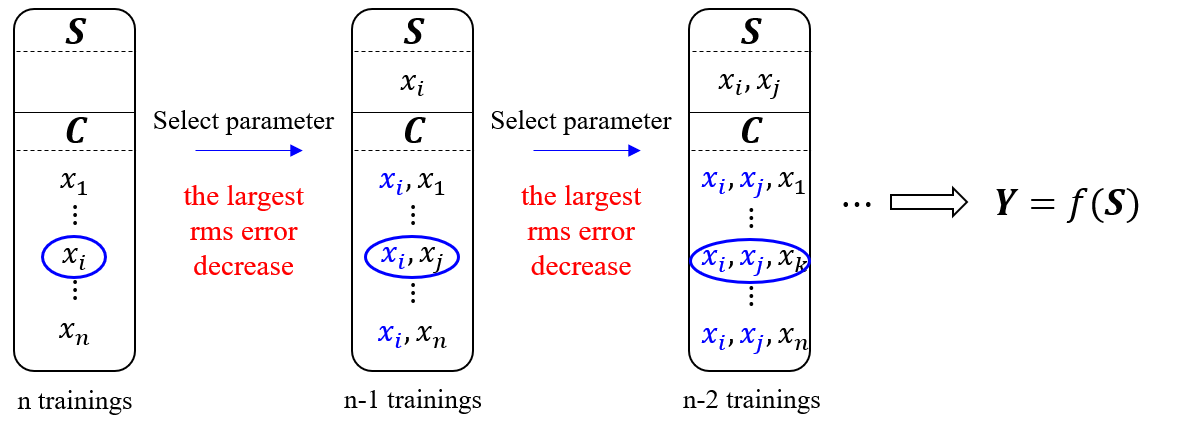}
\par\end{centering}
\caption{A detailed process of sequential forward selection.\label{fig:Training-based selection process}}
\end{figure}

The SFS selects the input parameters sequentially based on a criterion.
It was discussed or applied in several previous studies in various
fields such as \citet{John1994,Kohavi1997,MarcanoCedeno2010,Ladha2011,Chandrashekar2014}.
ANN-based SFS was applied to turbulence modeling \citep{Zhang2015,Duraisamy2015}.

Fig.~\ref{fig:Training-based selection process} shows a detailed
process of the ANN-based SFS. In order to obtain a parameter subset
($S_{M_{x}}$), the process starts from an empty set (i.e. $S_{0}=\left(0\right)$).
Then, a group of the candidate subsets ($C$) is generated by adding
each parameter to the existing subset ($S_{n-1}$). After training
ANNs with all $C$ members and evaluating the subset selection indices
(criteria), the most competitive one is chosen as the new subset ($S_{n}$).
As the subset selection index, the model errors such as the RMSE are
used. Based on the RMSE, the objective function of SFS is written
as
\begin{equation}
J_{SFS}=\min\left[\Delta\mathrm{RMSE}\right]=\min\left[\mathrm{RMSE}\left(S_{n}\right)-\mathrm{RMSE}\left(S_{n-1}\right)\right],\label{eq: J_SFS}
\end{equation}
where $\Delta\mathrm{RMSE}$ is negative. This process is iterated
until $\left|J_{SFS}\right|$ becomes sufficiently small.

In the literature, it was argued that SFS can produce the optimal
parameter subset for ANN-based modeling (e.g. \citep{Molina2002,pmlr-v4-janecek08a,Khalid2014,Biswas2016}).
Compared to the genetic algorithm that searches the optimal subset
in a random manner, it is computationally less intensive \citep{Gheyas2010,Thambi2017}.
In complex inter-parameter relationships, however, it can not remove
previously selected parameters that become unnecessary after subsequent
selections \citep{Ladha2011,Chandrashekar2014}. Thus, the final subset
may be suboptimal. Also, it may produce an inappropriate subset when
the model overfitting occurs \citep{Kohavi1997,Chandrashekar2014},
which will be discussed later.

In the computational viewpoint, the number of ANN trainings to select
$M_{x}$ parameters from the original $N_{x}$ parameters is estimated
as
\begin{equation}
N_{training,SFS}=\sum_{i=1}^{M_{x}}\left(N_{x}-i+1\right)=\frac{1}{2}\left(2M_{x}N_{x}+M_{x}-M_{x}^{2}\right).\label{eq: N_ANN add}
\end{equation}

\subsubsection{Sequential backward elimination (SBE)}

\begin{figure}[H]
\begin{centering}
\includegraphics[scale=0.48]{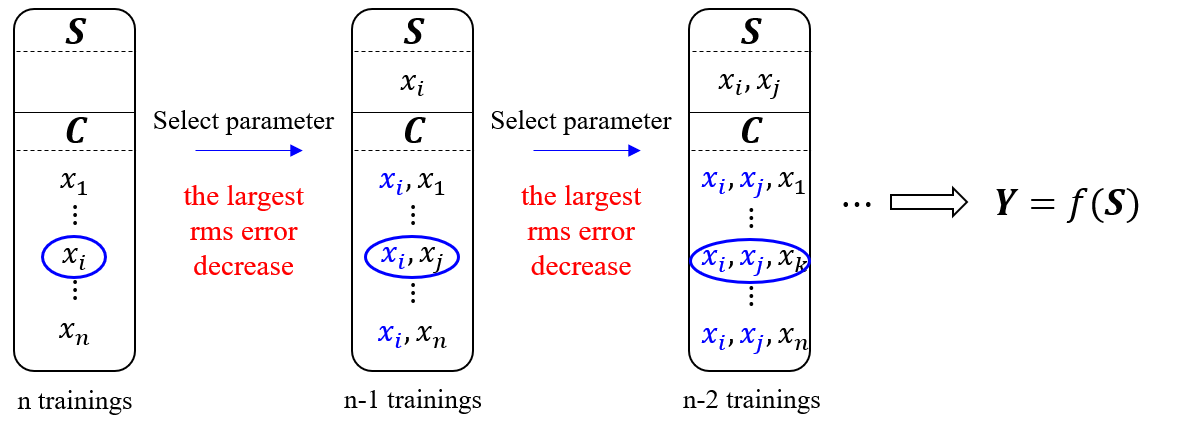}
\par\end{centering}
\centering{}\caption{A detailed process of sequential backward elimination. Grey parameters
present the removed parameters.\label{fig:Training-based removal process}}
\end{figure}

The SBE in previous studies \citep{John1994,Kohavi1997,Ladha2011,Chandrashekar2014}
takes a similar approach to the SFS but in the way of reducing dimensionality
from the full parameter model, as illustrated in Fig.~\ref{fig:Training-based removal process}.
It starts from the full parameter set ($S=X$) and removes less useful
parameters one-by-one. At each iteration, a group of the subsets ($C$)
is generated by removing a parameter from the existing subset ($S_{n+1}$).
After ANN trainings and error evaluation, new subset ($S_{n}$) is
chosen via the following objective function:
\begin{equation}
J_{SBE}=\min\left[\Delta\mathrm{RMSE}\right]=\min\left[\mathrm{RMSE}\left(S_{n}\right)-\mathrm{RMSE}\left(S_{n+1}\right)\right].\label{eq: J_SBE}
\end{equation}
While SBE shares several characteristics with SFS, it was mentioned
that interaction among the input parameters are better captured than
SFS \citep{Kohavi1997,Guyon2003}, which can be useful for a high-dimensional
physics modeling. However, it tends to be computationally more expensive
and also prone to an overfitting issue for a high-dimensional parameter
set \citep{Kohavi1997,Guyon2003}.

In the computational viewpoint, the number of ANN trainings to select
$M_{x}$ parameters from the original $N_{x}$ parameters is estimated
as
\begin{equation}
N_{training,SBE}=\sum_{i=1}^{N_{x}-M_{x}}\left(N_{x}-i+1\right)=\frac{1}{2}\left(N_{x}^{2}+N_{x}-M_{x}^{2}-M_{x}\right).\label{eq: N_ANN elim}
\end{equation}
If $M_{x}<N_{x}/2$, SBE trains a larger number of ANNs than SFS.

\subsubsection{Sequential gradient-based elimination (SGE)\label{subsec:Wrapper-method-=0000234}}

In this study, a new wrapper method is suggested to resolve an issue
of the ANN-wrapper methods in the previous sections. A few previous
studies \citep{Kohavi1997,Guyon2003,Chandrashekar2014} mentioned
a weakness of SFS and SBE under an overfitting situation of ANN models.
In the present study, we observe that SFS and SBE show reduced CoT
due to the overfitting and randomness of ANN training, which will
be presented in the next sections. To resolve this issue and achieve
more stable parameter selection (e.g., higher PEM-CoT1 (Eq.~\eqref{eq: metric of stability 1})
and PEM-CoT2 (Eq.~\eqref{eq:metric of stability 2})), we consider
two subset selection indices based on the gradient of the model function,
i.e. $\nabla f=\partial f/\partial x_{i}$. An idea to remove redundant
parameters by minimizing the change of $\nabla f$ between the new
and old subsets is presented here.

\begin{figure}[H]
\centering{}\includegraphics[width=10cm]{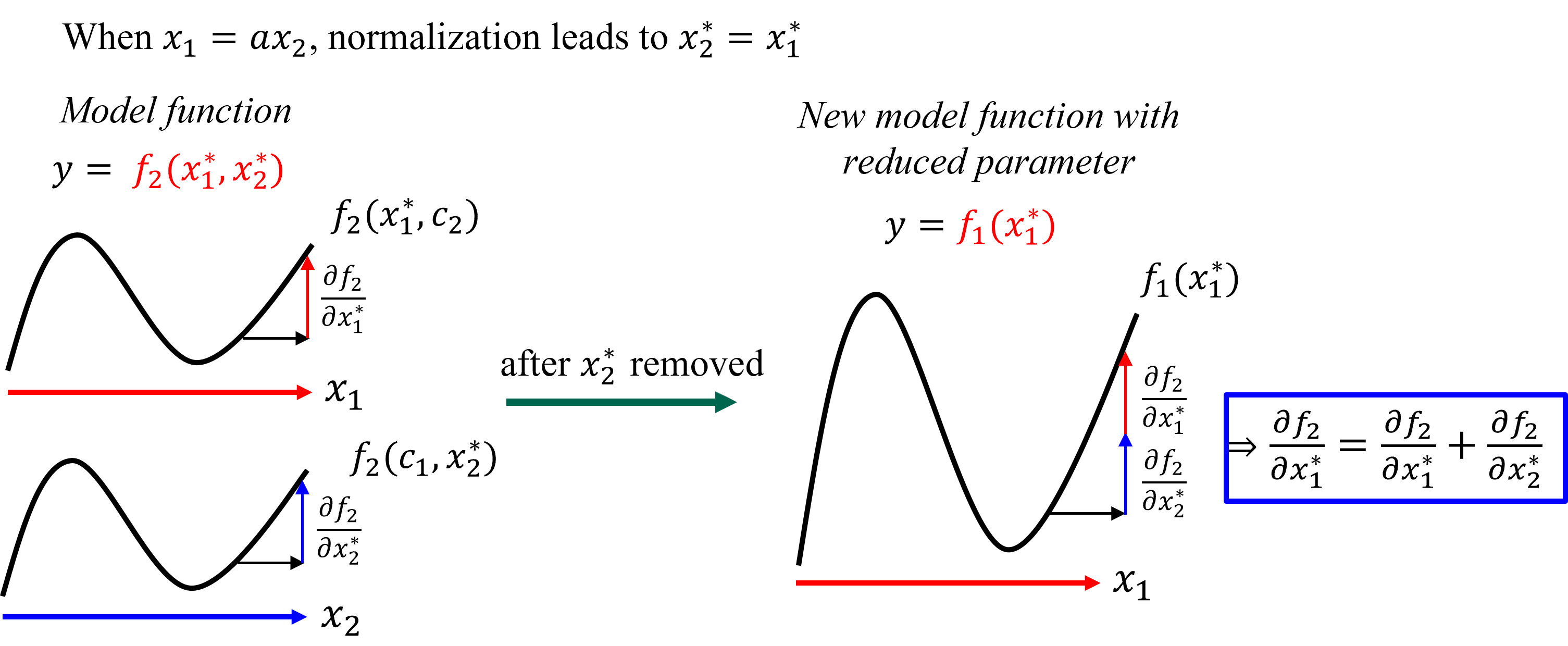}\caption{A schematic diagram of the minimum loss of \textcolor{black}{the}
total derivative to remove redundant parameters.\label{fig:minimum loss of total derivative}}
\end{figure}
The first alternative selection index is to minimize the loss of the
total derivative between the new ($S_{n}$) and old ($S_{n+1}$) subsets
during a parameter removal. Against $S_{n+1}$, it is desirable to
preserve the parametric variation of the function with $S_{n}$. In
Fig.~\ref{fig:minimum loss of total derivative}, two parameters
$x_{1}$ and $x_{2}$ satisfy $x_{2}=ax_{1}$ and have nearly identical
effects. After the normalization, $x_{2}^{*}=x_{1}^{*}$. If a model
function $Y^{*}=f_{2}\left(x_{1}^{*},x_{2}^{*}\right)$ is assumed,
it leads to $\frac{\partial f_{2}}{\partial x_{1}^{*}}=\frac{\partial f_{2}}{\partial x_{2}^{*}}$.
Then, the total derivative of $f_{2}(x_{1}^{*},x_{2}^{*})$ becomes
\begin{equation}
\frac{df_{2}}{dx^{*}}=\frac{\partial f_{2}}{\partial x_{1}^{*}}+\frac{\partial f_{2}}{\partial x_{2}^{*}}=2\frac{\partial f_{2}}{\partial x_{1}^{*}}=2\frac{\partial f_{2}}{\partial x_{2}^{*}}.
\end{equation}
When a new model function $f_{1}(x_{1}^{*})$ is built by removing
$x_{2}^{*}$, $x_{1}^{*}$ takes over the effect of $x_{2}^{*}$.
Then, the the total derivative of $f_{1}(x_{1}^{*})$ leads to
\begin{equation}
\frac{df_{1}}{dx^{*}}=\frac{\partial f_{1}}{\partial x_{1}^{*}}=\frac{\partial f_{2}}{\partial x_{1}^{*}}+\frac{\partial f_{2}}{\partial x_{2}^{*}}.
\end{equation}
This idea can be extended to an arbitrary input dimension and an objective
function of the SGE method to minimize the loss of the total derivative
(SGE-TD, hereafter) is written as

\begin{equation}
J_{TD}=\min\left[\overline{\Delta G_{TD}}\right]=\min\left[\overline{\left|\sum_{i=1}\frac{\partial f_{n}}{\partial x_{i}^{*}}\left(S_{n}\right)-\sum_{i=1}\frac{\partial f_{n+1}}{\partial x_{i}^{*}}\left(S_{n+1}\right)\right|}\right].\label{eq:minimum loss of total derivative}
\end{equation}
In most problems, the relationship of $x_{2}^{*}=x_{1}^{*}$ does
not hold exactly but the objective function $J_{TD}$ can be still
useful. A perfectly uncorrelated parameter ($x_{u}^{*}$) to $Y$
also can be removed by $J_{TD}$, as $\overline{\partial f_{n+1}/\partial x_{u}^{*}}\approx0$
and $\overline{\Delta G_{TD}}\approx0$ between $S_{n+1}$ and $S_{n}$
without $x_{u}^{*}$. If an essential parameter to $Y$ is removed,
the partial derivatives of the trained model can exhibit a random
behavior while redistributing its contribution with a limited success,
which affects $\overline{\Delta G_{TD}}$ significantly.

\begin{figure}[H]
\centering{}\includegraphics[width=9cm]{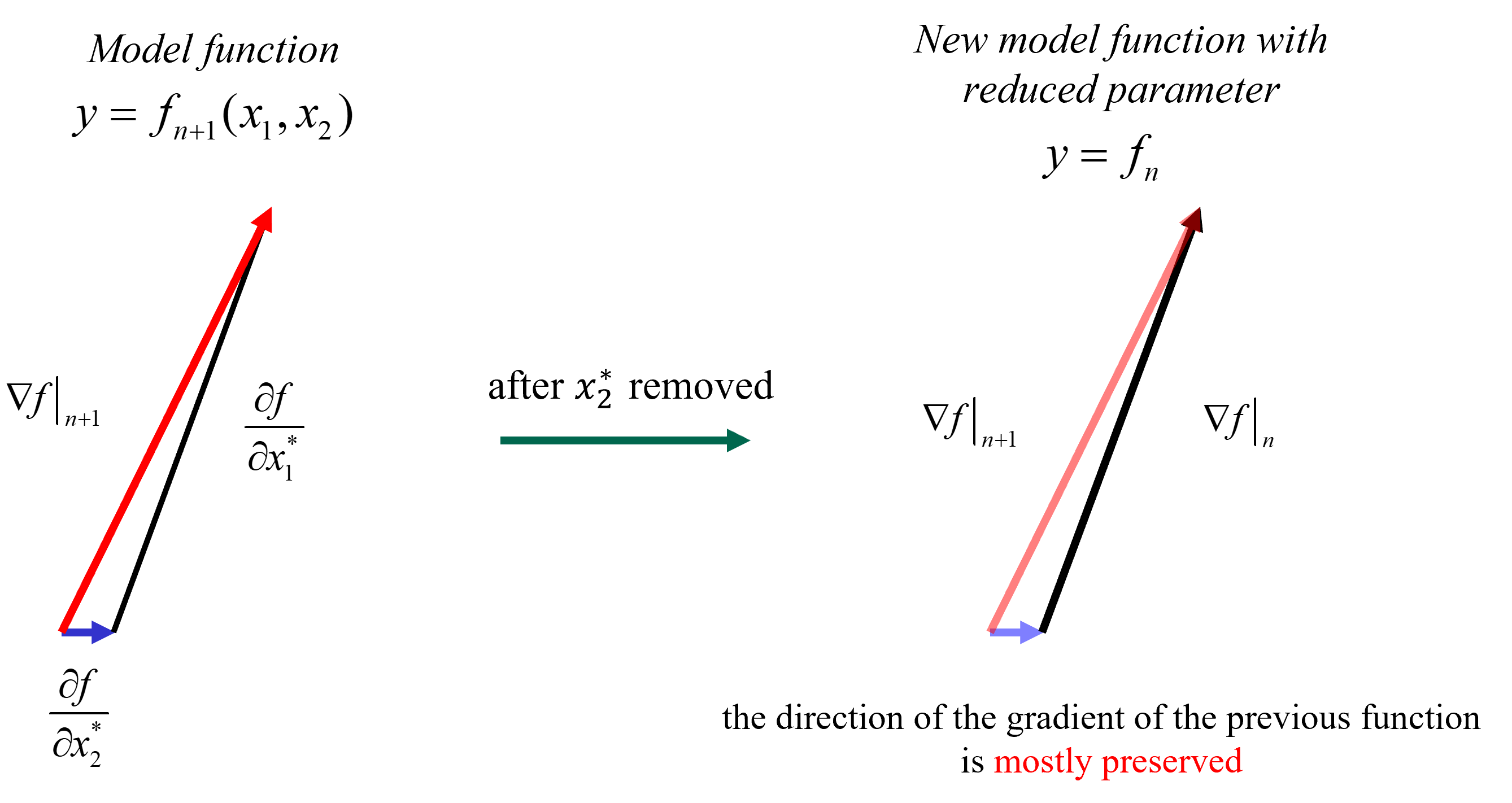}\caption{A schematic diagram of the directional consistency of \textcolor{black}{$\nabla f$
in case of unimportant parameter to $Y$}.\label{fig:directional consistency}}
\end{figure}

The second alternative selection index is to preserve the gradient
direction between the new and old subsets. As shown in Fig.~\ref{fig:directional consistency},
if a parameter $x_{2}$ has a trivial contribution to $Y$, $\nabla f_{n}$
and $\nabla f_{n+1}$ excluding $x_{2}$ will be similar. This idea
leads to an objective function as

\begin{equation}
J_{DC}=\min\left[\overline{1-\frac{\left|\nabla f_{n}\cdot\nabla f_{n+1\rightarrow n}\right|}{\left|\nabla f_{n}\right|\left|\nabla f_{n+1\rightarrow n}\right|}}\right],\label{eq:directional consistency}
\end{equation}
where $\nabla f_{n+1\rightarrow n}$ denotes $\nabla f_{n+1}$ reduced
to $n$-dimension by excluding a single component. This aims at preserving
the directional consistency of the gradient and called as SGE-DC hereafter.

The SGE methods (SGE-TD and SGE-DC) follow the same procedure as SBE
with the different objective function $J_{TD}$ or $J_{DC}$. Under
a circumstance of the overfitting, some neuron weights may become
excessive during training to keep reducing the RMSE, which increases
the sizes of the partial derivatives. The gradient-based indices can
impose a penalty in such a case and may improve the CoT compared to
SFS and SBE using the RMSE. The partial derivatives of an ANN are
computed in a semi-analytic way using a built-in feature of TensorFlow.

Also, it is possible to construct a mixed criterion of $\overline{\Delta G_{TD}}$
(or $\overline{\Delta G_{DC}}$) and $\Delta\mathrm{RMSE}$ (in Eq.~\eqref{eq: J_SBE})
to impose both criteria simultaneously, e.g.,
\begin{equation}
J_{mixed}=\min\left[\alpha_{SGE}\times\overline{\frac{\Delta G_{TD}}{G_{TD}(X)}}+(1-\alpha_{SGE})\times\frac{\Delta\mathrm{RMSE}}{\mathrm{RMSE}(X)}\right].\label{eq: J_mixed}
\end{equation}
Note that the present study focuses on testing SGE and a test on $J_{mixed}$
is reserved for a subsequent study.

\section{Application to a manufactured subset selection problem\label{sec: Analysis-of-parameter}}

In this section, the characteristics of ANN-wrapper methods introduced
in the previous section are analyzed using a manufactured subset selection
problem. We create an artificial database including a few randomly
distributed parameters and derived ones. We examine the performance
to remove redundant or irrelevant parameters and select a desirable
subset.

\subsection{Description of a manufactured problem and parameter set\label{subsec: manufactured problem}}

We consider three independent random parameters ($x_{1}$, $x_{4}$,
and $x_{7}$) with uniform distributions between 0 and 1. A few other
parameters are derived from them using the following equations:
\begin{equation}
\begin{cases}
x_{2}=x_{1}^{2},\;x_{3}=-\exp(x_{1}-0.5)\\
x_{5}=\exp(x_{4}),\;x_{6}=x_{4}^{3}
\end{cases}\label{eq: toy : params}
\end{equation}
where the first and second rows shows the parameters derived from
$x_{1}$ and $x_{4}$, respectively. The output parameter $y$ is
defined as
\begin{equation}
y=\frac{x_{3}x_{4}+(x_{2}-x_{5})+(x_{1}+x_{6})}{3},\label{eq:Definintion of toy problem}
\end{equation}
where $x_{7}$ is omitted on purpose and becomes an irrelevant parameter
to $y$. $x_{i}$ are categorized into $x_{1}$-group and $x_{4}$-group.
The parameters in each group are co-dependent among themselves.

\begin{figure}[H]
\centering{}\subfloat[Heat map of correlation coefficient]{\centering{}\includegraphics[height=5cm]{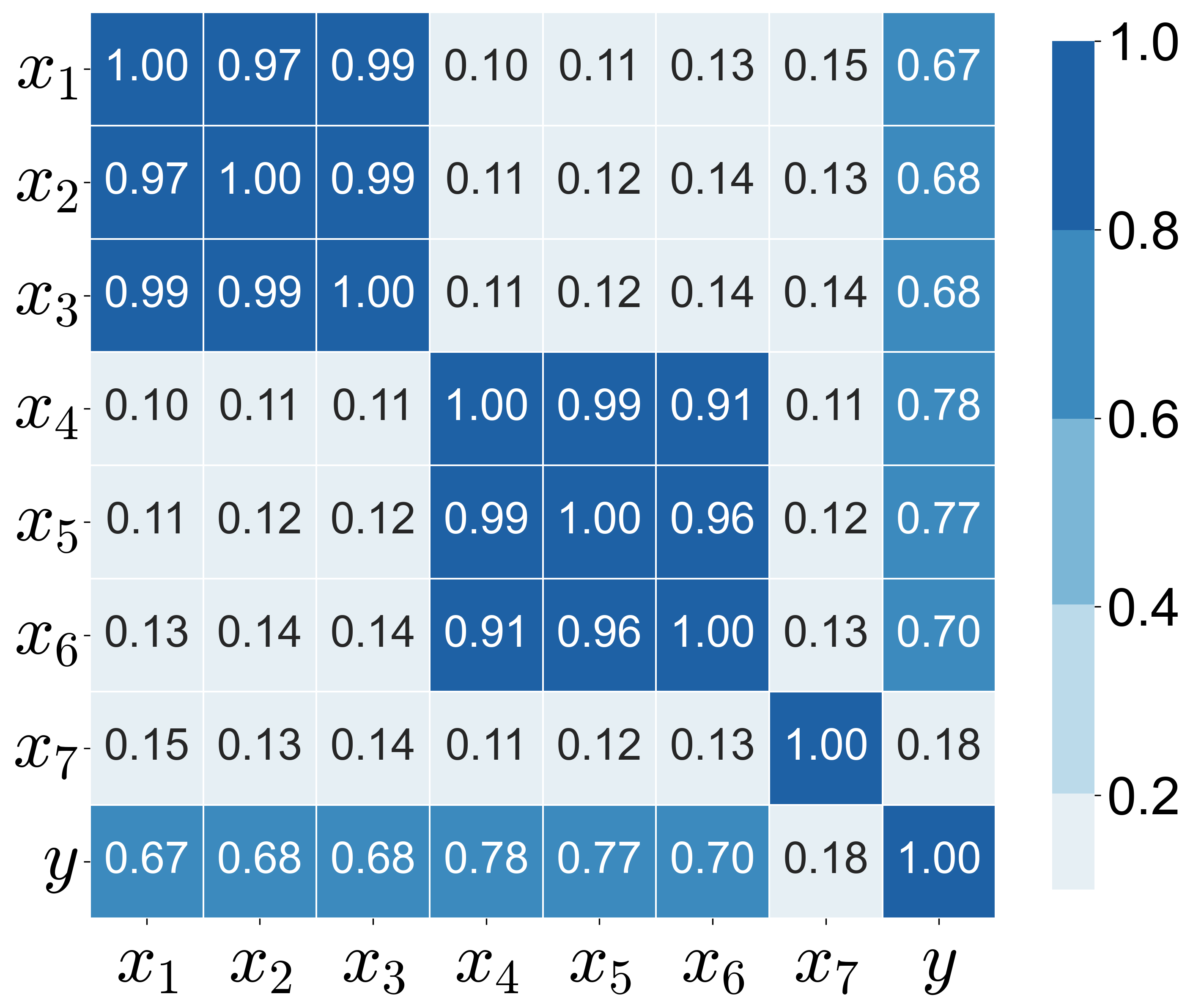}}\subfloat[Correlation between inputs and output]{\centering{}\includegraphics[height=5cm]{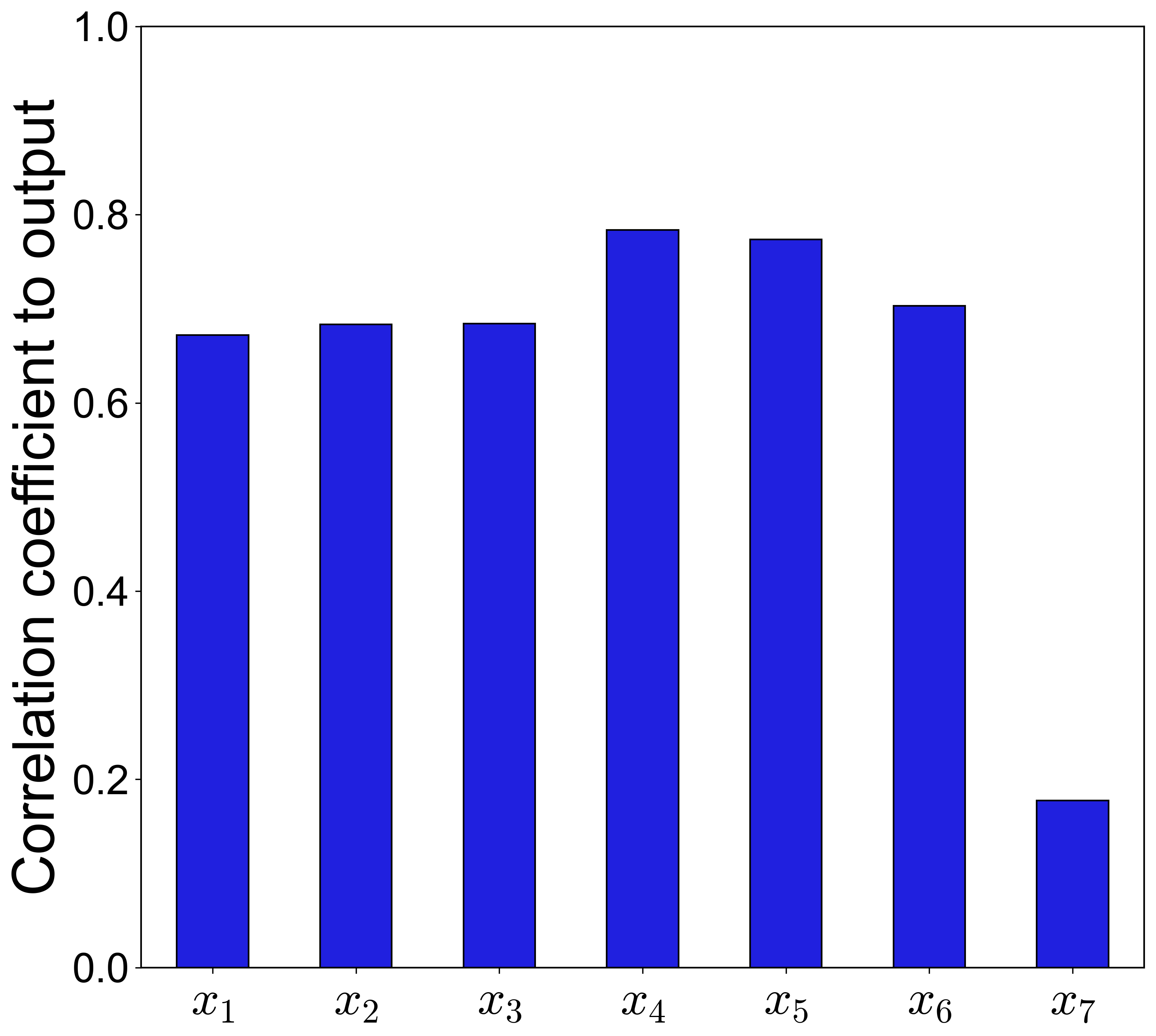}}\caption{\textcolor{black}{Pearson correlation coefficients} of the manufactured
problem.\label{fig:Correlation-of-manufactured}}
\end{figure}
Fig.~\ref{fig:Correlation-of-manufactured} shows the Pearson correlation
coefficients among the parameters. In Fig.~\ref{fig:Correlation-of-manufactured}(a),
the heat map shows that the parameters in the same group are relatively
highly correlated, denoting that they have similar effects. Fig.~\ref{fig:Correlation-of-manufactured}(b)
shows the correlation coefficients between each parameter and the
output $y$. It shows that all parameters except $x_{7}$ have similarly
high correlation coefficients with $y$. If the filter method (\citep{Guyon2003,Yu2004,Chandrashekar2014})
based on the correlation coefficient is used, $x_{4}$ and $x_{5}$
must be the first two parameters selected. Selecting two parameters
with similar effects is not ideal and leads to a suboptimal subset.

From $y=f\left(x_{1},\cdots,x_{7}\right)$, an ideal subset $S_{2}$
is assumed to have one parameter from each group. Such an outcome
based on the first two parameters selected for SFS or the last two
parameters for SBE and SGE is regarded as a successful selection.
This is used to evaluate the CoT of the ANN-wrapper methods in the
next section.

\subsection{Results of dimensionality reduction for manufactured problem}

\begin{figure}[H]
\centering{}\includegraphics[height=4cm]{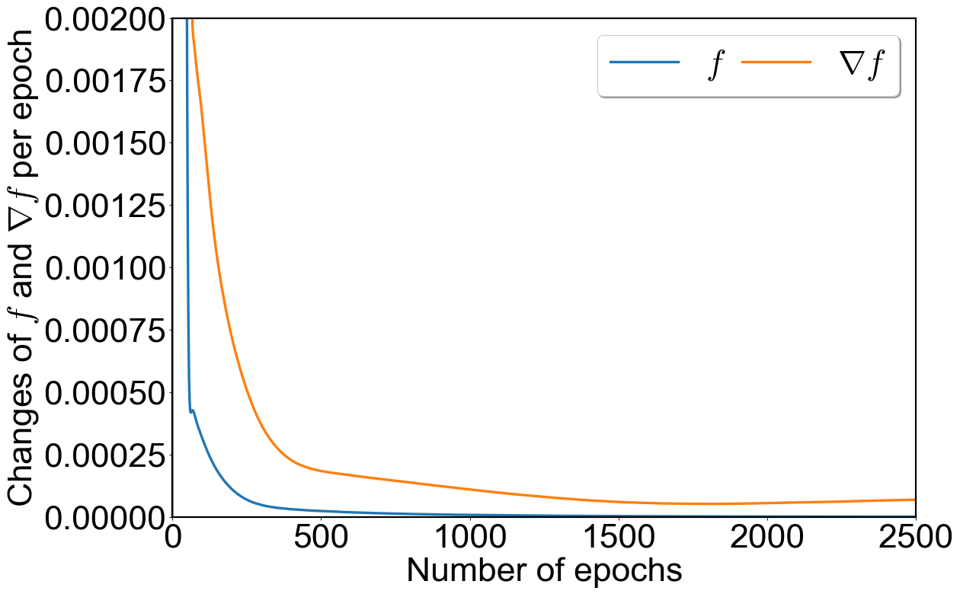}\caption{\textcolor{black}{Absolute changes of $f$ and $\nabla f$ per epoch
averaged over samples for 2L-10N/L ANN} for the manufactured problem.\label{fig:RMSE-and-gradient}}
\end{figure}
\begin{figure}[H]
\centering{}\includegraphics[height=4cm]{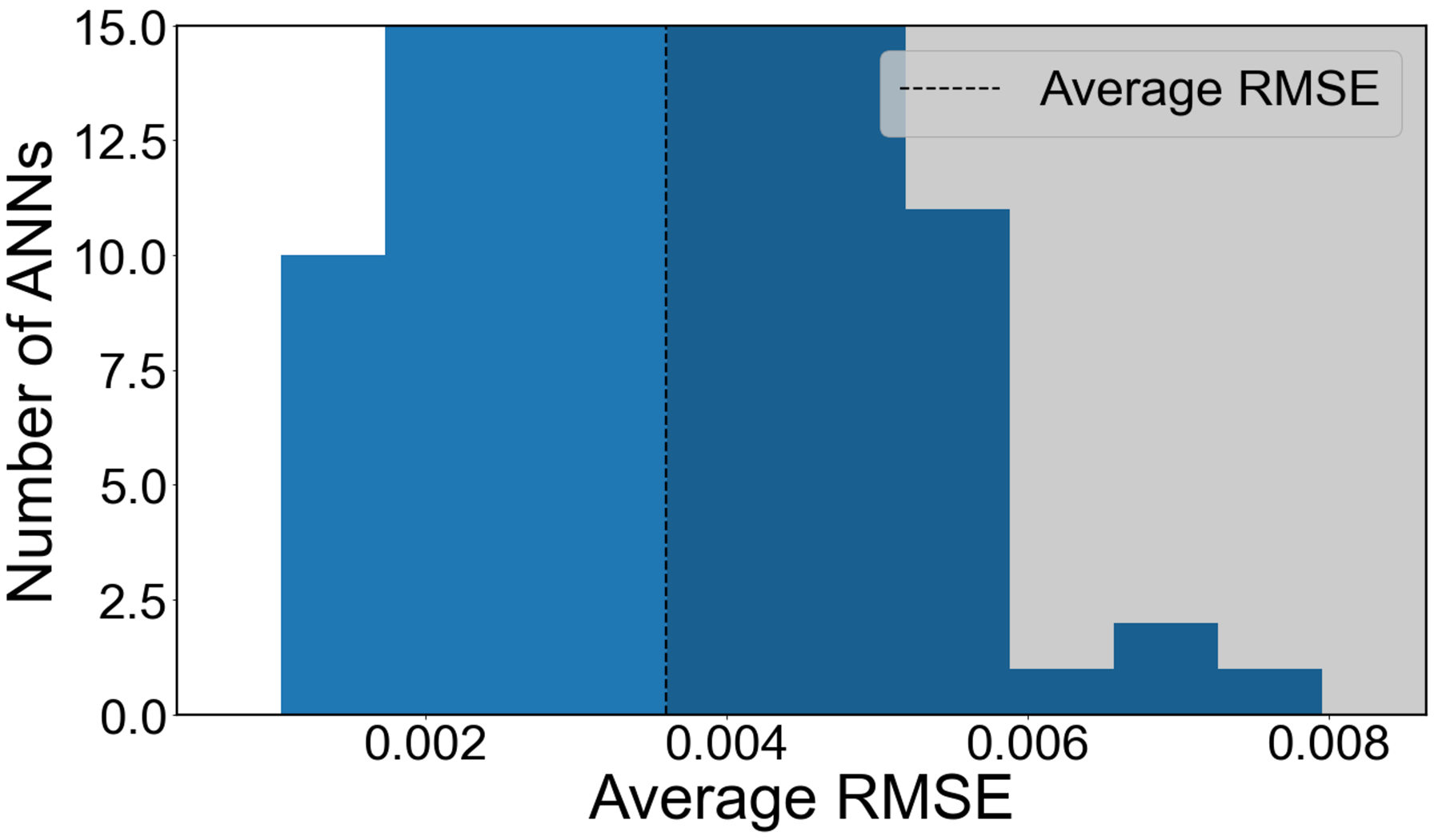}\caption{\textcolor{black}{Histogram of the RMSE from 200 trained ANN models
}for the manufactured problem.\label{fig:RMSE-histogram}}
\end{figure}
The ANN-wrapper methods introduced in Sec.~2 are applied to the manufactured
problem. The number of training epochs is set to 2500 times. To evaluate
the success (and failure) rate of each method, the parameter selections
are tried for 200 times.

Fig.~\ref{fig:RMSE-and-gradient} shows the \textcolor{black}{absolute}
changes of $f$ and \textcolor{black}{$\nabla f$} per epoch averaged
over samples for 2L-10N/L (2 layers, 10 neurons/layer) ANN. In Fig.~\ref{fig:RMSE-and-gradient},
the change of \textcolor{black}{$\nabla f$} per epoch decreases much
more slowly compared to $f$. Therefore, at the same training epochs,
a relatively larger error may exist in the gradient compared to the
function. A large error in the gradient can make the SGE selection
inconsistent. For a solution to this issue, we consider that the ANNs
with smaller RMSE at the same training epochs also have smaller errors
in their gradient. Then, we can train the ANNs multiple ($N$) times
and utilize only the ANNs with 50\% below-average RMSE (well-trained
ANNs, hereafter) during the selection. While this can increase the
overall training time by $N$ times, we found from a test that selecting
only the final subset (e.g. $S_{2}$) using well-trained ANNs similarly
improves the success rate (or CoT) of SGE significantly. Then, the
increase of the computational cost is minimal. Fig.~\ref{fig:RMSE-histogram}
shows the RMSE histogram from \textcolor{black}{200 trained ANN}s.
The case using well-trained ANNs utilizes the ANNs belonging to the
left (bright) side of the histogram. To validate the present approach,
the results using well-trained ANNs are compared with those using
all trained ANNs.

\begin{table}[H]
\centering{}%
\begin{tabular}{|c|>{\centering}p{0.7cm}|>{\centering}p{0.7cm}|>{\centering}p{0.7cm}|>{\centering}p{0.7cm}|>{\centering}p{0.7cm}|>{\centering}p{0.7cm}|>{\centering}p{0.7cm}|>{\centering}p{0.7cm}|>{\centering}p{0.7cm}|}
\hline 
 & \multicolumn{3}{c|}{1L-30N/L} & \multicolumn{3}{c|}{2L-10N/L} & \multicolumn{3}{c|}{4L-30N/L}\tabularnewline
\hline 
Method & S & R & F & S & R & F & S & R & F\tabularnewline
\hline 
SFS & 1 & 0 & 0 & 1 & 0 & 0 & 1 & 0 & 0\tabularnewline
\hline 
SBE & 1 & 0 & 0 & 1 & 0 & 0 & 1 & 0 & 0\tabularnewline
\hline 
SGE-TD & 0.975 & 0.025 & 0 & 0.99 & 0.01 & 0 & 1 & 0 & 0\tabularnewline
\hline 
SGE-DC & 0.88 & 0.12 & 0 & 0.995 & 0.005 & 0 & 1 & 0 & 0\tabularnewline
\hline 
\end{tabular}\caption{\textcolor{black}{The success (S), redundancy (R), and failure (F)
rates of $S_{2}$ over 200 selection trials using} all trained ANNs
for the manufactured problem.\label{tab:Table-of-stability-1-1}}
\end{table}
\begin{table}[H]
\centering{}%
\begin{tabular}{|c|>{\centering}p{0.7cm}|>{\centering}p{0.7cm}|>{\centering}p{0.7cm}|>{\centering}p{0.7cm}|>{\centering}p{0.7cm}|>{\centering}p{0.7cm}|>{\centering}p{0.7cm}|>{\centering}p{0.7cm}|>{\centering}p{0.7cm}|}
\hline 
 & \multicolumn{3}{c|}{1L-30N/L} & \multicolumn{3}{c|}{2L-10N/L} & \multicolumn{3}{c|}{4L-30N/L}\tabularnewline
\hline 
Method & S & R & F & S & R & F & S & R & F\tabularnewline
\hline 
SFS & 1 & 0 & 0 & 1 & 0 & 0 & 1 & 0 & 0\tabularnewline
\hline 
SBE & 1 & 0 & 0 & 1 & 0 & 0 & 1 & 0 & 0\tabularnewline
\hline 
SGE-TD & 1 & 0 & 0 & 1 & 0 & 0 & 1 & 0 & 0\tabularnewline
\hline 
SGE-DC & 1 & 0 & 0 & 1 & 0 & 0 & 1 & 0 & 0\tabularnewline
\hline 
\end{tabular}\caption{\textcolor{black}{The success (S), redundancy (R), and failure (F)
rates of $S_{2}$ over 100 selection trials using} well-trained ANNs
for the manufactured problem.\label{tab:Table-of-stability-1-2}}
\end{table}
Tables~\ref{tab:Table-of-stability-1-1} and \ref{tab:Table-of-stability-1-2}
show the success (S), redundancy (R), and failure (F) rates of $S_{2}$
over 100-200 trials of parameter selection using all trained ANNs
and well-trained ANNs, respectively. Redundancy means selecting two
parameters from a single group. Failure means one of two selected
parameters is irrelevant one ($x_{7}$). When all trained ANNs are
used, SFS and SBE produce perfectly successful parameter selection.
On the other hand, the two SGE methods show a less success rate for
relatively lower hyper-parameters. In contrast, when using well-trained
ANNs, the SGE methods also yield perfectly successful parameter selection
regardless of hyper-parameters. It is known that $\nabla f$ is more
sensitive to the numerical error than the model function $f$. The
present results show that SGE is more sensitive than SFS and SBE to
the training error of ANN. It is also shown that SGE using $\nabla f$
can successfully remove redundant and irrelevant parameters. Only
the results using well-trained ANNs are presented below.

\begin{figure}[H]
\centering{}\subfloat[1L-30N/L]{\centering{}\includegraphics[width=4cm]{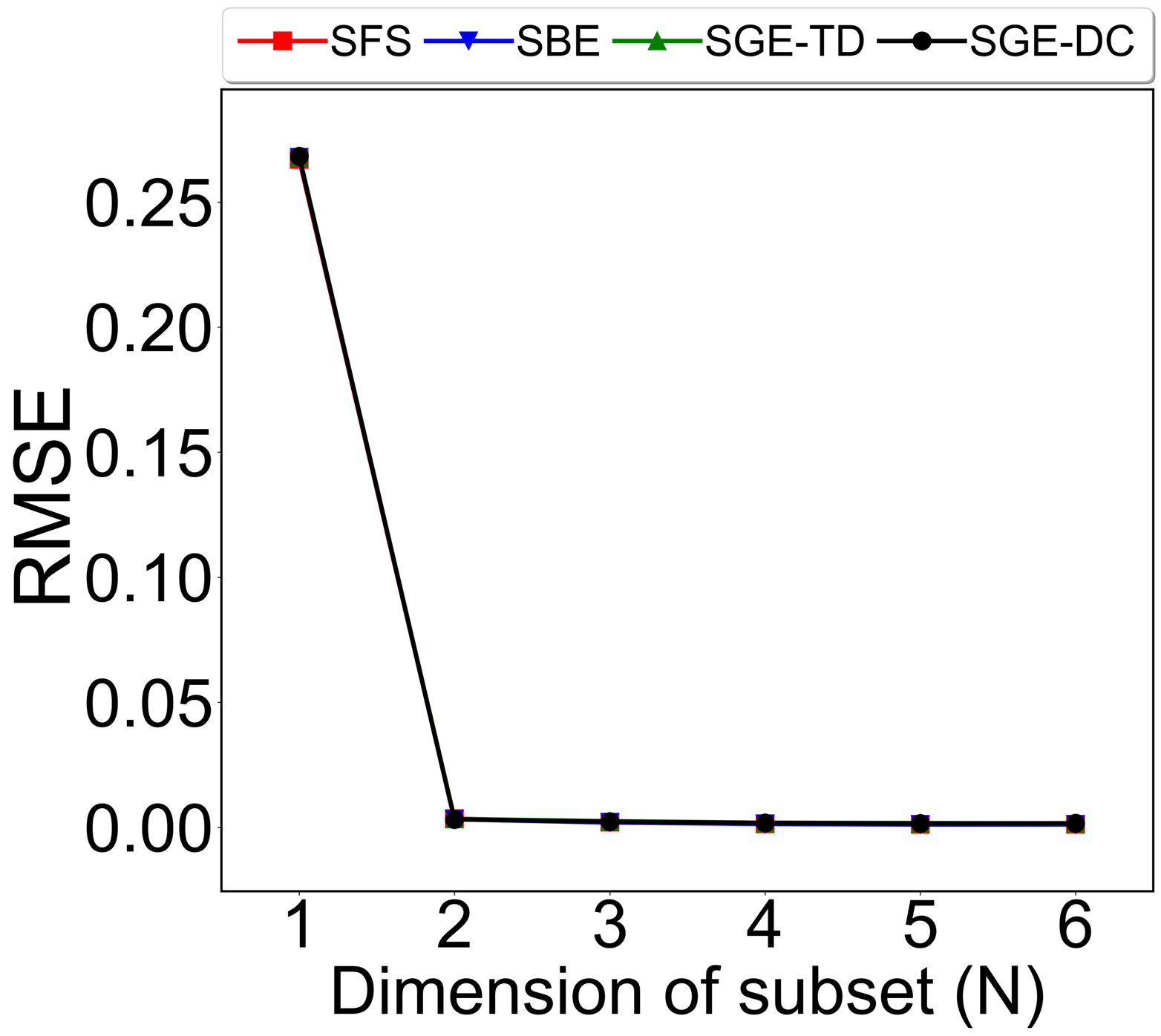}}\subfloat[2L-10N/L]{\centering{}\includegraphics[width=4cm]{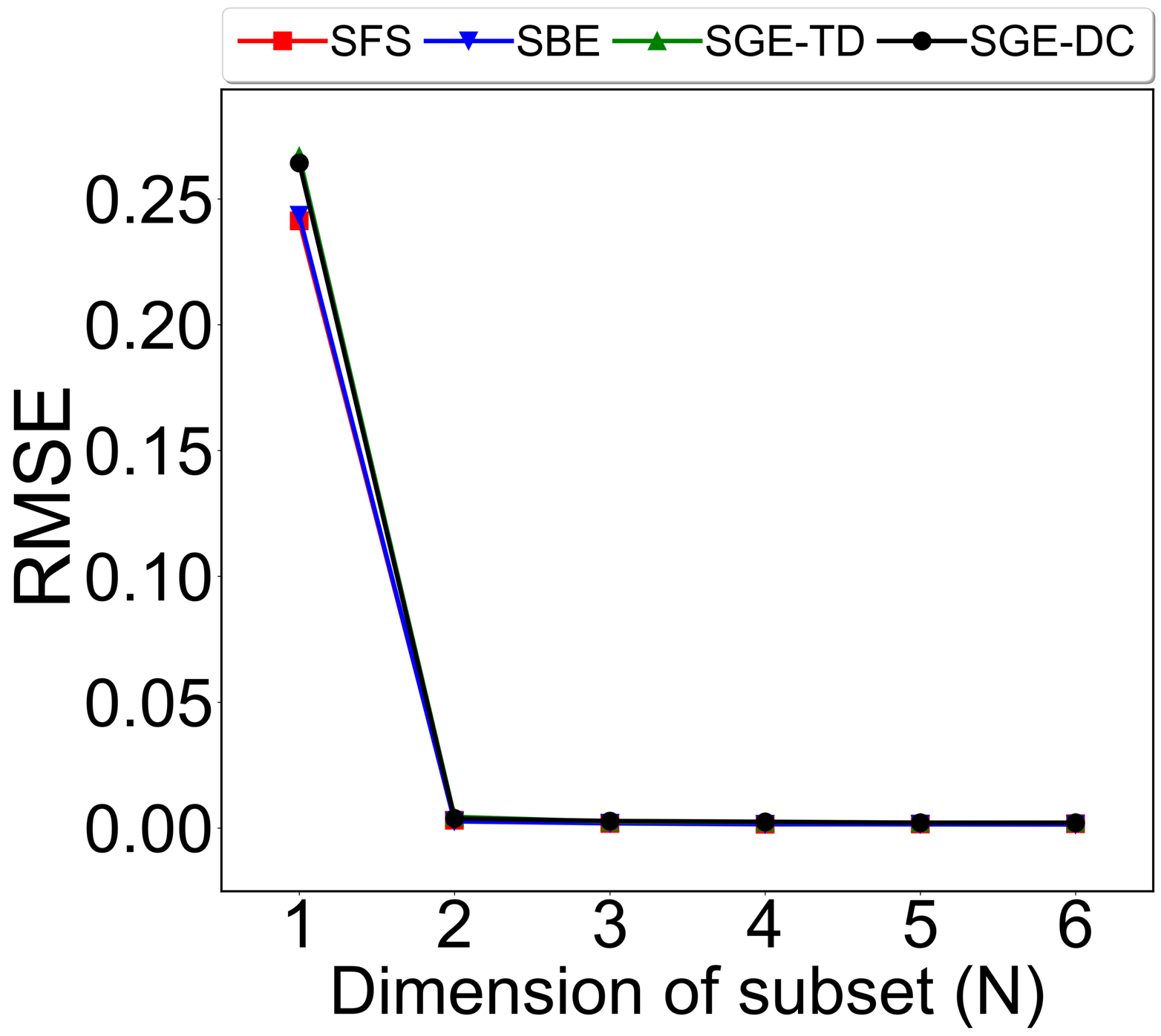}}\subfloat[4L-30N/L]{\centering{}\includegraphics[width=4cm]{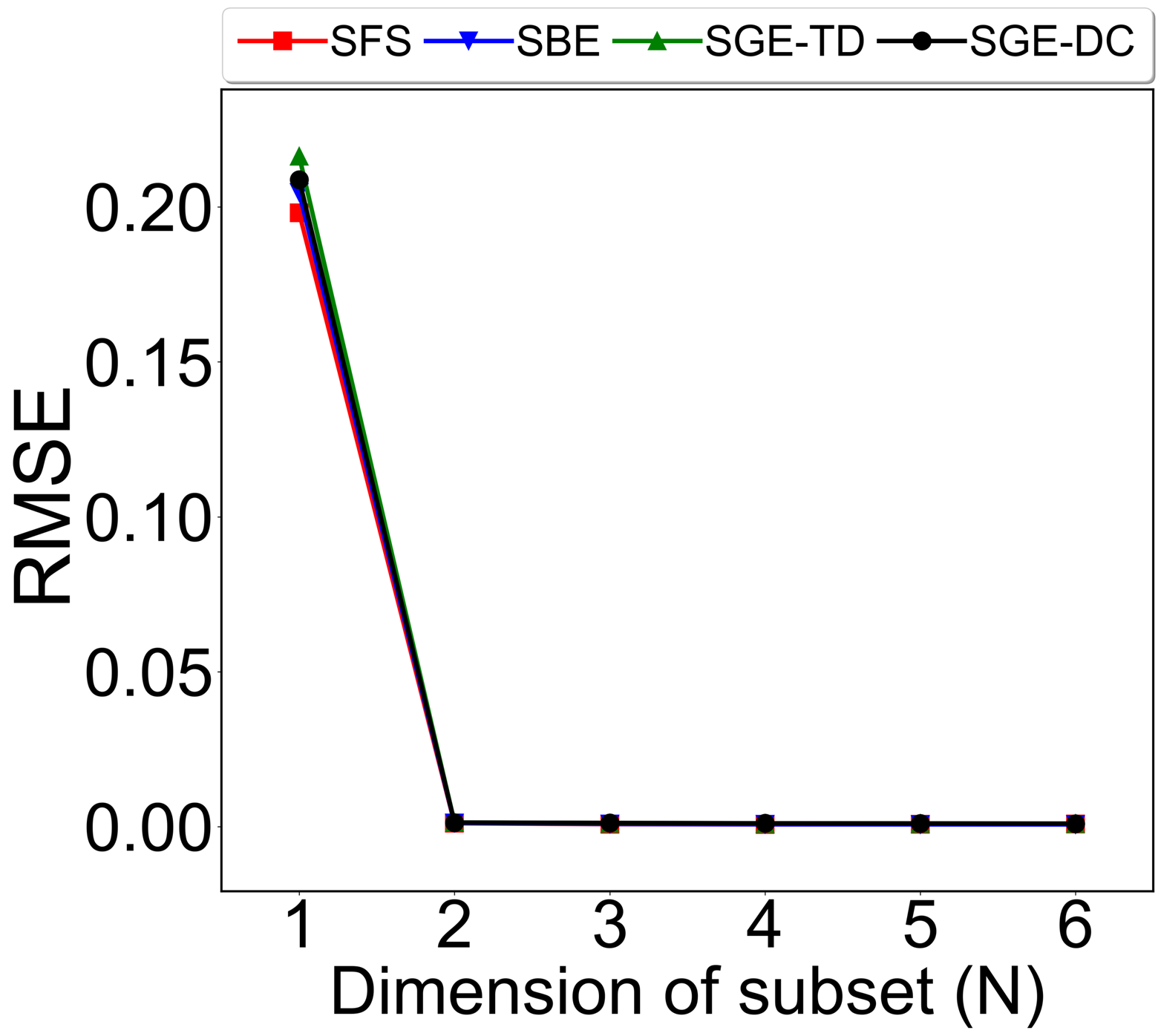}}\caption{RMSE averaged overs trials according to the size of $S_{N}$ for the
manufactured problem.\label{fig:RMSE-comparision}}
\end{figure}
\begin{figure}[H]
\centering{}\subfloat[1L-30N/L]{\centering{}\includegraphics[width=4cm]{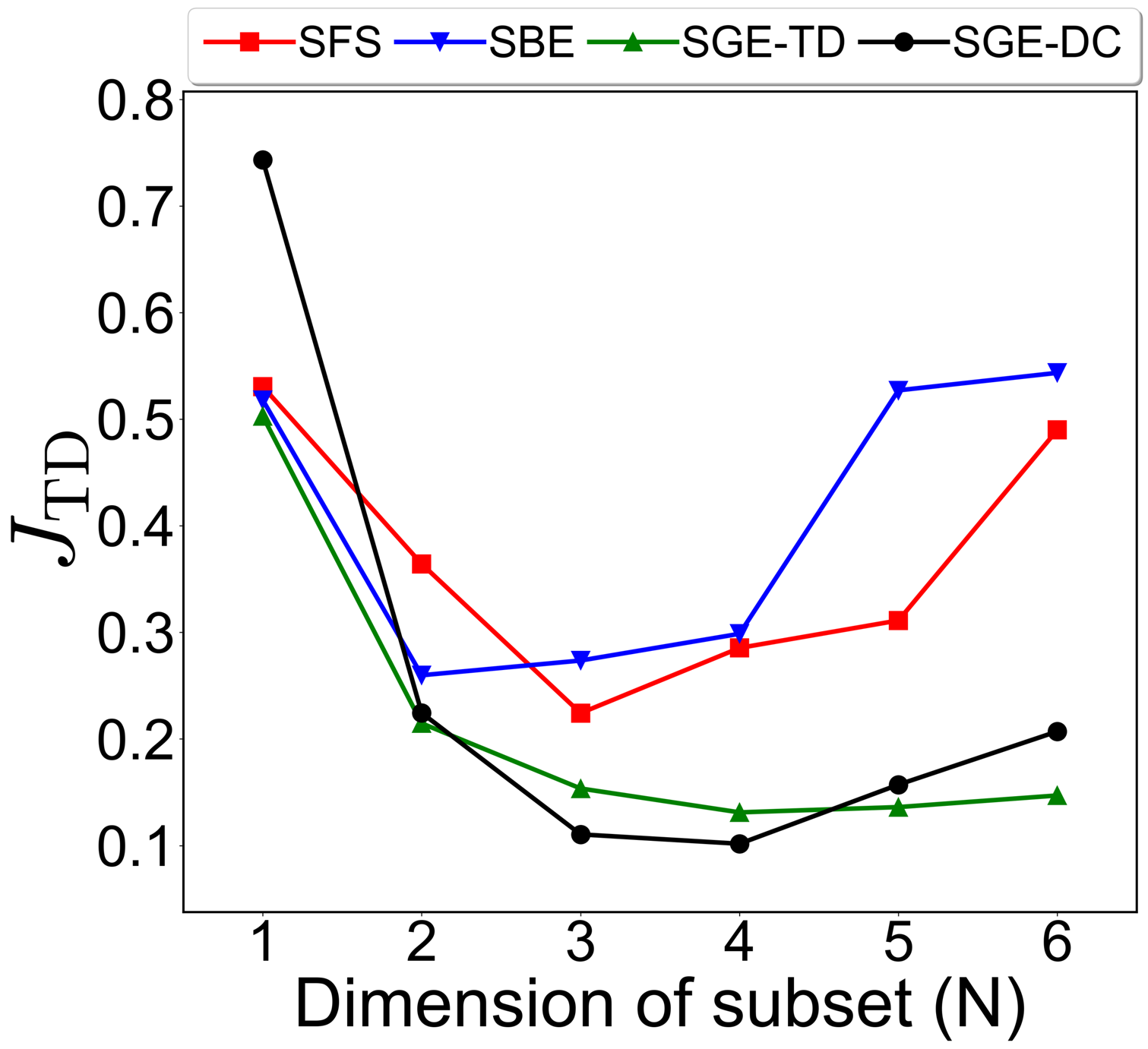}}\subfloat[2L-10N/L]{\centering{}\includegraphics[width=4cm]{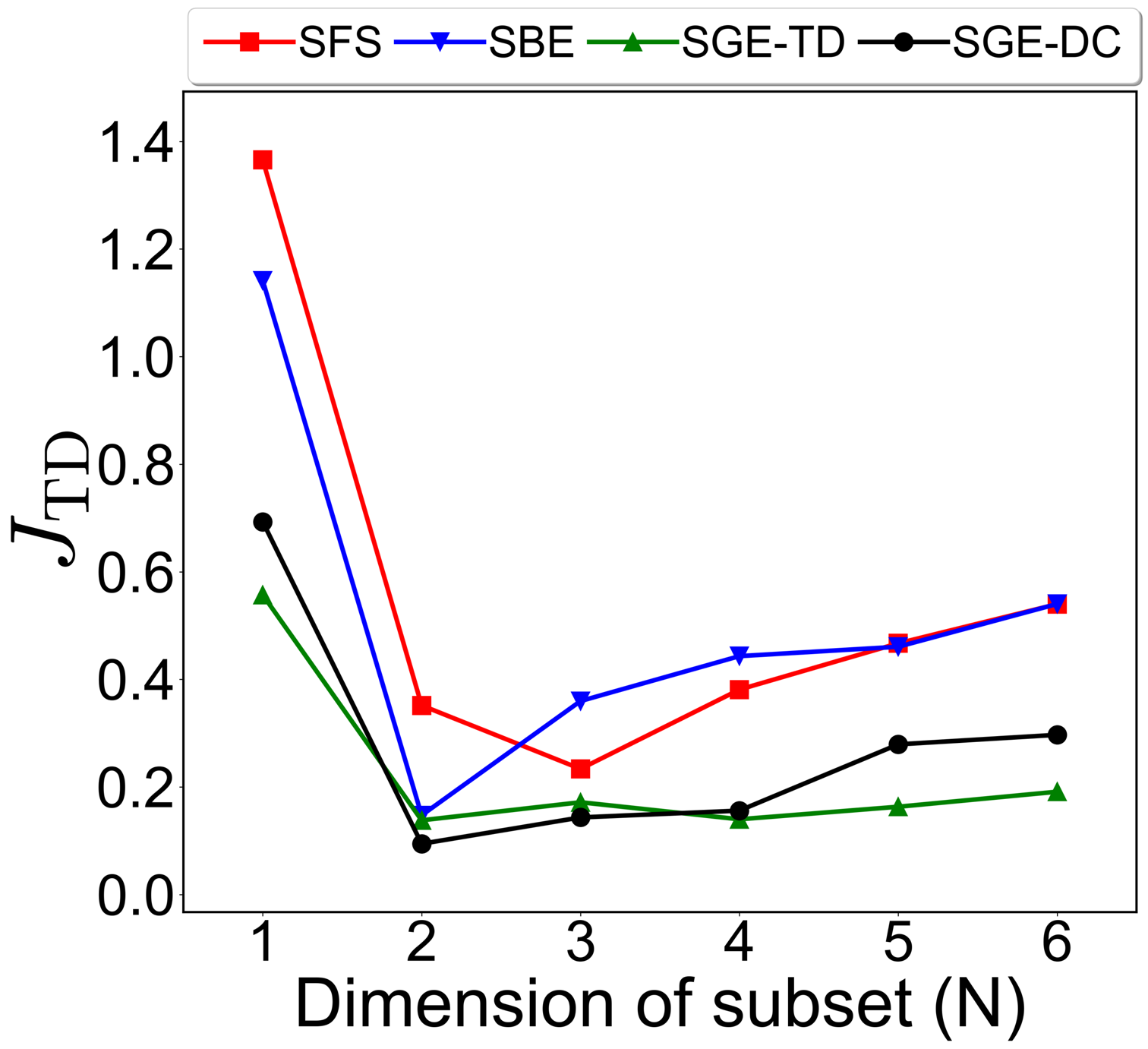}}\subfloat[4L-30N/L]{\centering{}\includegraphics[width=3.9cm]{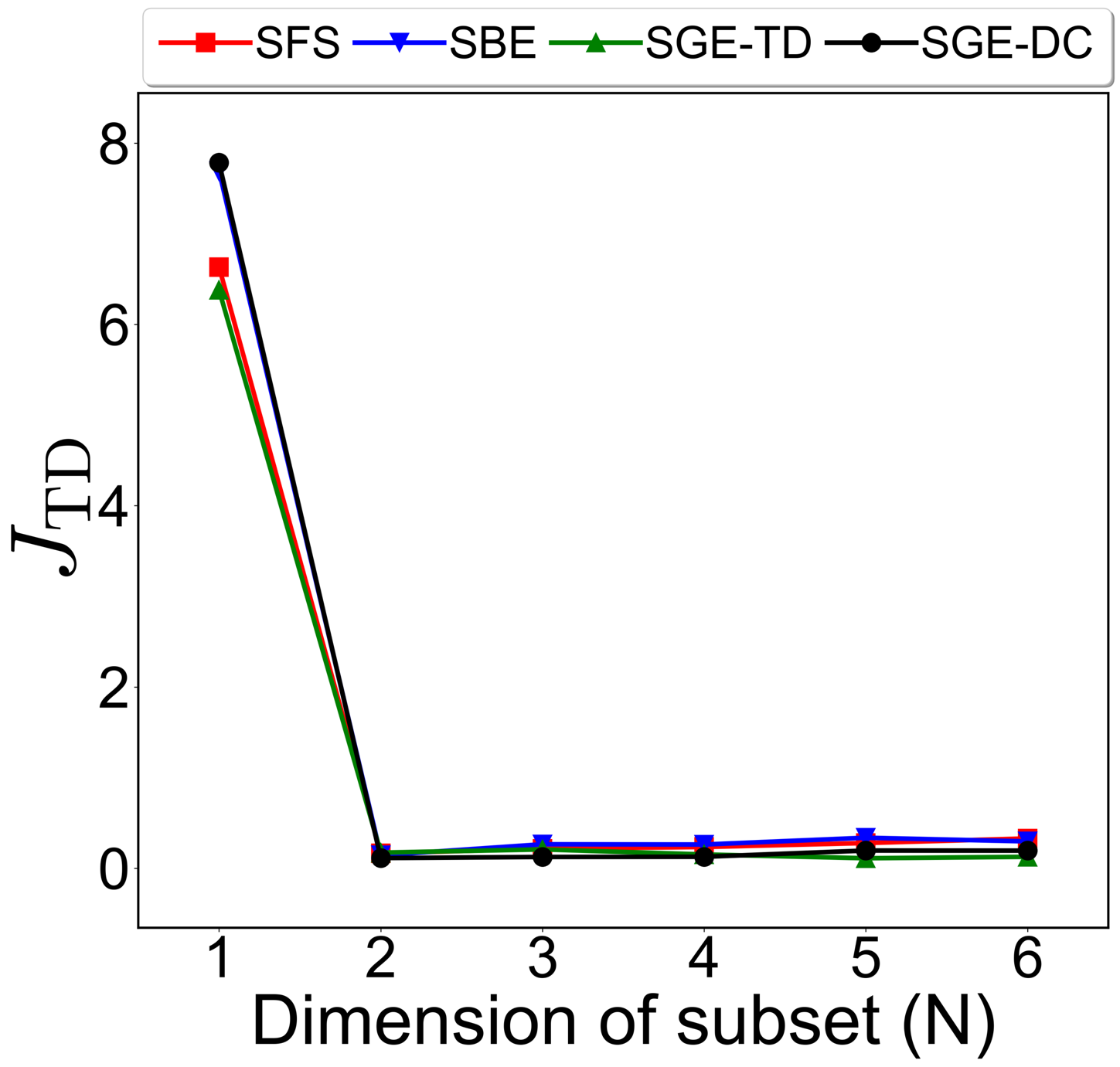}}\caption{$J_{TD}$ averaged overs trials according to the size of $S_{N}$
for the manufactured problem.\label{fig:Minimum-loss-of total derivative toy}}
\end{figure}
\begin{figure}[H]
\centering{}\subfloat[1L-30N/L]{\centering{}\includegraphics[width=4cm]{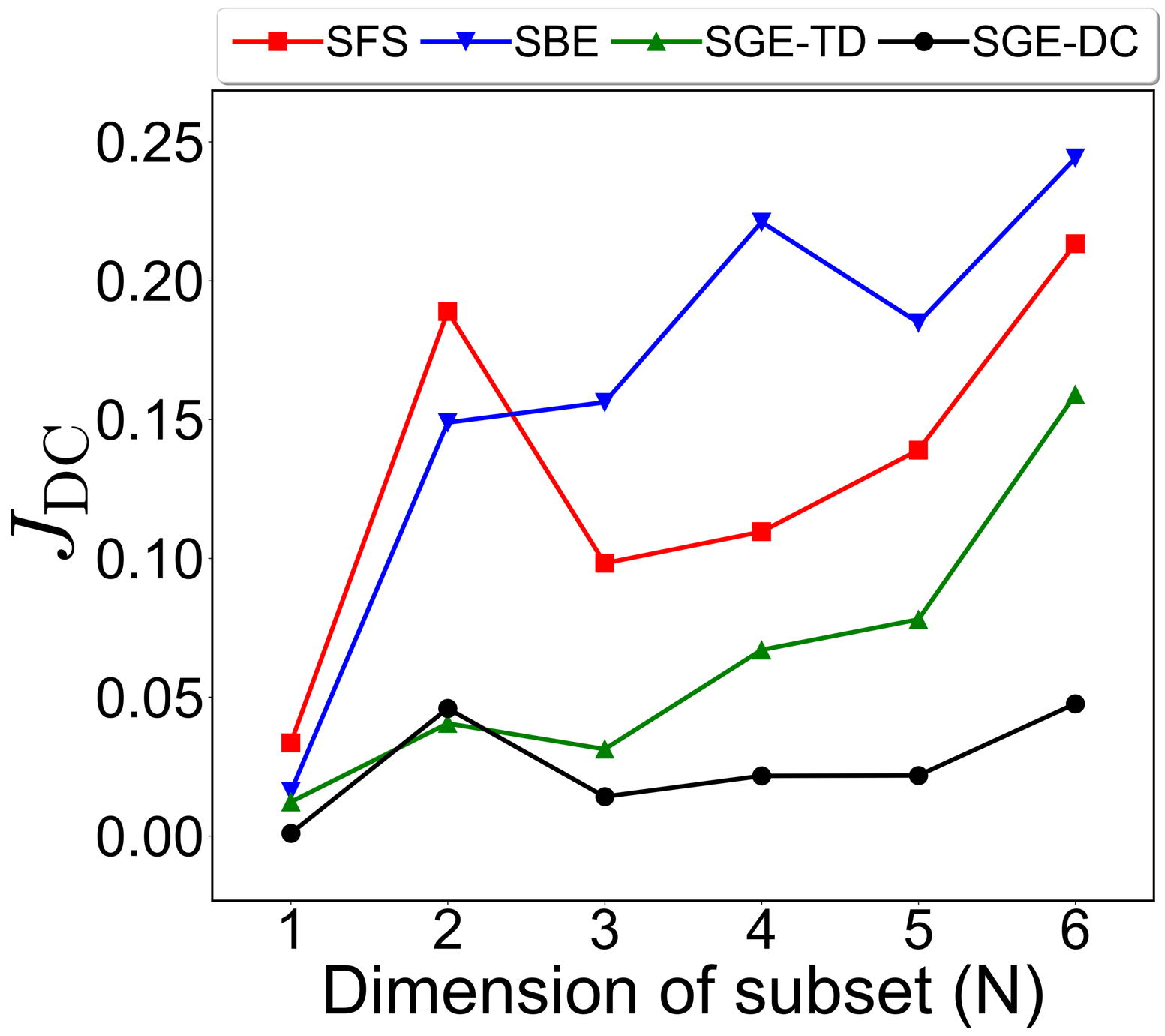}}\subfloat[2L-10N/L]{\centering{}\includegraphics[width=4cm]{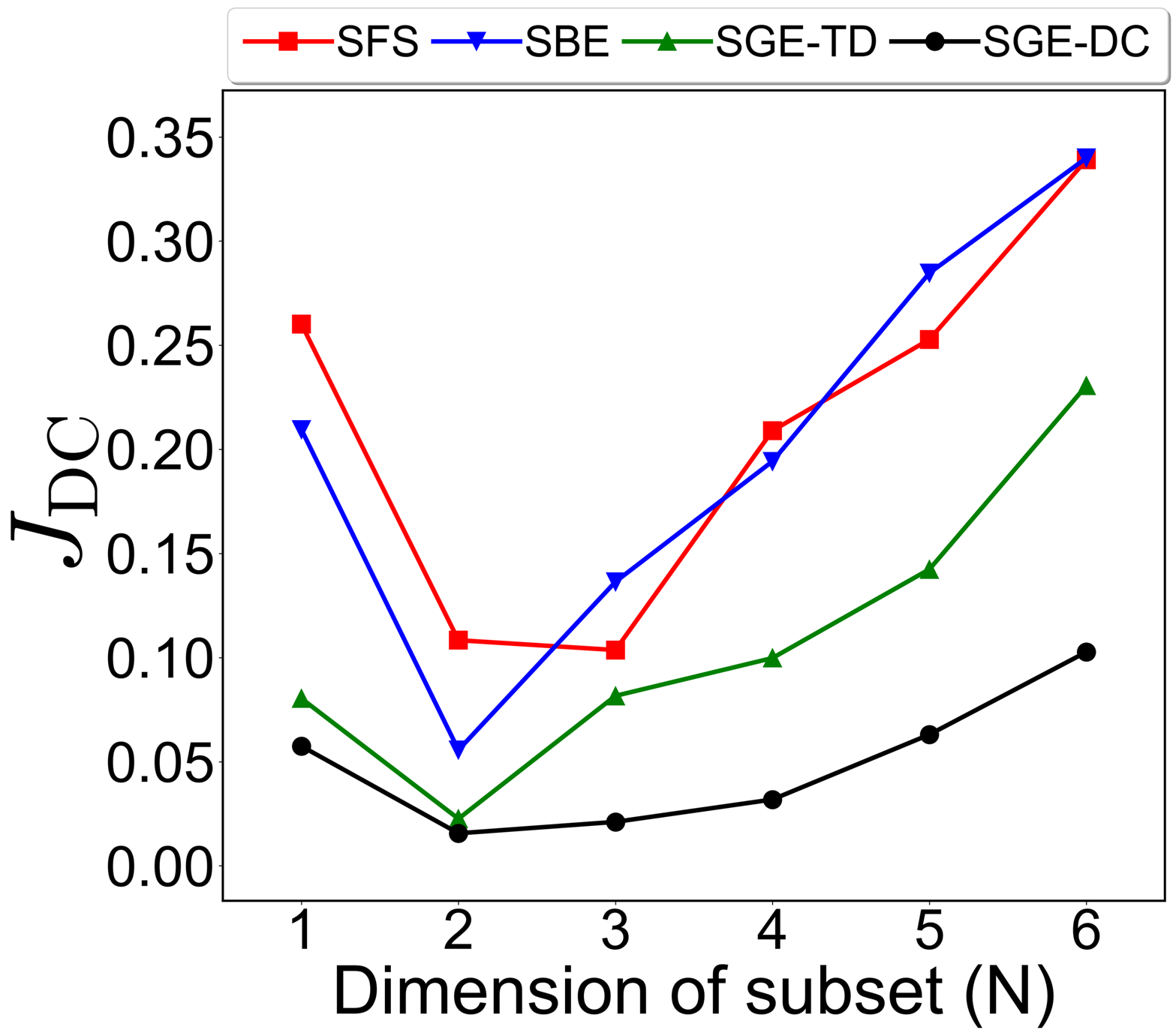}}\subfloat[4L-30N/L]{\centering{}\includegraphics[width=4cm]{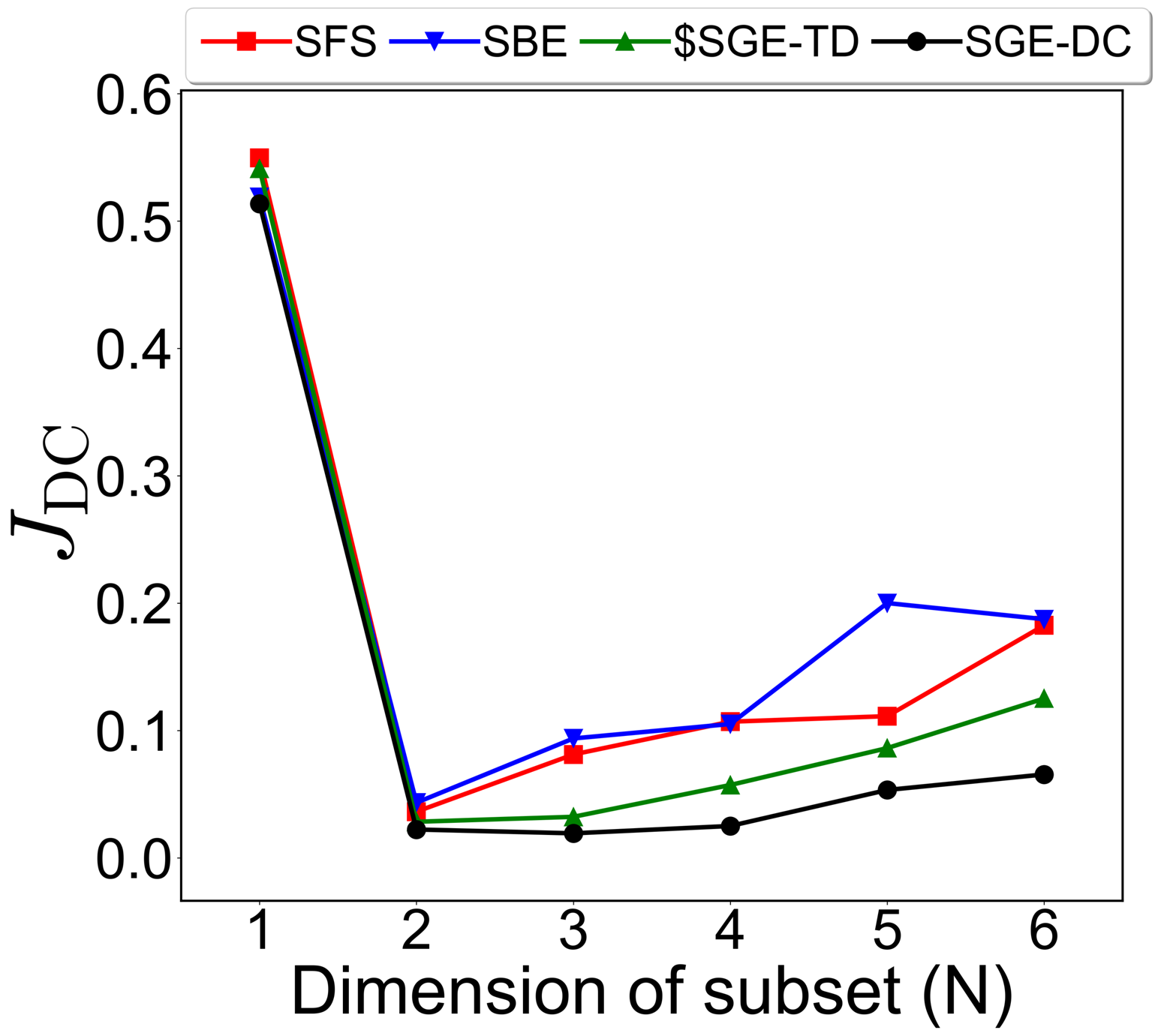}}\caption{$J_{DC}$ averaged overs trials according to the size of $S_{N}$
for the manufactured problem.\label{fig:Directional-consistency toy}}
\end{figure}
Fig.~\ref{fig:RMSE-comparision} shows the RMSE averaged overs trials
according to the size of the subset $S_{N}$ for each method using
well-trained ANNs. The RMSE is minimized from $S_{2}$ regardless
of hyper-parameters for all methods, including SGEs that do not use
the RMSE. Figs.~\ref{fig:Minimum-loss-of total derivative toy} and
\ref{fig:Directional-consistency toy} show $J_{TD}$ (Eq.~\eqref{eq:minimum loss of total derivative})
and $J_{DC}$ (Eq.~\eqref{eq:directional consistency}) averaged
over trials according to the size of $S_{N}$. In most cases, SGE-TD
shows the lowest $J_{TD}$ among all methods, and SGE-DC shows the
lowest $J_{DC}$. The changes of $J_{TD}$ and $J_{DC}$ of SFS and
SBE look similar to those of SGE, but their convergence looks less
clear especially for lower hyper-parameters.

\section{Data-driven modeling of bubble size in turbulent bubbly flows in
a pipe\label{sec: Applications-bubbly}}

In this section, the analysis of the ANN-wrapper methods is extended
to data-driven modeling for turbulent bubbly flows in a pipe. A data-driven
model of the Sauter mean diameter ($D_{sm}$) important in this problem
is considered.

\subsection{Description of physical problem and data-driven modeling\label{subsec:Description-RANS-two phase}}

Turbulent bubbly flows are observed frequently in several engineering
problems such as the boilers, heat exchangers, nuclear reactors, and
various chemical processes \citep{Colombo2015}. In engineering applications,
an Eulerian-Eulerian approach called as the two-fluid model (\citep{Ishii1984,Kim2014},
etc.) is used widely. In an iso-thermal and low Mach number condition,
the governing equations for a RANS simulation are written as 
\begin{equation}
\frac{\partial(\alpha_{q}\rho_{q})}{\partial t}+\nabla\cdot(\alpha_{q}\rho_{q}\vec{u}_{q})=0,\label{eq:eulerian mass conservation-1}
\end{equation}
\begin{equation}
\frac{\partial(\alpha_{q}\rho_{q}\vec{u}_{q})}{\partial t}+\nabla\cdot(\alpha_{q}\rho_{q}\vec{u}_{q}\vec{u}_{q})=-\alpha_{q}\nabla P+\alpha_{q}\rho_{q}\vec{g}+\nabla\cdot(\alpha_{q}\widetilde{\tau_{q}}+\alpha_{q}\widetilde{\tau_{q}}^{t})+\vec{M}_{q},\label{eq:eulerian momentum coservation-1}
\end{equation}
where the subscript $q$ represents either the liquid (L, water) or
gas (G, air) phases, respectively. The symbols $\alpha$, $\rho$,
$\vec{u}$, $\widetilde{\tau}$ and $\widetilde{\tau}^{t}$ denote
the volume fraction, density, velocity vector, molecular and turbulent
stress tensors of a phase, respectively. The symbols $P$ and $\vec{g}$
are the pressure and acceleration gravity vector, respectively. $\vec{M}_{q}$
includes the interfacial momentum exchange terms such as the drag,
shear- and wall-induced lifts, and turbulent dispersion force. Details
on their physical models used in this study are found in \citet{Ishii1979,Sato1981,Ishii1984,Tomiyama2002,Kim2014,Kim2015}
as summarized in \citet{Jung2019}. As noted in \citet{Colombo2016},
a primary factor to model the interfacial momentum exchange terms
is the bubble size called as the Sauter mean diameter ($D_{sm}$).
Due to its physical and practical importance, there exist numerous
studies (e.g. \citep{Hibiki2000,Hibiki2002a,Cheung2007,Lo2012,Nguyen2012,Lin2014,Chuang2015,Schlegel2015},
etc.) on its modeling. It is difficult to develop a model predicting
the bubble size accurately due to complex interaction between the
turbulence and bubbles \citep{Liao2015}. Although there have been
a few efforts on data-driven modeling such as \citet{Shaikh2003},
\citet{MonrosAndreu2013} and \citet{Ma2016}, there exist very limited
studies on machine learning for a closure model to simulate a bubbly
flow. Recently for turbulent flows in a vertical pipe, \citet{Jung2019}
presented modeling of $D_{sm}$ using ANN with experimental database
and a closure approach for RANS simulations. Based on this approach,
we aim to produce a reduced model for $D_{sm}$ from the full parameter
model of \citep{Jung2019}.

\begin{figure}[H]
\begin{centering}
\includegraphics[height=4.5cm]{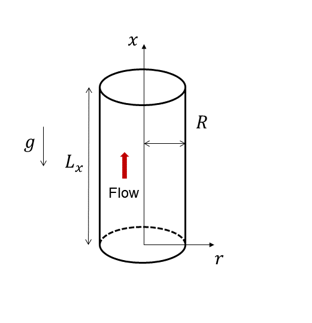}
\par\end{centering}
\caption{A schematic diagram of a turbulent bubbly flow in a vertical round
pipe.\label{fig:Schematic diagram of experiments}}
\end{figure}
\begin{table}[H]
\centering{}%
\begin{tabular}{|>{\centering}p{1.8cm}|>{\centering}p{1.8cm}|>{\centering}p{1.8cm}|>{\centering}p{1.8cm}|>{\centering}p{1.8cm}|}
\hline 
Symbol & $\rho_{L}$, $\rho_{G}$ /

$\mu_{L}$, $\mu_{G}$ & $u_{L}$, $u_{G}$ (m/s) & $J_{L}$, $J_{G}$ (m/s) & $\alpha_{G}$\tabularnewline
\hline 
Meaning & Density / Viscosity & Local mean axial velocity & Inlet superficial velocity & Local void fraction\tabularnewline
\hline 
Value & Constants for water and air & 0.41-2.795,

0.647-3.098 & 0.491-2.01,

0.0275-0.321 & 0.049-0.259\tabularnewline
\hline 
\end{tabular}\caption{Variables in the database of turbulent bubbly flows in a pipe.\label{tab:database variables of bubbly}}
\end{table}
There are a few experimental studies on turbulent bubbly flow in vertical
pipes (\citep{Hibiki1998,Hibiki2001,MonrosAndreu2013,Doup2013}) with
a schematic diagram in Fig.~\ref{fig:Schematic diagram of experiments}.
The data from these studies include a few common variables listed
as
\begin{itemize}
\item Material properties : $\rho_{G}$, $\rho_{L}$, $\mu_{G}$, $\mu_{L}$
\item Flow variables : $u_{G}$, $u_{L}$, $J_{G}$, $J_{L}$, $\alpha_{G}$
\end{itemize}
The physical meanings and values of these variables are presented
in Table~\ref{tab:database variables of bubbly}. A database is built
using the results of 15 flow cases in the literature. Further details
on the database are available in \citet{Jung2019}. We employ the
non-dimensionalized $D_{sm}$ model in \citep{Jung2019} as the full
parameter model, i.e.,
\begin{equation}
\frac{D_{sm}}{R}=f\left(Re_{L},\frac{J_{G}}{J_{L}},\alpha_{G},\frac{u_{L}}{J_{L}},\frac{u_{G}}{J_{L}}\right),\label{eq:Final model - bubble}
\end{equation}
which is regarded as $Y=f\left(X\right)$. It is noted that ($\alpha_{G}$,
$u_{L}/J_{L}$, $u_{G}/J_{L}$) are the Reynolds-averaged spatially
varying (local) variables. ($Re_{L}$, $J_{G}/J_{L}$) are the conditions
of each flow case.

\begin{figure}[H]
\centering{}\subfloat[Heat map of the correlation coefficient]{\centering{}\includegraphics[height=5cm]{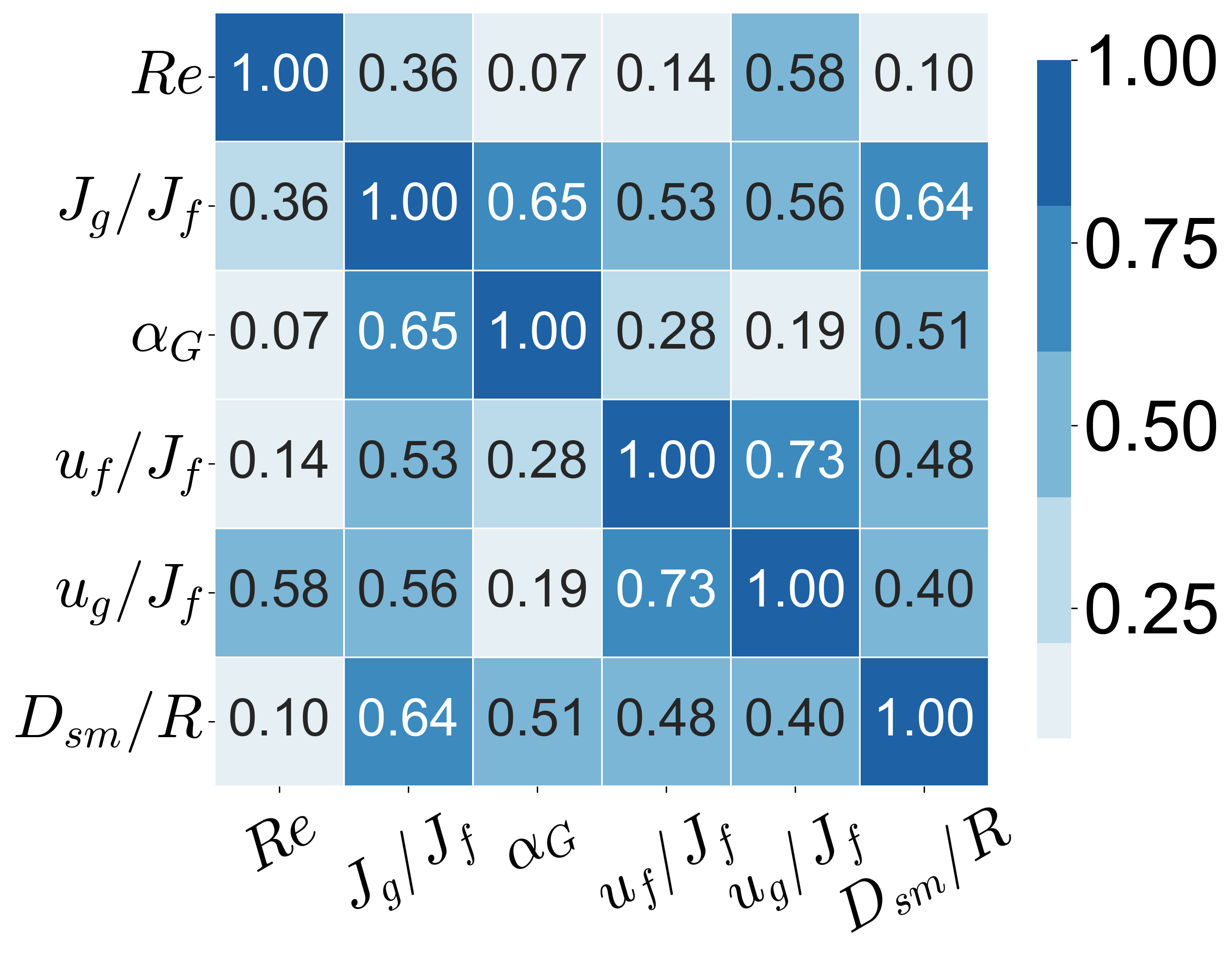}}\subfloat[Correlation between inputs and $D_{sm}$]{\begin{centering}
\includegraphics[height=5cm]{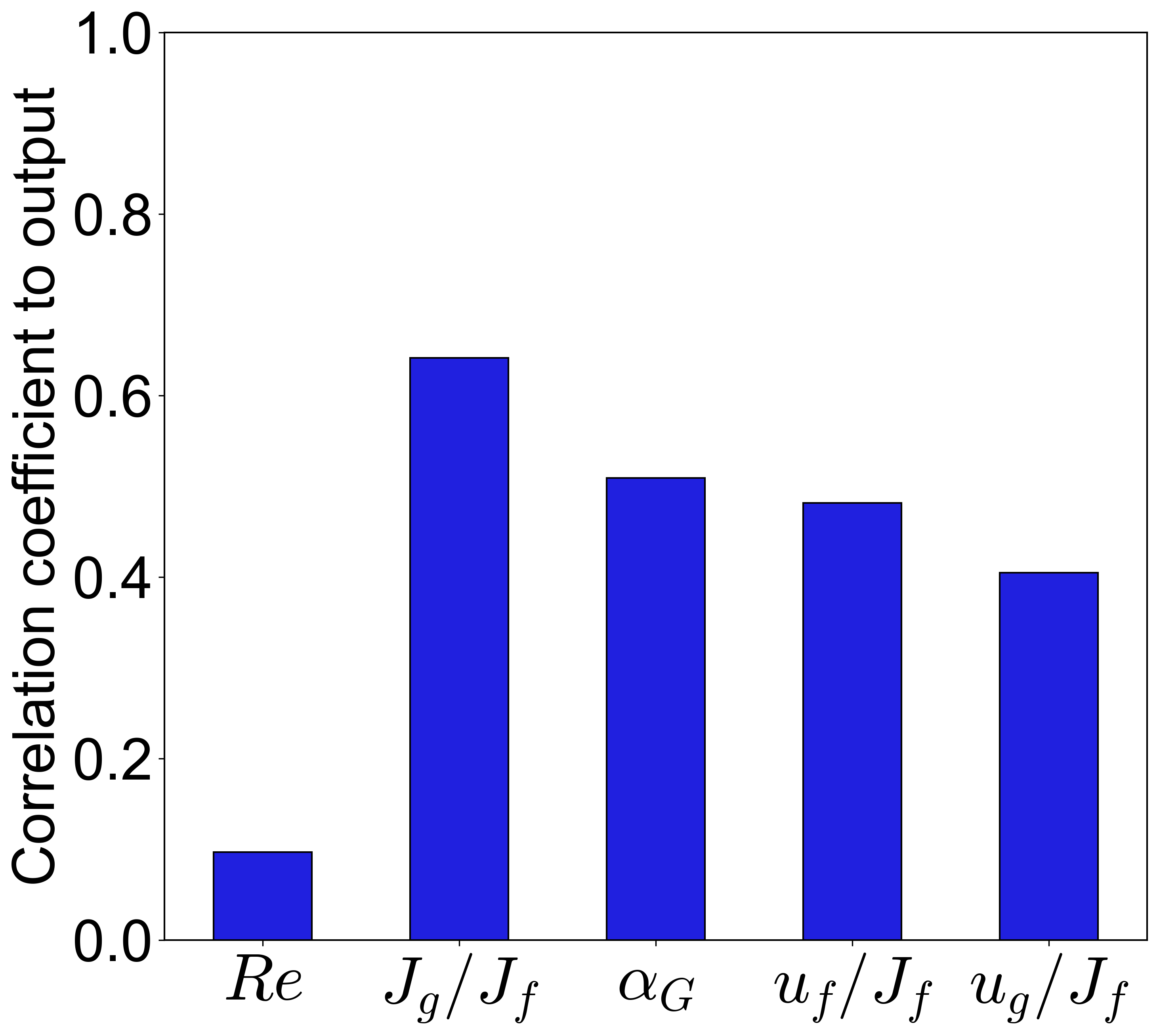}
\par\end{centering}
}\caption{\textcolor{black}{Pearson correlation coefficients} of turbulent bubbly
flows in a pipe.\label{fig:Correlation - bubble}}
\end{figure}
Fig.~\ref{fig:Correlation - bubble} shows the Pearson correlation
coefficients among the parameters from the database. In Fig.~\ref{fig:Correlation - bubble}(a),
it is shown that relatively strong inter-dependency between $u_{L}/J_{L}$
and $u_{G}/J_{L}$ as well as $J_{G}/J_{L}$ and $\alpha_{G}$. It
is anticipated that two-phase velocities are correlated closely. A
relatively large gas-phase velocity at the inlet (i.e. a large $J_{G}/J_{L}$)
leads to an increased void fraction in the downstream. From these,
the overlapping effects are assumed inside each pair of ($u_{L}/J_{L}$,
$u_{G}/J_{L}$) and ($J_{G}/J_{L}$, $\alpha_{G}$). Fig.~\ref{fig:Correlation - bubble}(b)
shows that $D_{sm}$ shows the highest correlation with $J_{G}/J_{L}$
followed by $\alpha_{G}$.

It is noted that this problem has a relatively small number of the
parameters and may not be ideal to investigate the effects of dimensionality
reduction methods. This database has a range of operating conditions
from several accredited sources but the data resolution is rather
coarse. Thus, it can be useful to test practical performance of the
ANN-wrapper methods including SGE for a sparse database.

\subsection{Results of dimensionality reduction for $D_{sm}$ model\label{subsec: parameter selection - bubble}}

The ANN-wrapper methods are applied to the full parameter $D_{sm}$
model (Eq.~\eqref{eq:Final model - bubble}) to obtain reduced parameter
subsets. The number of training epochs is determined using the K-fold
technique \citep{Breiman1992} with 5-fold cross validation \citep{Kohavi1995}.
To evaluate the averaged performance and the consistency over selection
trials (e.g., PEM-CoT1 and PEM-CoT2), the parameter selections are
tried for 20-35 times.

\begin{figure}[H]
\centering{}\includegraphics[height=5cm]{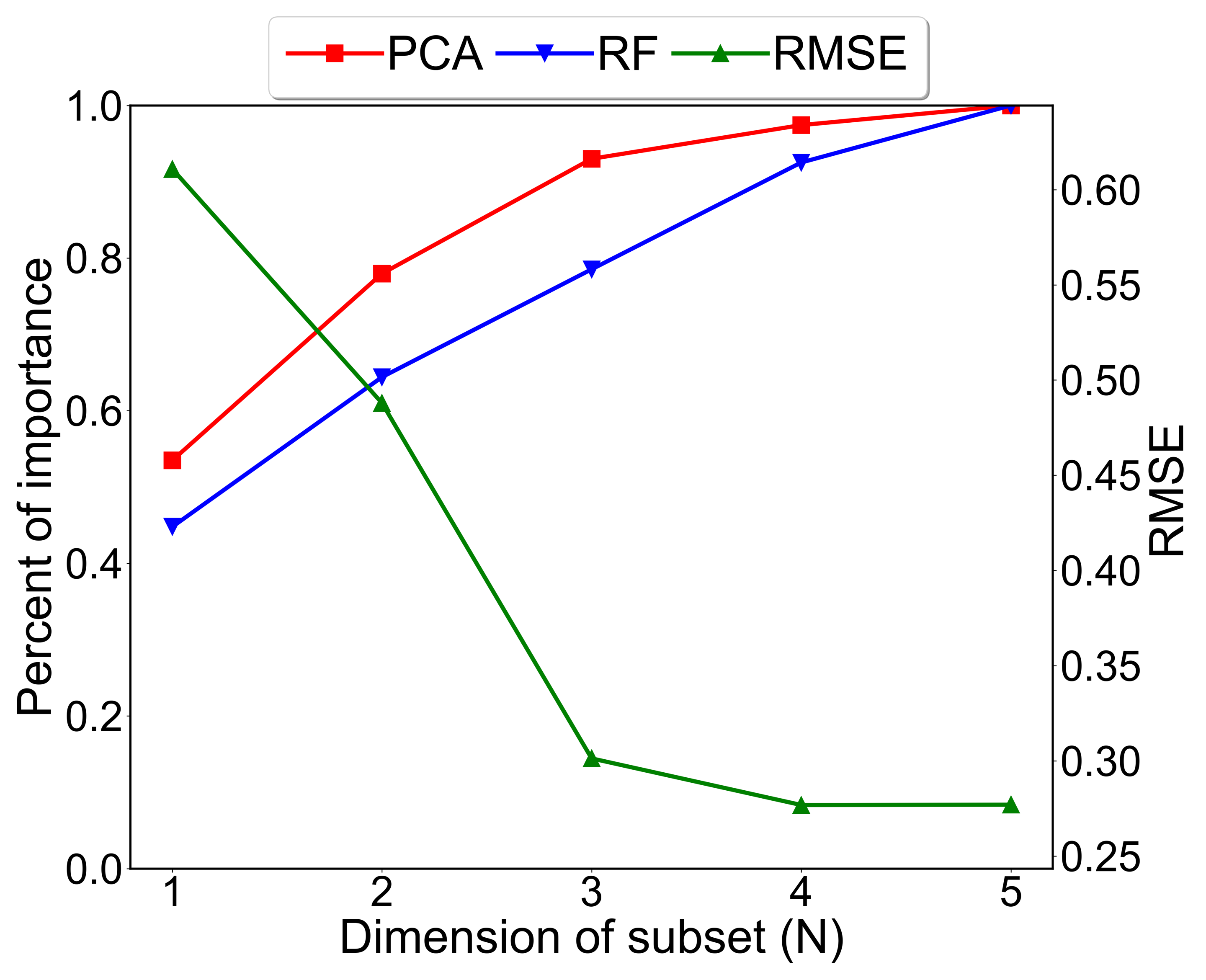}\caption{\textcolor{black}{Accumulated feature importance of the RF method
and PCA compared with the minimum RMSE (validation) of parameter combinations
over the number of components}.\label{fig:percent-of-information}}
\end{figure}
To estimate the necessary dimension of a reduced $D_{sm}$ model,
Fig.~\ref{fig:percent-of-information} shows the accumulated feature
importance of the random forest (RF) method and PCA over the number
of components. The green line shows the minimum RMSE of all parameter
combinations with a specific number of input parameters. It is shown
that the RF method and PCA needs 3-4 parameters for sufficiently converged
(> 90\% in importance) model prediction, similarly to the RMSE trend.
Thus, reduced parameter subsets of 3 parameters (i.e. $S_{3}$) are
considered below.

\begin{table}[H]
\centering{}%
\begin{tabular}{|c|c|c|c|c|c|c|}
\hline 
 & \multicolumn{2}{c|}{1L-30N/L} & \multicolumn{2}{c|}{2L-30N/L} & \multicolumn{2}{c|}{4L-30N/L}\tabularnewline
\hline 
Method & A-T & W-T & A-T & W-T & A-T & W-T\tabularnewline
\hline 
SFS & 1 & 1 & 0.883 & 0.889 & 1 & 1\tabularnewline
\hline 
SBE & 1 & 1 & 1 & 1 & 1 & 1\tabularnewline
\hline 
SGE-TD & 1 & 1 & 1 & 1 & 1 & 1\tabularnewline
\hline 
SGE-DC & 1 & 1 & 0.994 & 1 & 1 & 1\tabularnewline
\hline 
\end{tabular}\caption{PEM-CoT1 of \textcolor{black}{$S_{3}$} over \textcolor{black}{20-35
selection trials }using different hyper-parameters for turbulent bubbly
flows in a pipe (A-T: all trained ANNs (35), W-T: well-trained ANNs
(20)).\label{tab: bubbly: CoT1}}
\end{table}

\begin{table}[H]
\centering{}%
\begin{tabular}{|c|c|c|c|c|c|c|}
\hline 
 & \multicolumn{2}{c|}{1L-30N/L} & \multicolumn{2}{c|}{2L-30N/L} & \multicolumn{2}{c|}{4L-30N/L}\tabularnewline
\hline 
Method & A-T & W-T & A-T & W-T & A-T & W-T\tabularnewline
\hline 
SFS & 1 & 1 & 0.657 & 0.85 & 1 & 1\tabularnewline
\hline 
SBE & 1 & 1 & 1 & 1 & 1 & 1\tabularnewline
\hline 
SGE-TD & 1 & 1 & 1 & 1 & 1 & 1\tabularnewline
\hline 
SGE-DC & 1 & 1 & 0.971 & 1 & 1 & 1\tabularnewline
\hline 
\end{tabular}\caption{PEM-CoT2 of \textcolor{black}{$S_{3}$} over \textcolor{black}{20-35
selection trials }using different hyper-parameters for turbulent bubbly
flows in a pipe (A-T: all trained ANNs (35), W-T: well-trained ANNs
(20)).\label{tab: bubbly: CoT2}}
\end{table}

To examine the CoT of the parameter selection, Tables~\ref{tab: bubbly: CoT1}
and \ref{tab: bubbly: CoT2} show PEM-CoT1 (Eq.~\eqref{eq: metric of stability 1})
and PEM-CoT2 (Eq.~\eqref{eq:metric of stability 2}) metrics of $S_{3}$
over 20-35 selection trials using different ANN-wrapper methods, hyper-parameters,
and trained ANNs. With 2L-30N/L, SFS show relatively low CoT metrics
even with well-trained ANNs. The CoT metrics of SGE-DC looks slightly
reduced with all trained ANNs but becomes ideal (i.e. 1) with well-trained
ANNs. SBE and SGE-TD show the perfect CoT for all cases.

\begin{table}[H]
\centering{}%
\begin{tabular}{|c|c|c|c|}
\hline 
Method & 1L-30N/L & 2L-30N/L & 4L-30N/L\tabularnewline
\hline 
SFS & $Re,\ J_{g}/J_{f},\ u_{f}/J_{f}$ & $Re,\ J_{g}/J_{f},\ u_{g}/J_{f}$ & $Re,\ J_{g}/J_{f},\ u_{g}/J_{f}$\tabularnewline
\hline 
SBE & $\alpha_{G},\ Re,\ u_{g}/J_{f}$ & $\alpha_{G},\ Re,\ u_{g}/J_{f}$ & $\alpha_{G},\ Re,\ u_{g}/J_{f}$\tabularnewline
\hline 
SGE-TD & $\alpha_{G},\ Re,\ u_{g}/J_{f}$ & $\alpha_{G},\ Re,\ u_{g}/J_{f}$ & $\alpha_{G},\ Re,\ u_{f}/J_{f}$\tabularnewline
\hline 
SGE-DC & $\alpha_{G},\ Re,\ u_{g}/J_{f}$ & $\alpha_{G},\ Re,\ u_{g}/J_{f}$ & $\alpha_{G},\ Re,\ u_{f}/J_{f}$\tabularnewline
\hline 
\end{tabular}\caption{List of 3 parameters with the highest selection possibilities from
$S_{3}$ over\textcolor{black}{{} selection trials} for turbulent bubbly
flows in a pipe.\label{tab:Table-of-selected parameters}}
\end{table}
Table~\ref{tab:Table-of-selected parameters} show the list of 3
parameters with the highest selection possibilities from $S_{3}$.
SFS does not select $\alpha_{G}$ that is known to be a primary factor
for $D_{sm}$. Instead of $\alpha_{G}$, SFS selects $J_{G}/J_{L}$
correlated to $\alpha_{G}$, but $J_{G}/J_{L}$ is not a spatially
varying parameter and may be less effective to model $D_{sm}$. For
all cases, either $u_{f}/J_{f}$ or $u_{g}/J_{f}$ is selected due
to their close similarity.

\begin{figure}[H]
\centering{}\subfloat[1L-30N/L]{\centering{}\includegraphics[width=4cm]{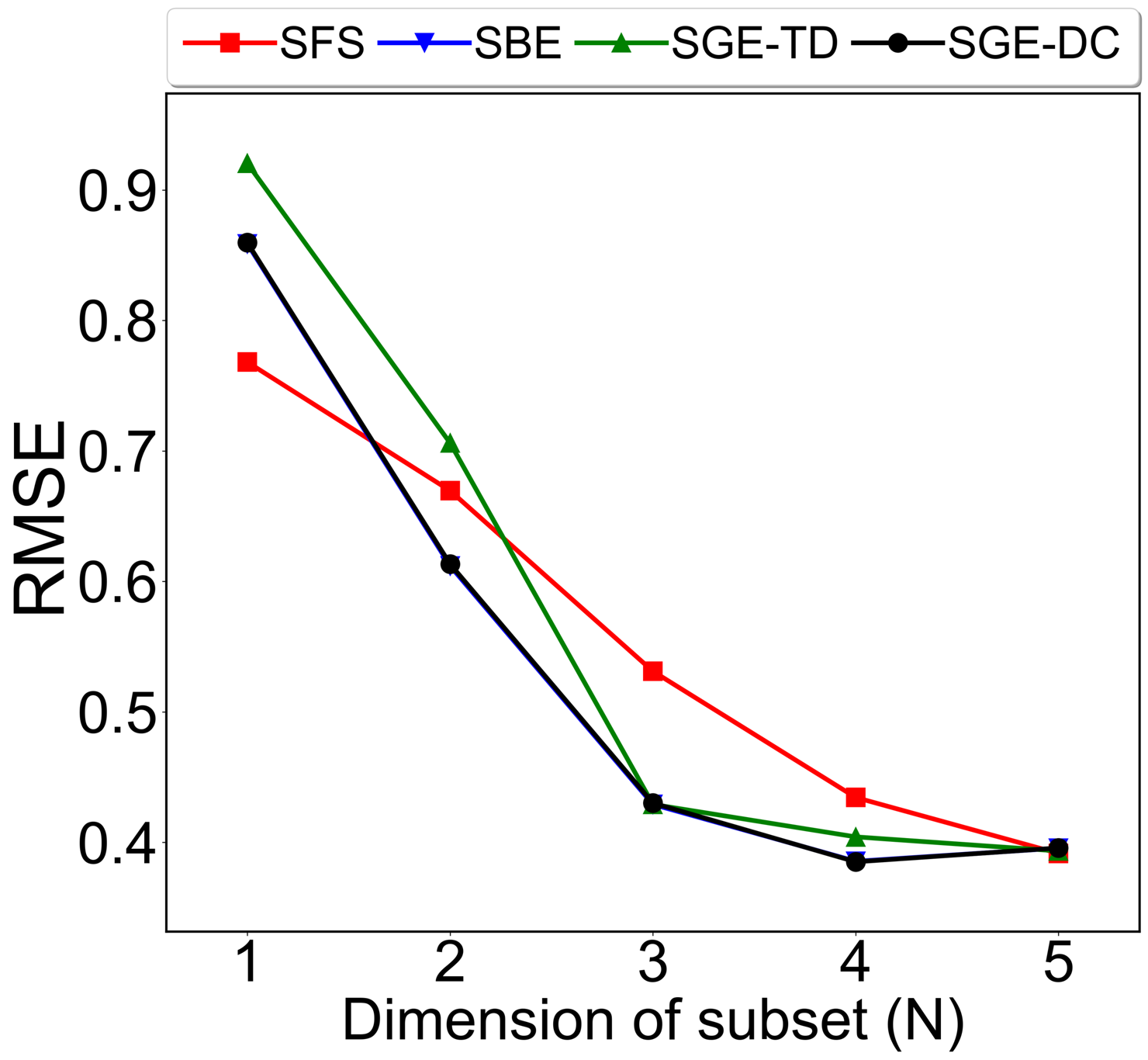}}\subfloat[2L-30N/L]{\centering{}\includegraphics[width=4cm]{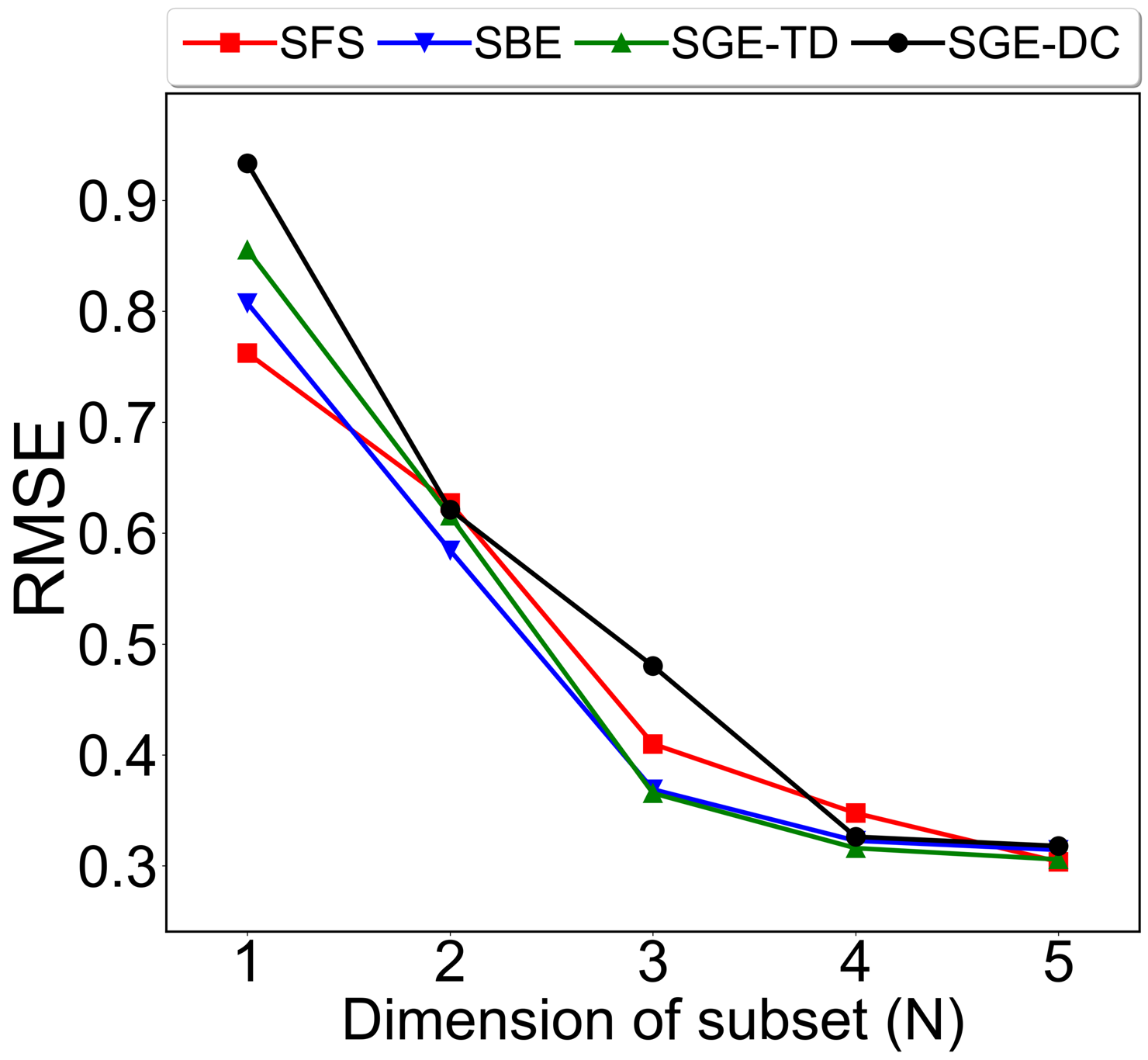}}\subfloat[4L-30N/L]{\centering{}\includegraphics[width=4cm]{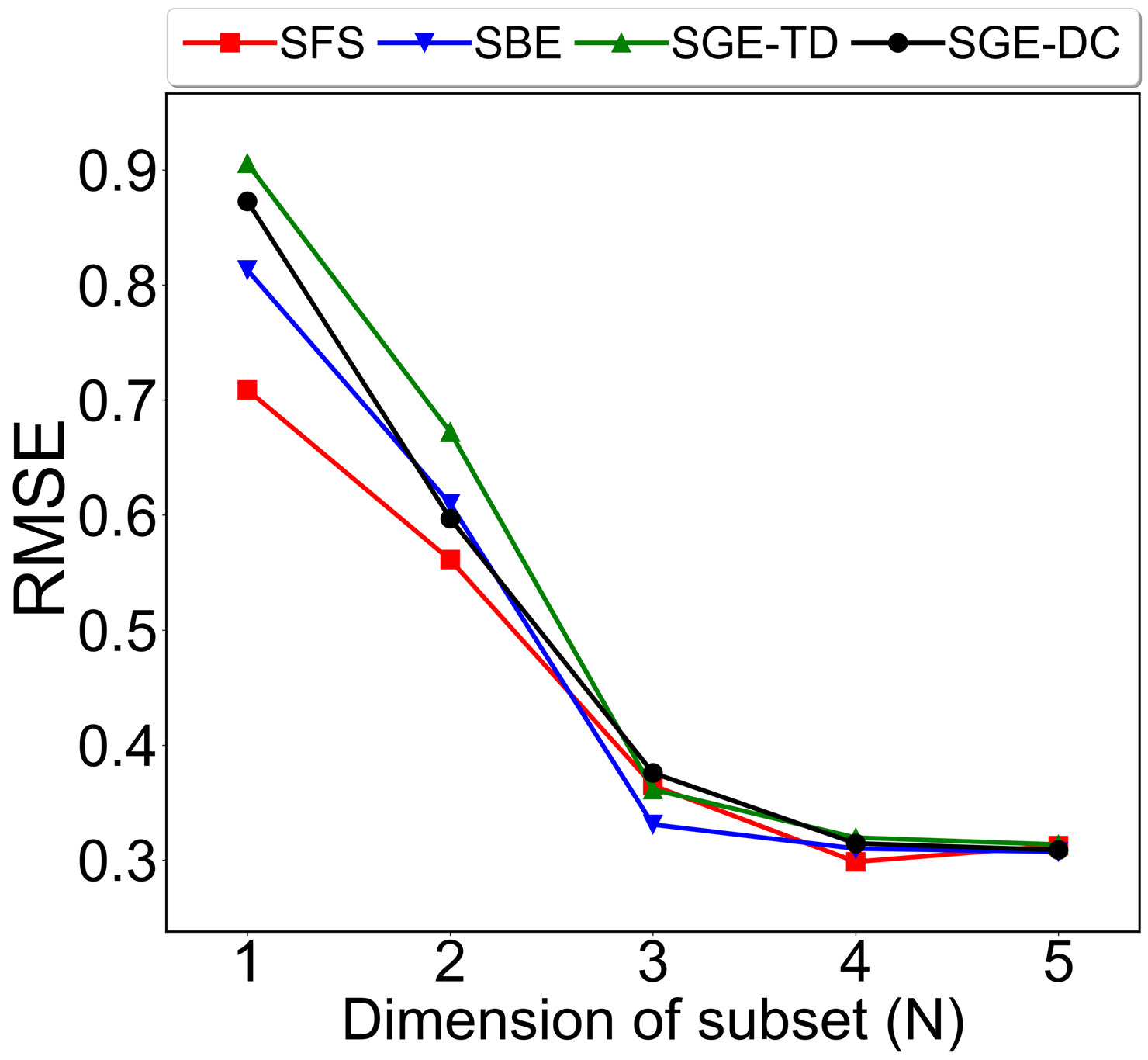}}\caption{RMSE \textcolor{black}{(validation)} averaged overs trials according
to the size of $S_{N}$ for turbulent bubbly flows in a pipe.\label{fig:RMSE-comparision-bubble}}
\end{figure}
\begin{figure}[H]
\centering{}\subfloat[1L-30N/L]{\centering{}\includegraphics[width=4cm]{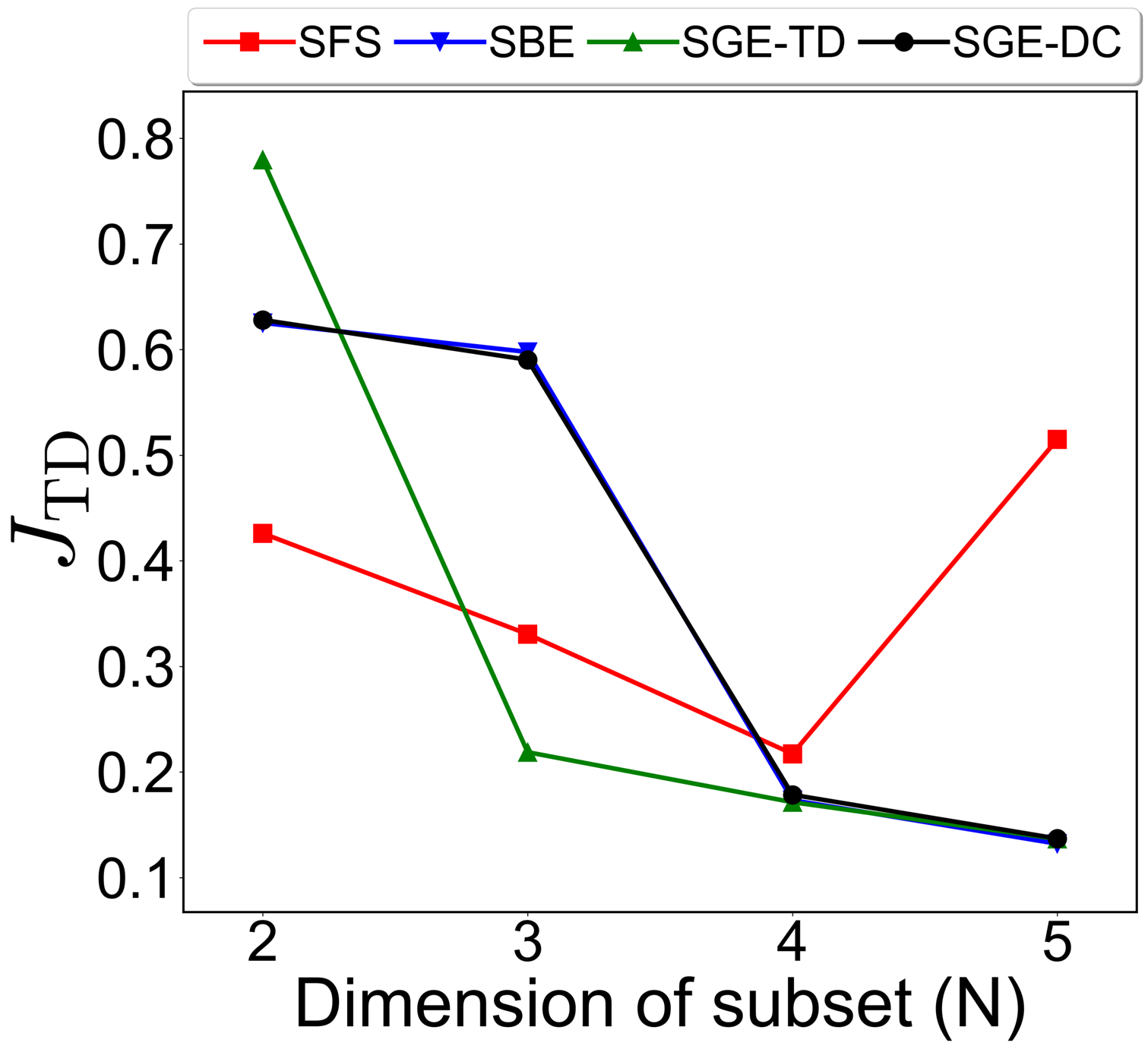}}\subfloat[2L-30N/L]{\centering{}\includegraphics[width=4cm]{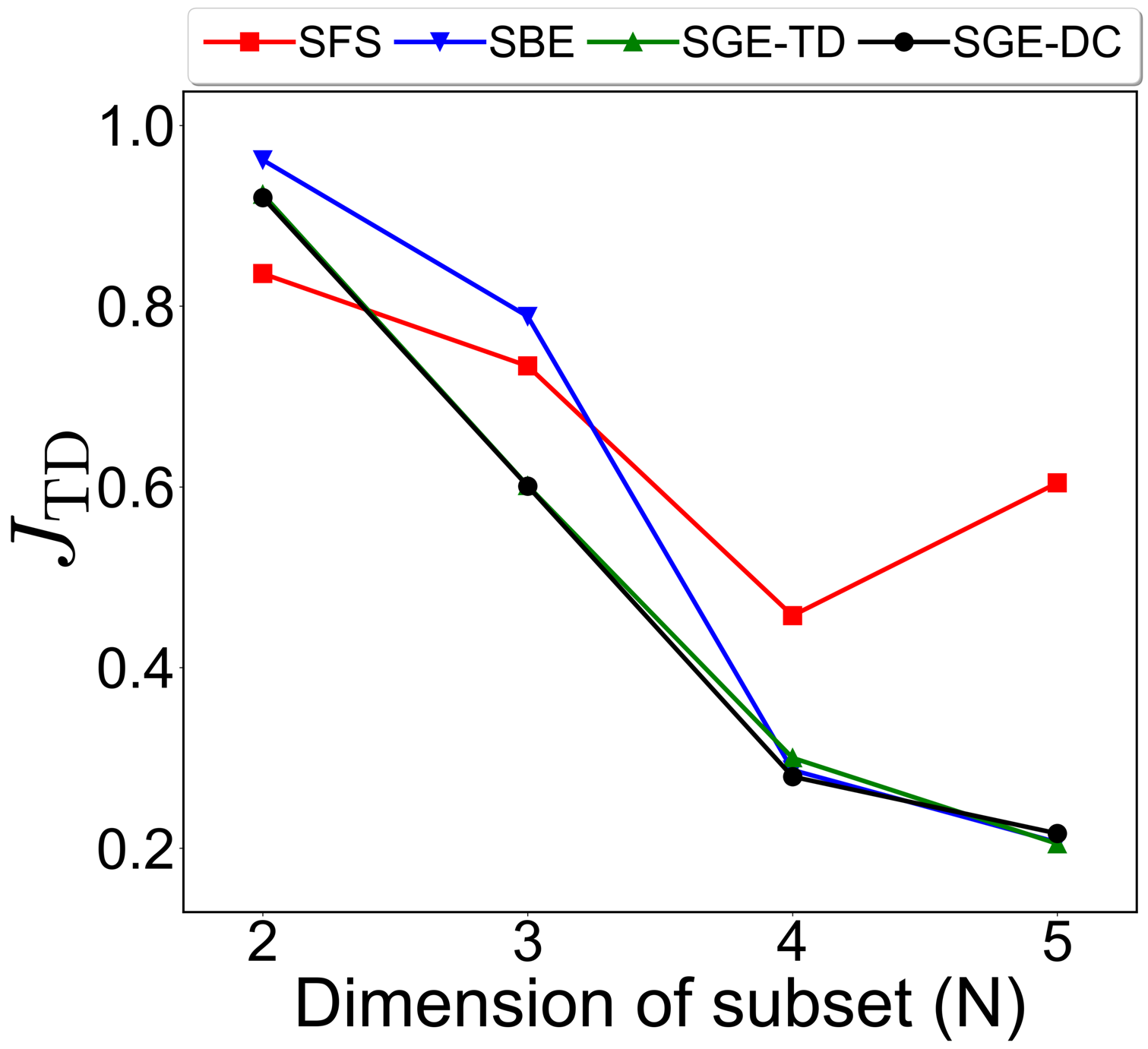}}\subfloat[4L-30N/L]{\centering{}\includegraphics[width=4cm]{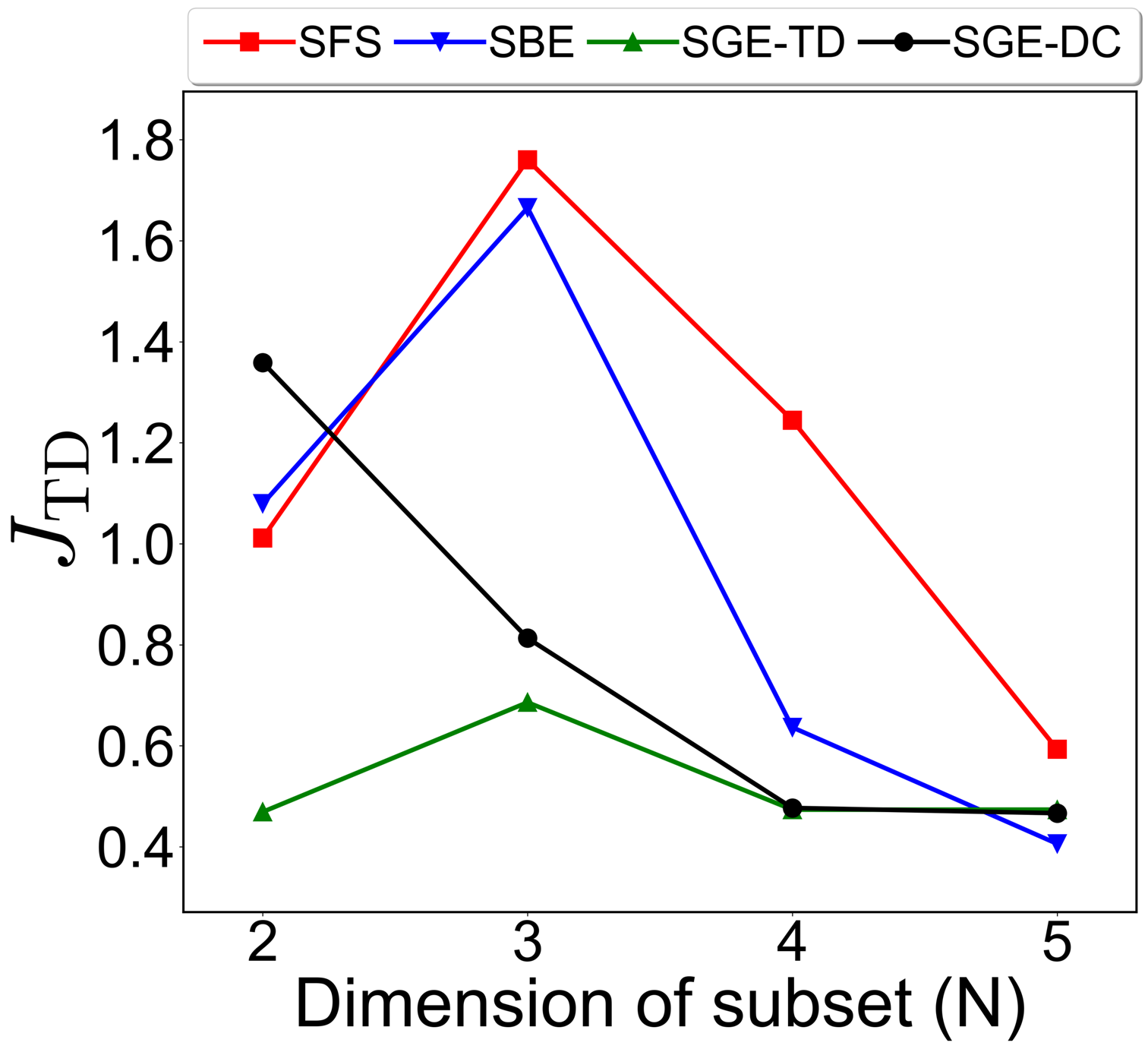}}\caption{$J_{TD}$ averaged overs trials according to the size of $S_{N}$
for turbulent bubbly flows in a pipe.\label{fig:Minimum-loss-of bubble}}
\end{figure}
\begin{figure}[H]
\centering{}\subfloat[1L-30N/L]{\centering{}\includegraphics[width=4cm]{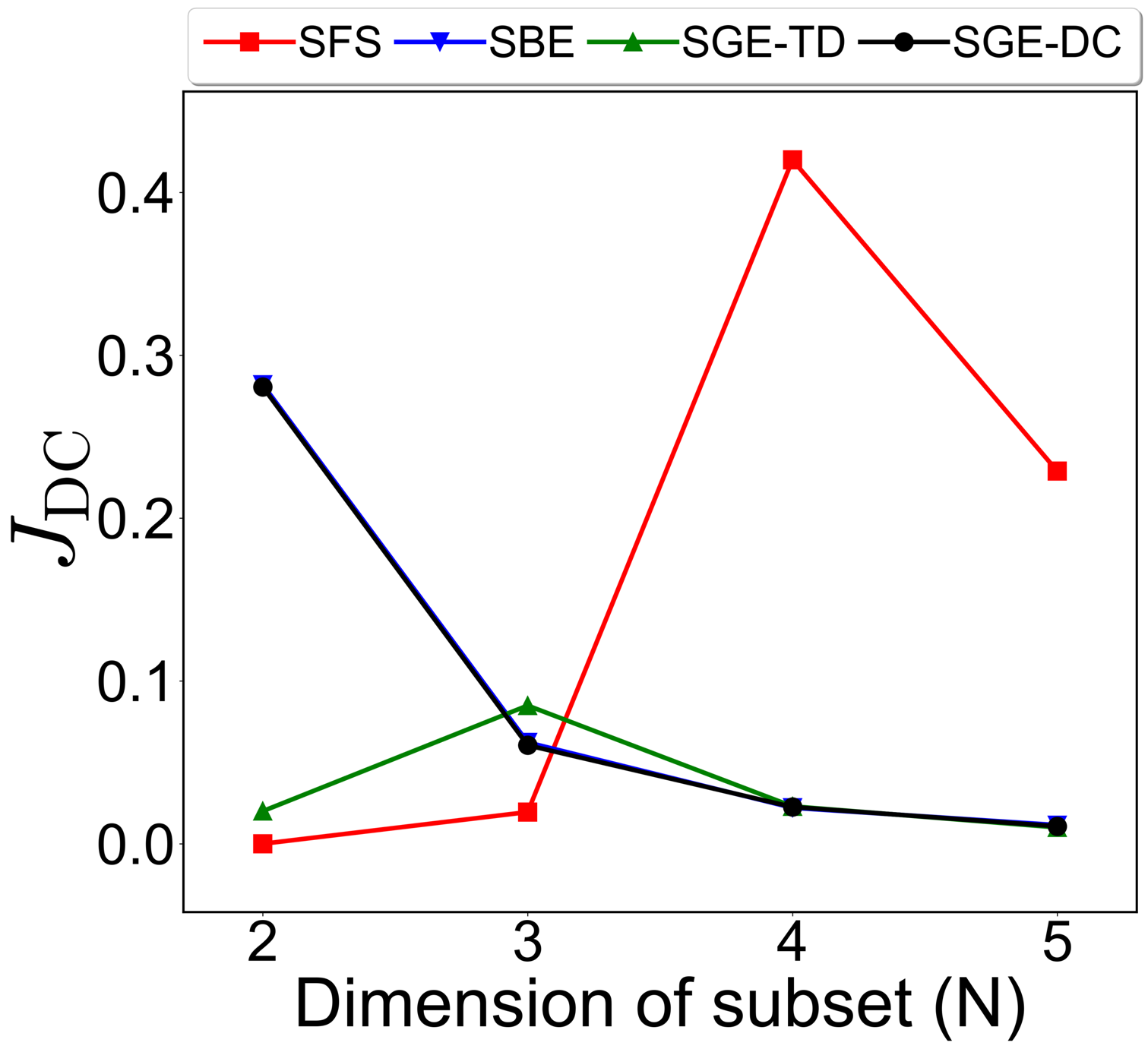}}\subfloat[2L-30N/L]{\centering{}\includegraphics[width=4cm]{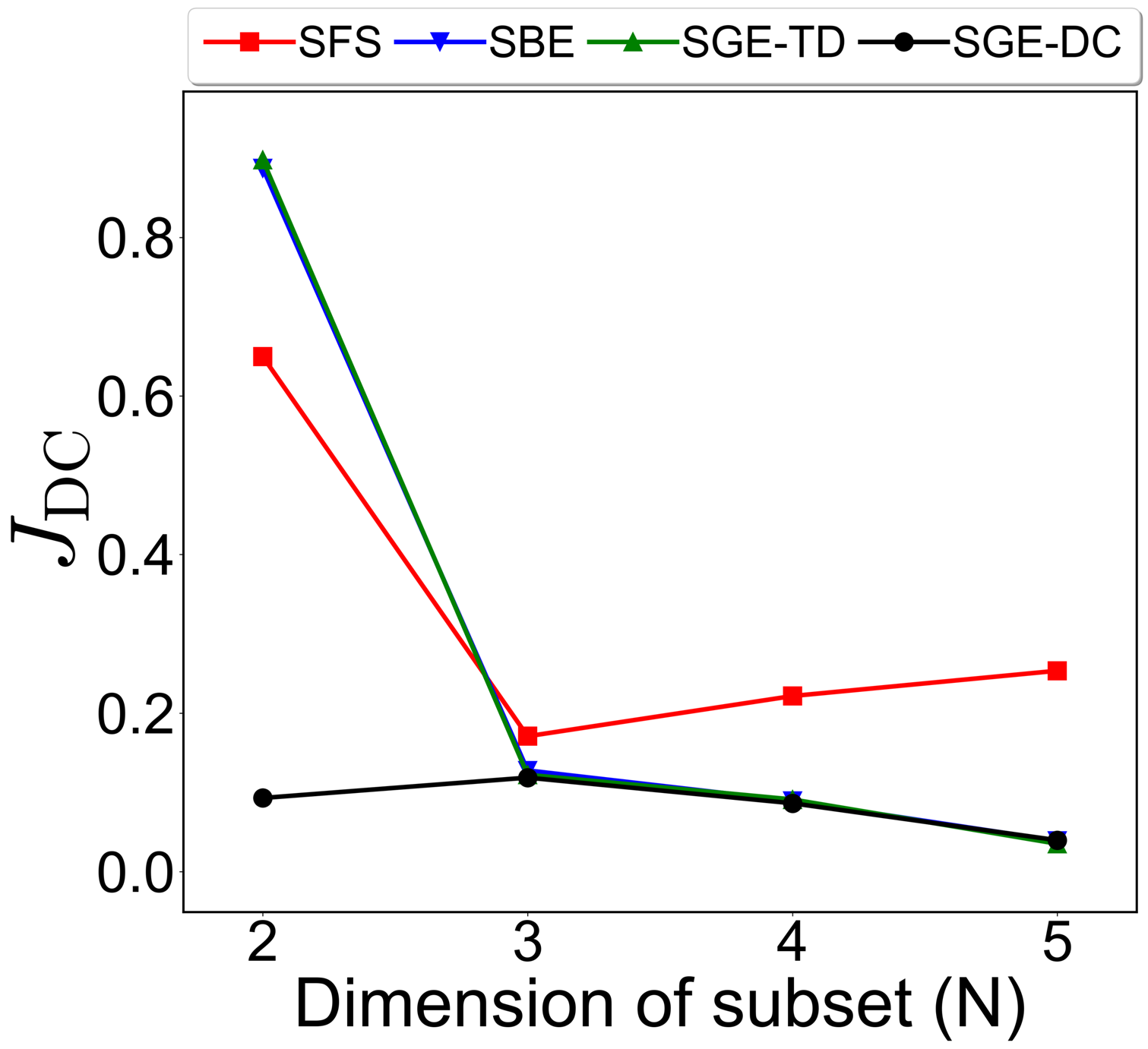}}\subfloat[4L-30N/L]{\centering{}\includegraphics[width=4cm]{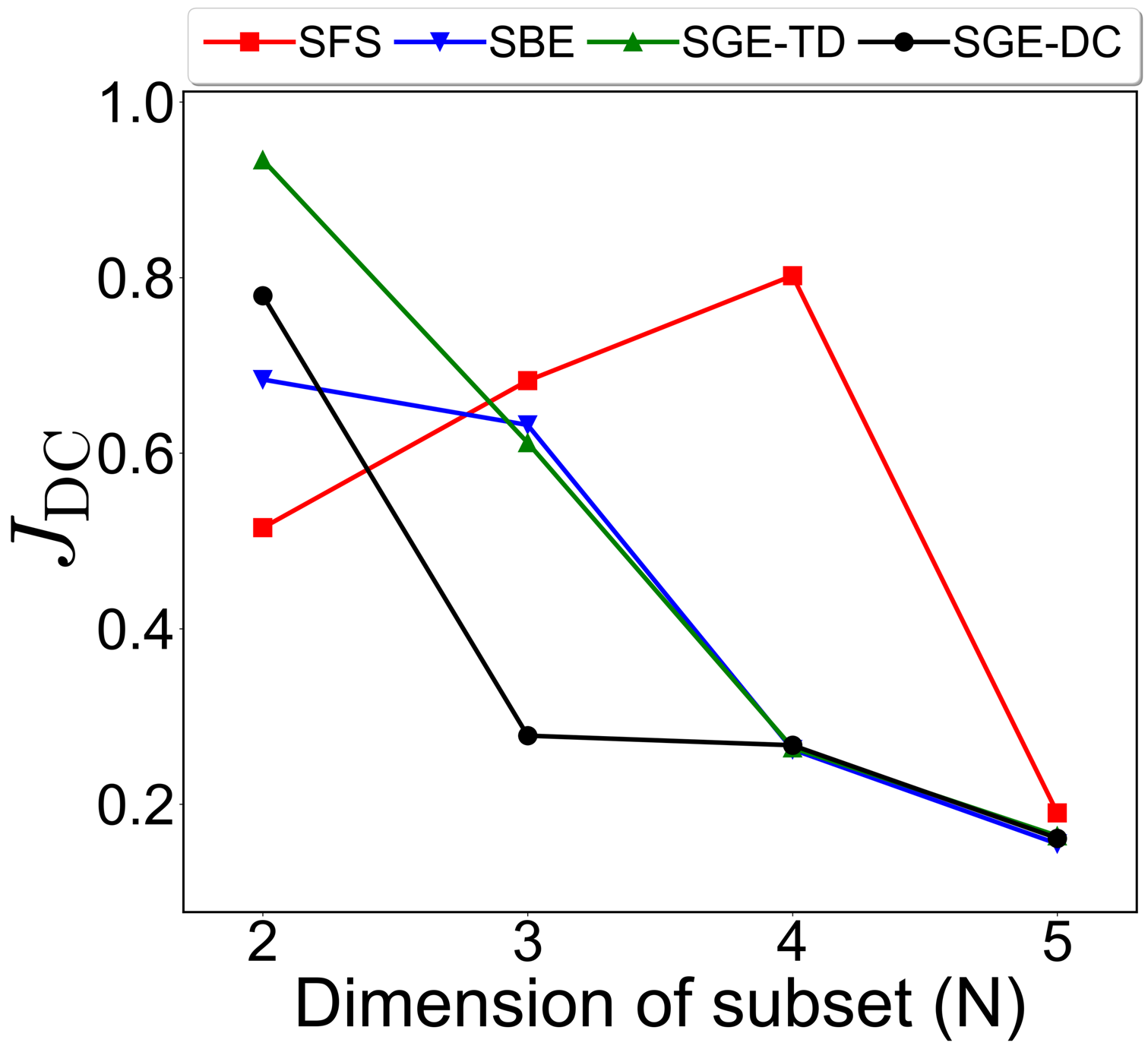}}\caption{$J_{DC}$ averaged overs trials according to the size of $S_{N}$
for turbulent bubbly flows in a pipe.\label{fig:Directional-consistency-bubble}}
\end{figure}

Fig.~\ref{fig:RMSE-comparision-bubble} shows the RMSE averaged overs
trials according to the size of the subset $S_{N}$ for each method
using well-trained ANNs. Figs.~\ref{fig:Minimum-loss-of bubble}
and \ref{fig:Directional-consistency-bubble} show $J_{TD}$ and $J_{DC}$
averaged over trials according to the size of $S_{N}$. Overall trends
look similar to those of the manufactured problem. It is shown that
the RMSE of SFS is somewhat worse than the others and the SGE methods
show the comparable RMSE to SBE. $J_{TD}$ and $J_{DC}$ are smallest
for the SGE methods.

\section{Data-driven modeling of turbulent Prandtl number in a duct flow\label{sec:Application-Prt}}

\subsection{Description of physical problem and data-driven modeling}

The turbulent Prandtl number ($Pr_{t}$) is an important parameter
to predict the heat transfer rate in RANS simulations by relating
the turbulent diffusivity ($\alpha_{t}$) and the turbulent viscosity
($\nu_{t}$). $Pr_{t}$ has been assumed often as a constant based
on the Reynolds analogy \citep{Reynolds1975}. For various turbulent
flows, however, several previous studies showed that $Pr_{t}$ varies
significantly over space, e.g., flows around a cylinder \citep{Antonia1987,Zhou2000a,Xu2005},
jet flows \citep{Chua1990,Chua2001}. Due to its importance, there
are many previous studies on modeling $Pr_{t}$ (\citep{Deissler1952,Aoki1963,Jischa1979,Rosen1995},
among others). However, there are only limited studies on spatially
varying $Pr_{t}$ models \citep{Kays1994,Tang2016}. Mostly, they
employ simple relationships using $Re$, $Pr$, the wall distance,
and $\nu_{t}$ to model only the wall-normal distribution, and the
similarity to a DNS result is marginal.

\begin{figure}[H]
\begin{centering}
\includegraphics[width=11cm]{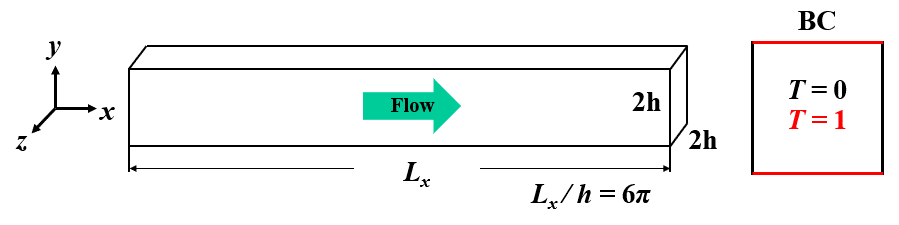}
\par\end{centering}
\caption{A physical domain of a turbulent flow inside a duct with heat transfer.\label{fig:Computational-domain}}
\end{figure}

Therefore, the present study aims at addressing this issue by applying
the ANN-wrapper method to a DNS database for a turbulent flow inside
a duct. The database is obtained from \citet{Nam2022}, and the problem
is briefly introduced. In the domain shown in Fig.~\ref{fig:Computational-domain},
the top and bottom walls are heated to a higher temperature which
leads to an incompressible flow with forced convection. The Reynolds
number based on the wall unit is $Re_{\tau}$=150. To avoid the overfitting
issue, the results of $Pr$=0.2, 2.0 are used for the train data and
those of $Pr$=0.7 are used for the validation data.

On the input parameters for the model $Pr_{t}=f\left(X\right)$, a
few parameters available from RANS simulations are utilized. We consider
a few Galilean-invariant variables derived from the velocity, and
pressure, and turbulent kinetic energy (TKE) as well as $\nu_{t}$
often used in the previous models. For a high-dimensional database,
a few variables in the mean budget equations are also considered,
supposing their close relationships to the TKE and temperature variance
budget equations that are crucial to turbulent transport. All candidate
parameters are listed in Table~\ref{tab: duct: parameter list} and
the definitions of the mean budget terms are presented in Appendix.
The dimension of the full parameter model is 12 that is larger than
5 of the turbulent bubbly flow. This possibly makes achieving a high
CoT more difficult. Considering the histogram of the database variables,
the following standardization (instead of normalization) is used to
pre-process the input and output parameters:

\begin{equation}
x^{*}=\frac{x-\overline{x}}{\sigma_{x}},\:\:Y^{*}=\frac{Y-\overline{Y}}{\sigma_{Y}},\label{eq:standardization}
\end{equation}
where $\sigma$ refers to the standard deviation.
\begin{table}[H]
\begin{centering}
\begin{tabular}{|c|>{\centering}m{1.8cm}|>{\centering}m{5.5cm}|}
\hline 
Global parameters & \centering{}$Pr$ & Prandtl number\tabularnewline
\hline 
\multirow{11}{*}{Local parameters} & $\nu_{t}$ & Turbulent viscosity\tabularnewline
\cline{2-3} \cline{3-3} 
 & $k$ & Turbulent kinetic energy\tabularnewline
\cline{2-3} \cline{3-3} 
 & $\Pi_{MKE}$ & Mean pressure transport\tabularnewline
\cline{2-3} \cline{3-3} 
 & $\varepsilon_{MKE}$ & Mean dissipation\tabularnewline
\cline{2-3} \cline{3-3} 
 & $D_{MKE}$ & Mean viscous diffusion\tabularnewline
\cline{2-3} \cline{3-3} 
 & $\varepsilon_{MTV}$ & Mean temperature dissipation\tabularnewline
\cline{2-3} \cline{3-3} 
 & $D_{MTV}$ & Mean molecular diffusion\tabularnewline
\cline{2-3} \cline{3-3} 
 & $\left|\partial\bar{P}/\partial x_{i}\right|$ & Magnitude of pressure gradient\tabularnewline
\cline{2-3} \cline{3-3} 
 & $\left|\partial\bar{T}/\partial x_{i}\right|$ & Magnitude of temperature gradient\tabularnewline
\cline{2-3} \cline{3-3} 
 & $\left|\partial k/\partial x_{i}\right|$ & Magnitude of turbulent kinetic energy gradient\tabularnewline
\cline{2-3} \cline{3-3} 
 & $\left|\bar{S}_{ij}\right|$ & Magnitude of strain rate tensor\tabularnewline
\hline 
\end{tabular}
\par\end{centering}
\caption{A list of input parameters ($X$) to model the turbulent Prandtl number
in a duct.\label{tab: duct: parameter list}}
\end{table}

\begin{figure}[H]
\begin{centering}
\subfloat[Heat map of the correlation coefficient]{\begin{centering}
\includegraphics[height=5cm]{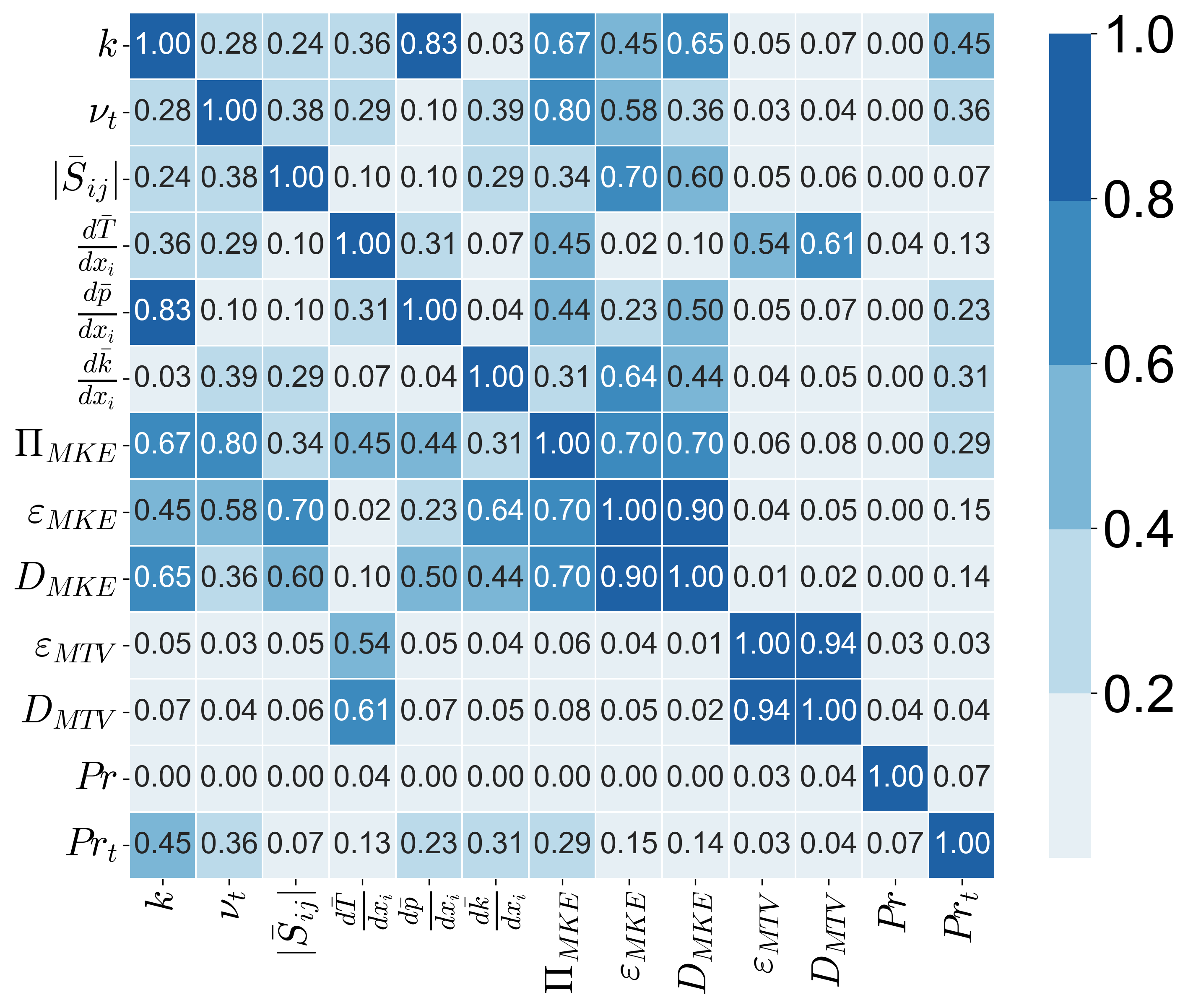}
\par\end{centering}
}\subfloat[Correlation between inputs and $Pr_{t}$]{\begin{centering}
\includegraphics[height=5cm]{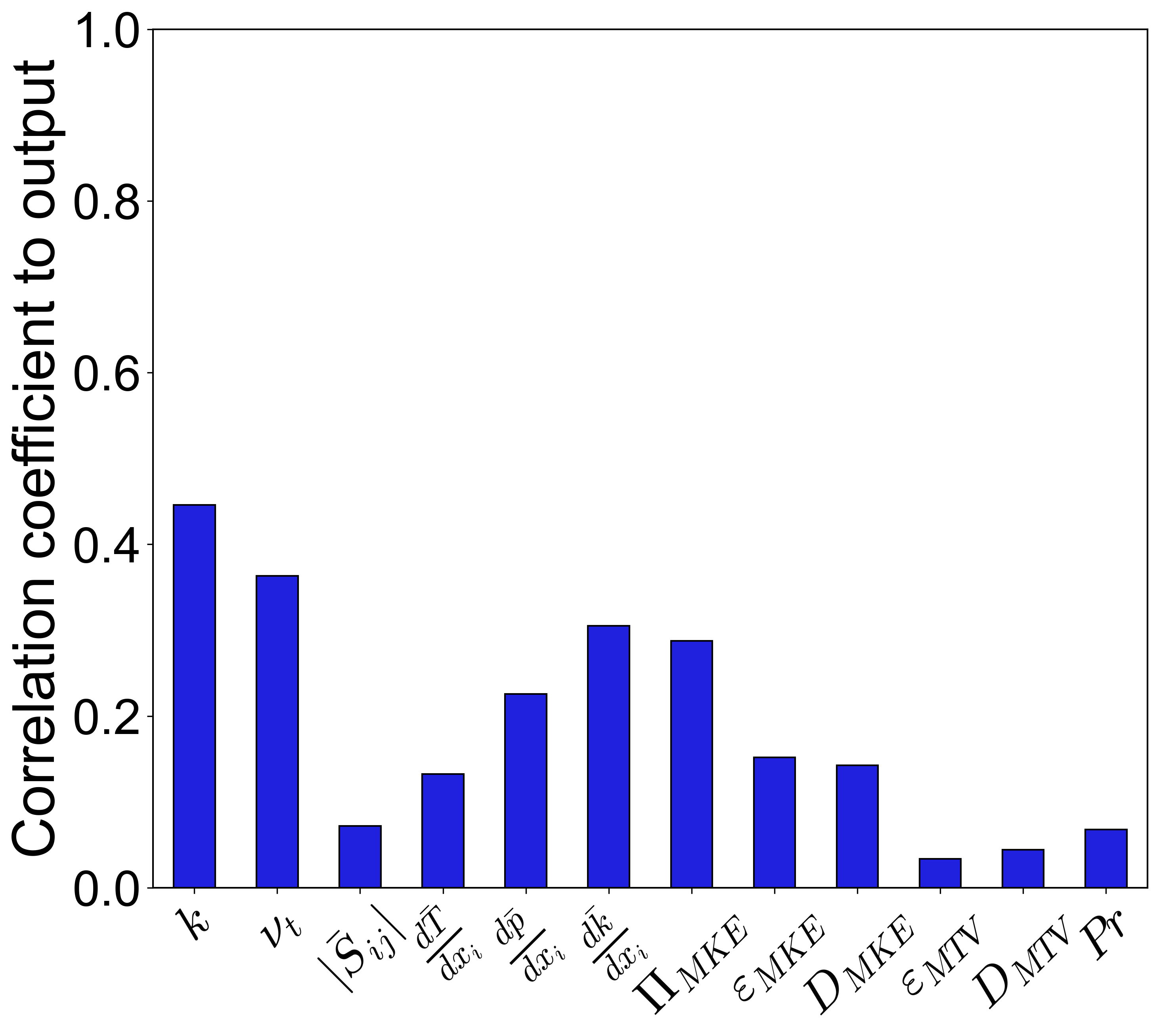}
\par\end{centering}
}
\par\end{centering}
\caption{\textcolor{black}{Pearson correlation coefficients} of turbulent heat
transfer in a duct.\label{fig:Correlation - channel}}
\end{figure}

Fig.~\ref{fig:Correlation - channel} shows the Pearson correlation
coefficients among the parameters and $Pr_{t}$. A difficulty in modeling
spatially varying $Pr_{t}$ is conjectured from complicated relationships
in Fig.~\ref{fig:Correlation - channel}(a). Fig.~\ref{fig:Correlation - channel}(b)
shows the correlation coefficients between $Pr_{t}$ and input parameters.
There are three variables (i.e. $k$, $\nu_{t}$ and $\left|\partial k/\partial x_{i}\right|$)
with the correlation coefficients of 0.3 or larger to $Pr_{t}$, but
overall the correlation coefficients are relatively much smaller compared
to the previous two applications. In Fig.~\ref{fig:Correlation - channel}(a),
there are about 10 additional variables with the correlation coefficients
of 0.3 or larger to ($k$, $\nu_{t}$ and $\left|\partial k/\partial x_{i}\right|$).
This implies the complex relationships among the parameters and possibly
overlapping effects among them.

\subsection{Training model and number of optimal parameters}

\begin{figure}[H]
\centering{}\subfloat[Using train data]{\begin{centering}
\includegraphics[height=5cm]{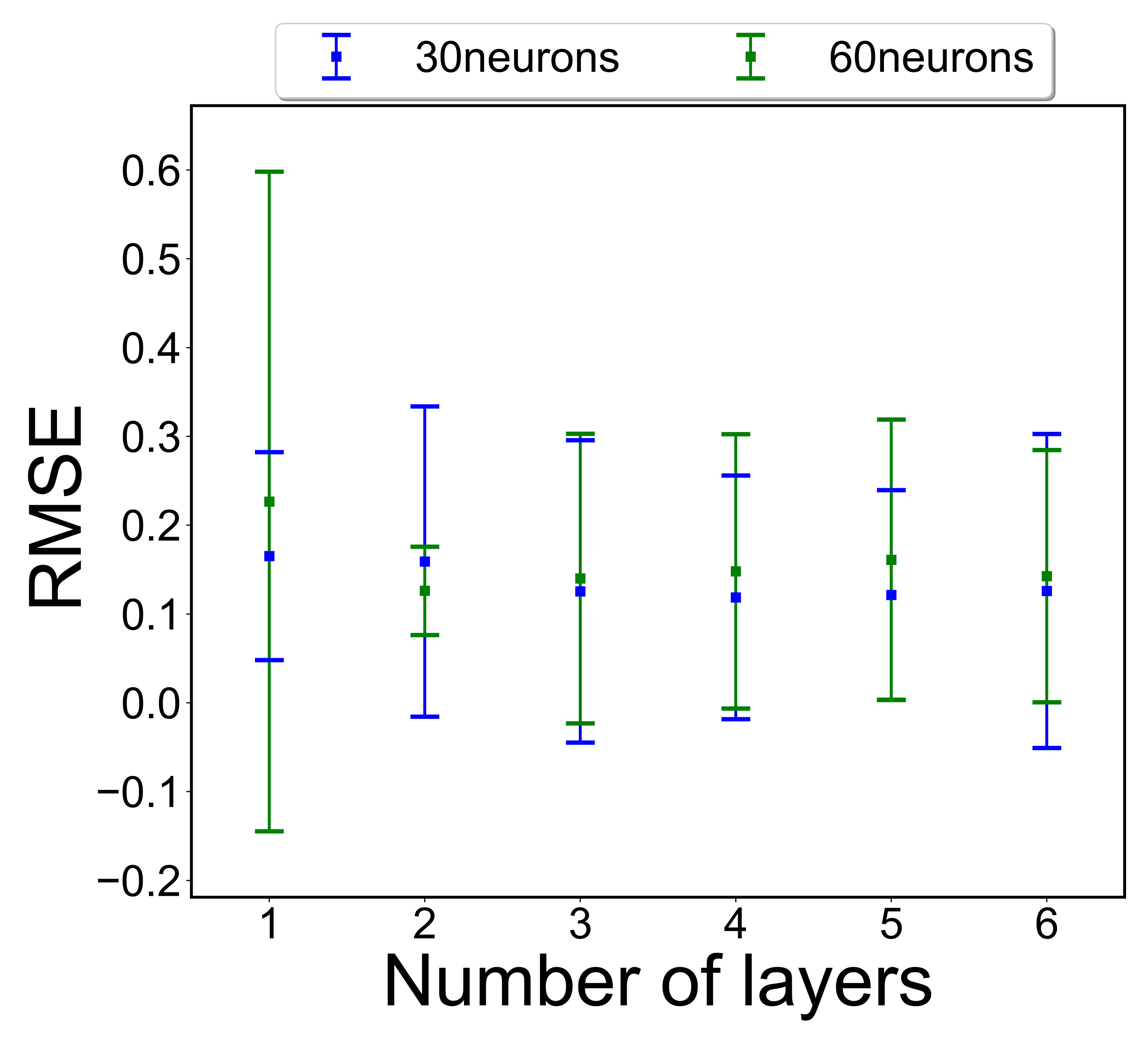}
\par\end{centering}
}\subfloat[Using validation data]{\centering{}\includegraphics[height=5cm]{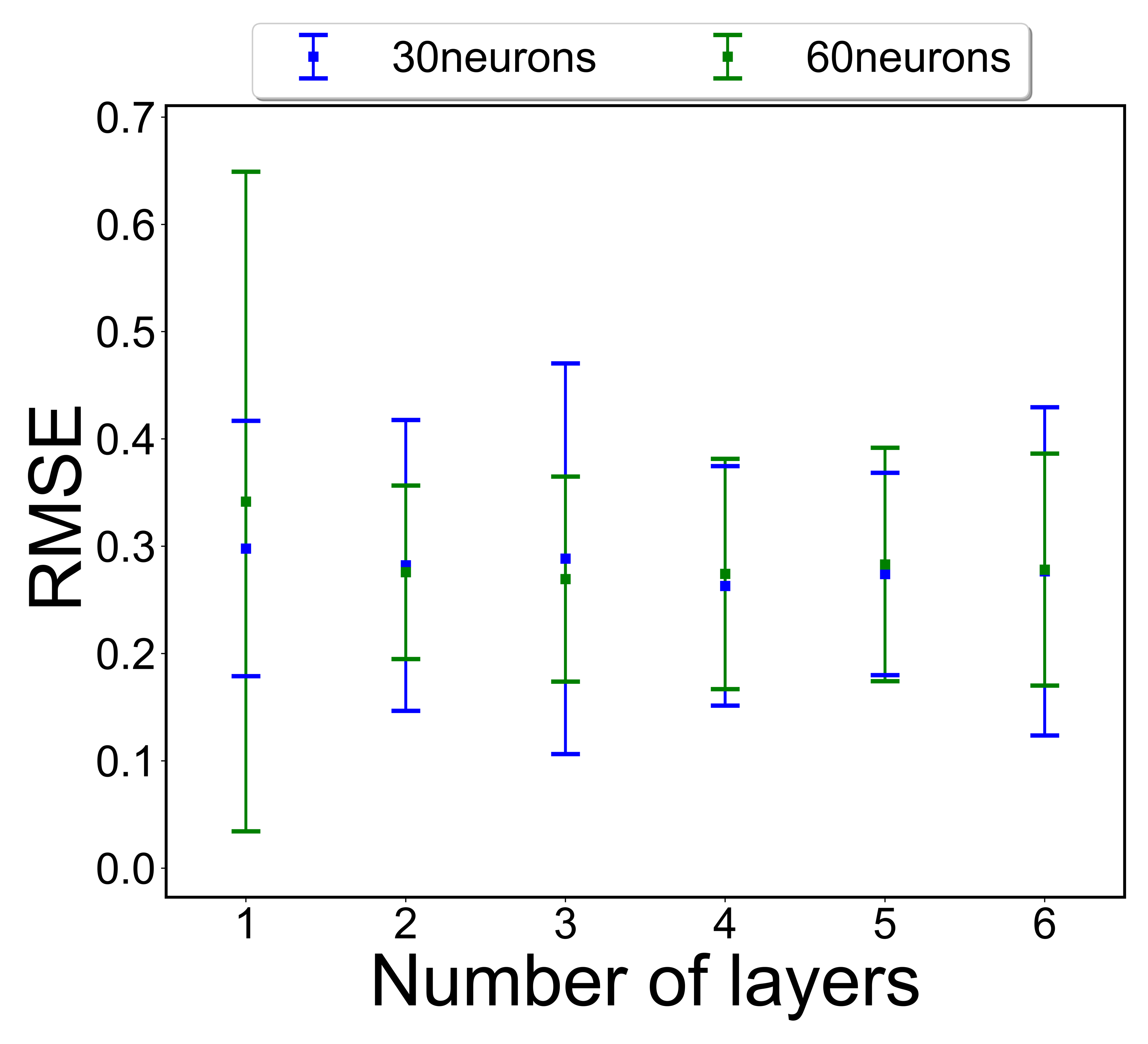}}\caption{\textcolor{black}{RMSE values with the different numbers of the layer
and neuron per layer} for turbulent heat transfer in a duct.\label{fig:RMSE-by-layers duct}}
\end{figure}

To train ANNs efficiently for the database of $Pr_{t}$, the cosine
annealing learning rate scheduler \citep{loshchilov2016sgdr} and
the ELU activation function \citep{Clevert2015} are used. The K-fold
technique \citep{Breiman1992} with 5-fold cross validation \citep{Kohavi1995}
was used to minimize the overfitting issue. We select the number of
the layers and neurons by a comparison study for hyper-parameters.
Fig.~\ref{fig:RMSE-by-layers duct} shows the RMSE values with the
different values of the layer and neuron per layer. Each RMSE value
is obtained from 120 trained ANNs. The error bars show 95\% confidence
intervals. The averaged RMSE show similar values over different conditions.
We select 2 layers and 60 neurons/layer (2L-60N/L) based on relatively
lower RMSE and better error bar range. To test dependency on the hyper-parameters,
1L-30N/L is also considered in the next section. Note that all RMSE
results are computed based on $Pr_{t}^{*}$. As $\sigma_{Pr_{t}}=\sigma_{Y}=0.334$,
0.3 for the RMSE of $Pr_{t}^{*}$ (as found in Fig.~\ref{fig:RMSE-by-layers duct}(b))
is equivalent to 0.1 for the RMSE of $Pr_{t}$.

\begin{figure}[H]
\centering{}\includegraphics[height=5cm]{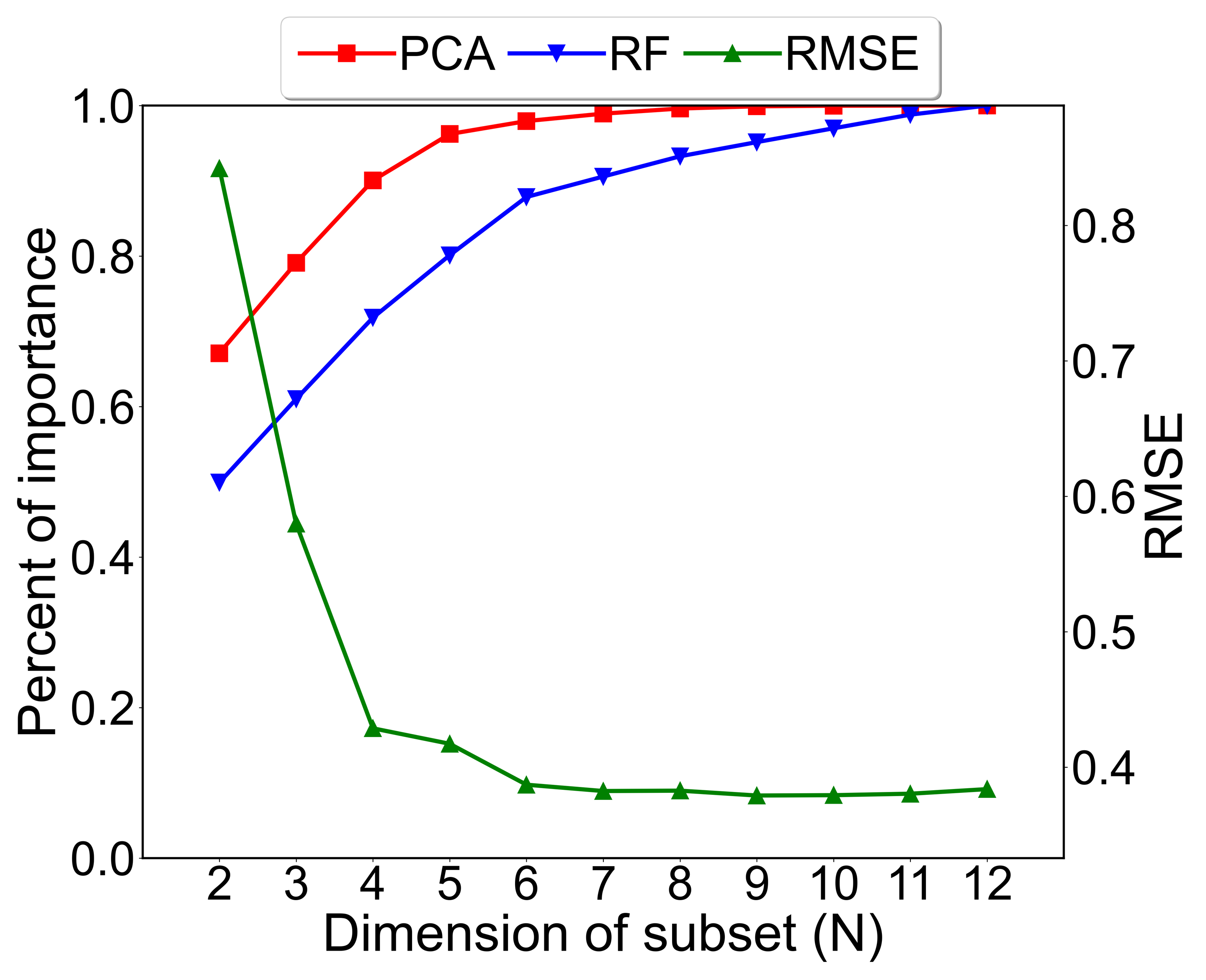}\caption{\textcolor{black}{Accumulated feature importance of the RF method
and PCA compared with the RMSE (validation) of SGE-TD over the number
of components} (including $Pr$).\label{fig:Percent-of-information duct}}
\end{figure}

To estimate the dimension of a reduced $Pr_{t}$ model, Fig.~\ref{fig:Percent-of-information duct}
shows the accumulated feature importance of the RF method and PCA
over the number of components. The green line shows the RMSE of SGE-TD
over the number of the input parameters. It is shown that RF needs
8-9 and PCA needs 5 parameters for sufficiently converged (> 90\%
in importance) model prediction. The RMSE of SGE-TD decreases rapidly
up to 4 parameters and fully converges from 6 parameters.

\subsection{Results of dimensionality reduction for $Pr_{t}$ model}

The ANN-wrapper methods are applied to the full parameter $Pr_{t}$
model. With respect to the trained ANNs, only the well-trained ANNs
(50\% below the RMSE median) are used.

Table~\ref{tab:Table-of-stability duct} shows PEM-CoT1 and PEM-CoT2
metrics of $S_{5}$ and $S_{6}$ over \textcolor{black}{60 }selection
trials of different methods. For brevity, the metrics were averaged
over different hyper-parameters (1L-30N/L and 2L-60N/L). Note that
$Pr$ is an important global parameter for $Pr_{t}$ and pre-included
in the parameter subset. For both $S_{5}$ and $S_{6}$, SGE-TD shows
better CoT than the other methods, especially for PEM-CoT2 that shows
at least 10\% increase. Higher PEM-CoT2 of SGE-TD implies obtaining
a specific parameter subset more stably. The second highest CoT is
obtained by SBE. SGE-DC is a little behind SBE, followed by SFS by
relatively large margins especially for PEM-CoT2.
\begin{table}[H]
\centering{}%
\begin{tabular}{|c|>{\centering}p{2cm}|>{\centering}p{2cm}|>{\centering}p{2cm}|>{\centering}p{2cm}|}
\hline 
 & \multicolumn{2}{c|}{5-parameter subset ($S_{5}$)} & \multicolumn{2}{c|}{6-parameter subset ($S_{6}$)}\tabularnewline
\hline 
Method & Average PEM-CoT1 & Average PEM-CoT2 & Average PEM-CoT1 & Average PEM-CoT2\tabularnewline
\hline 
SFS & 0.664 & 0.117 & 0.790 & 0.192\tabularnewline
\hline 
SBE & 0.947 & 0.800 & 0.853 & 0.350\tabularnewline
\hline 
SGE-TD & 0.973 & 0.892 & 0.877 & 0.425\tabularnewline
\hline 
SGE-DC & 0.873 & 0.675 & 0.810 & 0.292\tabularnewline
\hline 
\end{tabular}\caption{PEM-CoT1 and PEM-CoT2 over \textcolor{black}{60 selection trials}
for turbulent heat transfer in a duct.\label{tab:Table-of-stability duct}}
\end{table}

Table~\ref{tab:Table-of-selected duct} show the list of 6 parameters
with the highest selection possibilities from $S_{6}$ over\textcolor{black}{{}
60 selection trials}. All methods include $\nu_{t}$ and $k$ with
the exception to SFS. Notably, previously suggested two-equation models
\citep{Abe1995,Brinckman2007,Sommer1993,Deng2001}, which directly
solves transport equations of the dissipation rate and thermal variance
to modeling $Pr_{t}$, defines $\alpha_{t}$ as a function of $k$,
implying its importance in modeling $Pr_{t}$. With 2L-60N/L, the
most selected parameters becomes identical between SGE and SBE methods.
Note that the maximum inter-parameter correlation coefficients are
shown to be smallest for the SGE methods followed by SBE and SFS.
A lower maximum inter-parameter correlation may imply reduced overlapping
effects. 
\begin{table}[H]
\begin{centering}
\begin{tabular}{|>{\centering}m{1.5cm}|>{\centering}m{4cm}|>{\centering}m{4cm}|>{\centering}m{1.5cm}|>{\centering}m{1.5cm}|}
\hline 
\multirow{2}{1.5cm}{Method} & \multicolumn{2}{c|}{6 most selected parameters from $S_{6}$} & \multicolumn{2}{>{\centering}p{3cm}|}{Max. inter-parameter correlation}\tabularnewline
\cline{2-5} \cline{3-5} \cline{4-5} \cline{5-5} 
 & 1L-30N/L & 2L-60N/L & 1L-30N/L & 2L-60N/L\tabularnewline
\hline 
SFS & $\varepsilon_{MKE}$, $\varepsilon_{MTV},$$\Pi_{MKE}$, $D_{MTV}$,
$\left|\partial\bar{P}/\partial x_{i}\right|$, $Pr$ & $\varepsilon_{MKE}$, $\varepsilon_{MTV},$$\Pi_{MKE}$, $\left|\bar{S}_{ij}\right|$,
$\left|\partial\bar{P}/\partial x_{i}\right|$, $Pr$ & 0.70 & 0.70\tabularnewline
\hline 
SBE & $\nu_{t}$, $k$, $\left|\partial\bar{T}/\partial x_{i}\right|$,
$D_{MKE}$, $\left|\bar{S}_{ij}\right|$, $Pr$ & $\nu_{t}$, $k$, $\left|\partial\bar{T}/\partial x_{i}\right|$,
$\left|\partial k/\partial x_{i}\right|$, $D_{MKE}$, $Pr$ & 0.60 & 0.54\tabularnewline
\hline 
SGE-TD & $\nu_{t}$, $k$, $\left|\partial\bar{T}/\partial x_{i}\right|$,
$\varepsilon_{MTV}$, $D_{MKE}$, $Pr$ & $\nu_{t}$, $k$, $\left|\partial\bar{T}/\partial x_{i}\right|$,
$\left|\partial k/\partial x_{i}\right|$, $D_{MKE}$, $Pr$ & 0.54 & 0.54\tabularnewline
\hline 
SGE-DC & $\nu_{t}$, $k$, $\left|\partial\bar{T}/\partial x_{i}\right|$,
$\left|\partial k/\partial x_{i}\right|$, $D_{MKE}$, $Pr$ & $\nu_{t}$, $k$, $\left|\partial\bar{T}/\partial x_{i}\right|$,
$\left|\partial k/\partial x_{i}\right|$, $D_{MKE}$, $Pr$ & 0.54 & 0.54\tabularnewline
\hline 
\end{tabular}
\par\end{centering}
\centering{}\caption{List of 6 parameters with the highest selection possibilities from
$S_{6}$ over\textcolor{black}{{} 60 selection trials} for turbulent
heat transfer in a duct. Maximum inter-parameter correlation coefficients
are shown in the right.\label{tab:Table-of-selected duct}}
\end{table}

\begin{figure}[H]
\centering{}\subfloat[RMSE]{\centering{}\includegraphics[width=4cm]{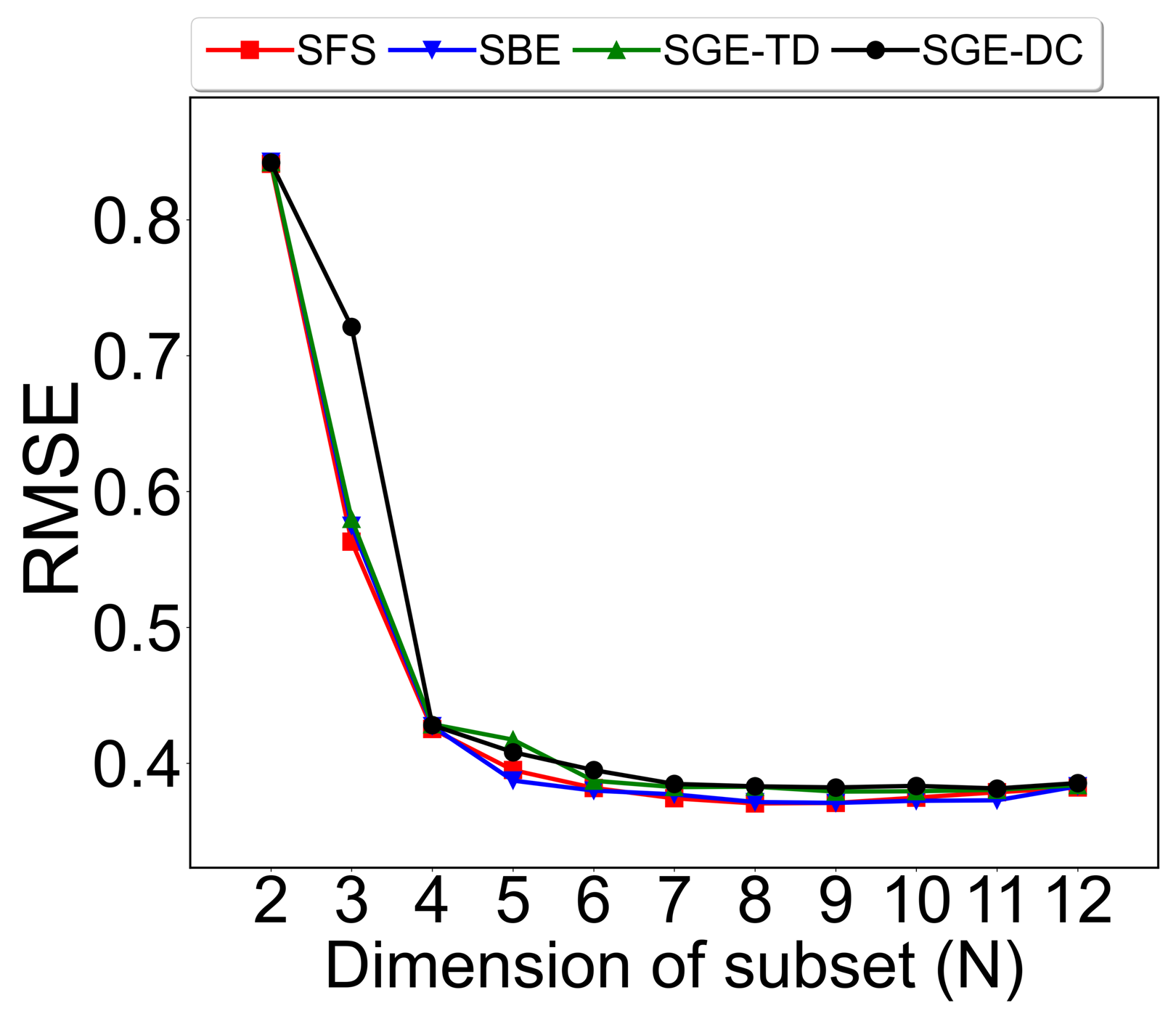}}\subfloat[$J_{TD}$]{\centering{}\includegraphics[width=4cm]{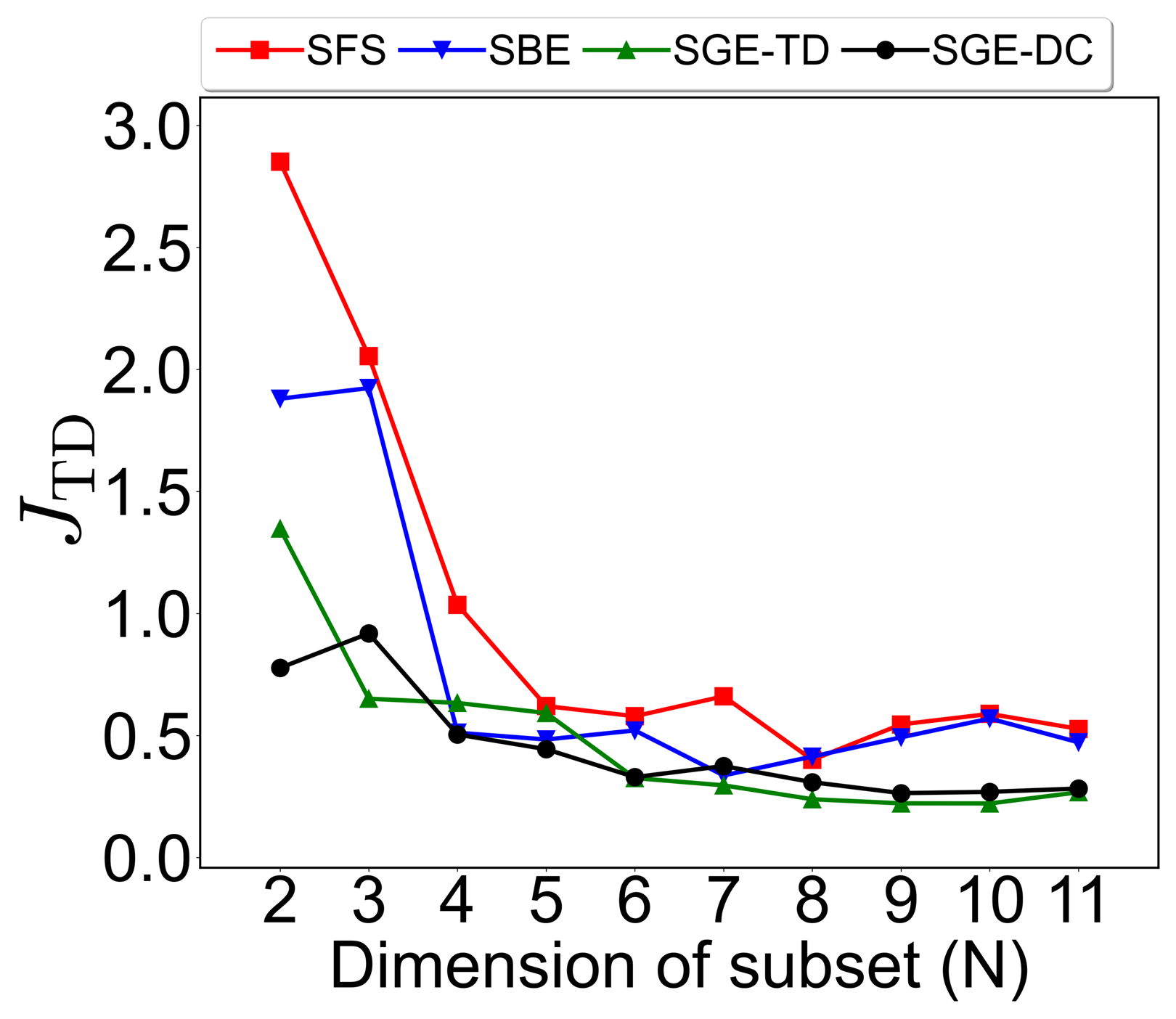}}\subfloat[$J_{DC}$]{\centering{}\includegraphics[width=4cm]{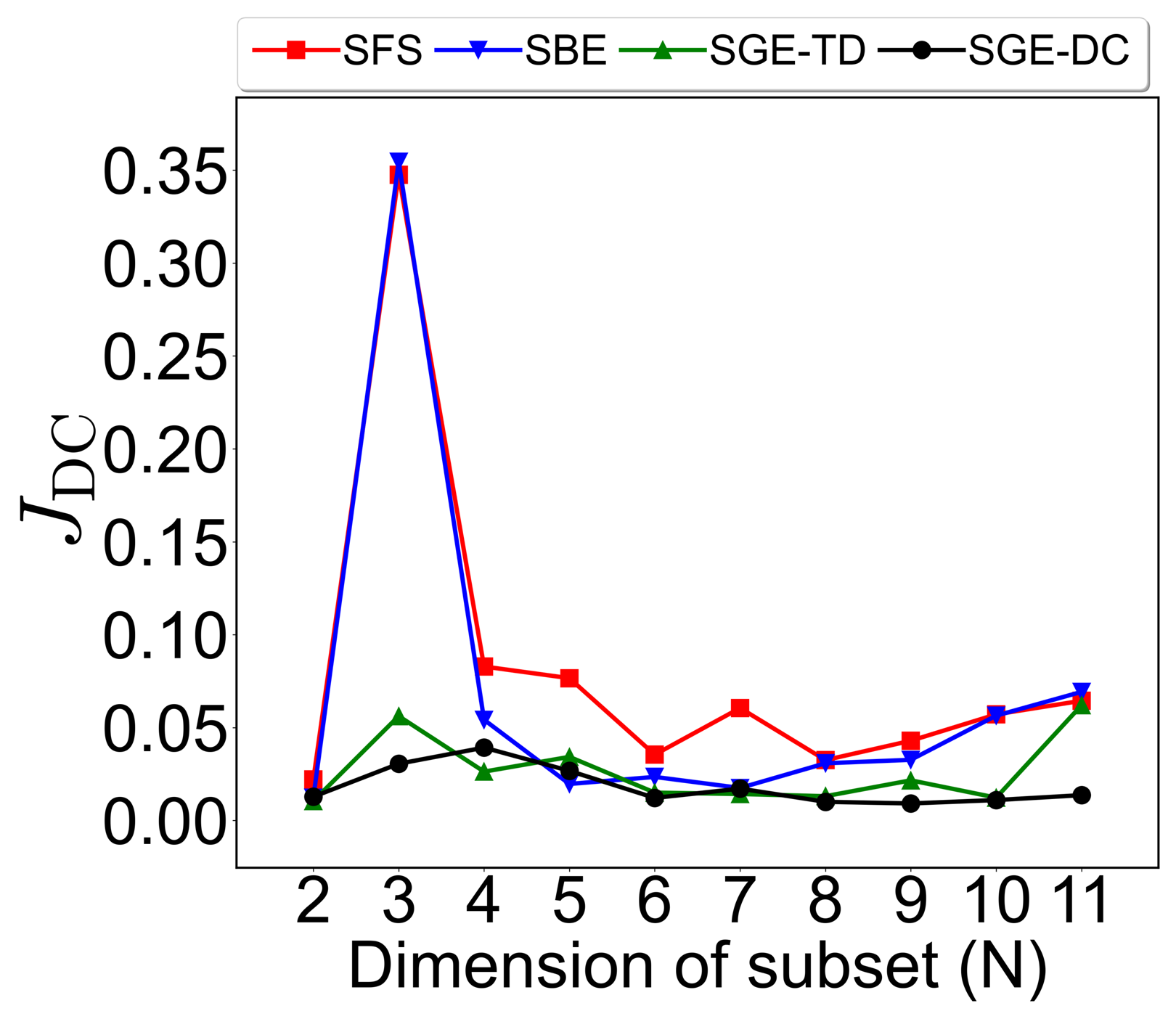}}\caption{RMSE \textcolor{black}{(validation)}, $J_{TD}$, and $J_{DC}$ averaged
overs trials according to the size of $S_{N}$ for turbulent heat
transfer in a duct (\textcolor{black}{2L-60N/L}).\label{fig:Metrics-for-wrapper duct}}
\end{figure}

\textcolor{black}{Fig.~}\ref{fig:Metrics-for-wrapper duct}\textcolor{black}{{}
}shows the RMSE, \textcolor{black}{$J_{TD}$, }and\textcolor{black}{{}
$J_{DC}$} averaged overs trials according to the size of the subset
$S_{N}$ for each method. Overall trends \textcolor{black}{look similar
to those of the }previous applications\textcolor{black}{.} The RMSE
is smallest for SFS and SBE by a relatively small margin, and \textcolor{black}{$J_{TD}$
}and\textcolor{black}{{} $J_{DC}$ are smallest for the SGE methods.}
The RMSE shows that the reduced model converges with 5-6 parameters,
similarly to the PCA result in Fig.~\ref{fig:Percent-of-information duct}.

\begin{figure}[H]
\begin{centering}
\subfloat[DNS]{\centering{}\includegraphics[width=3.6cm]{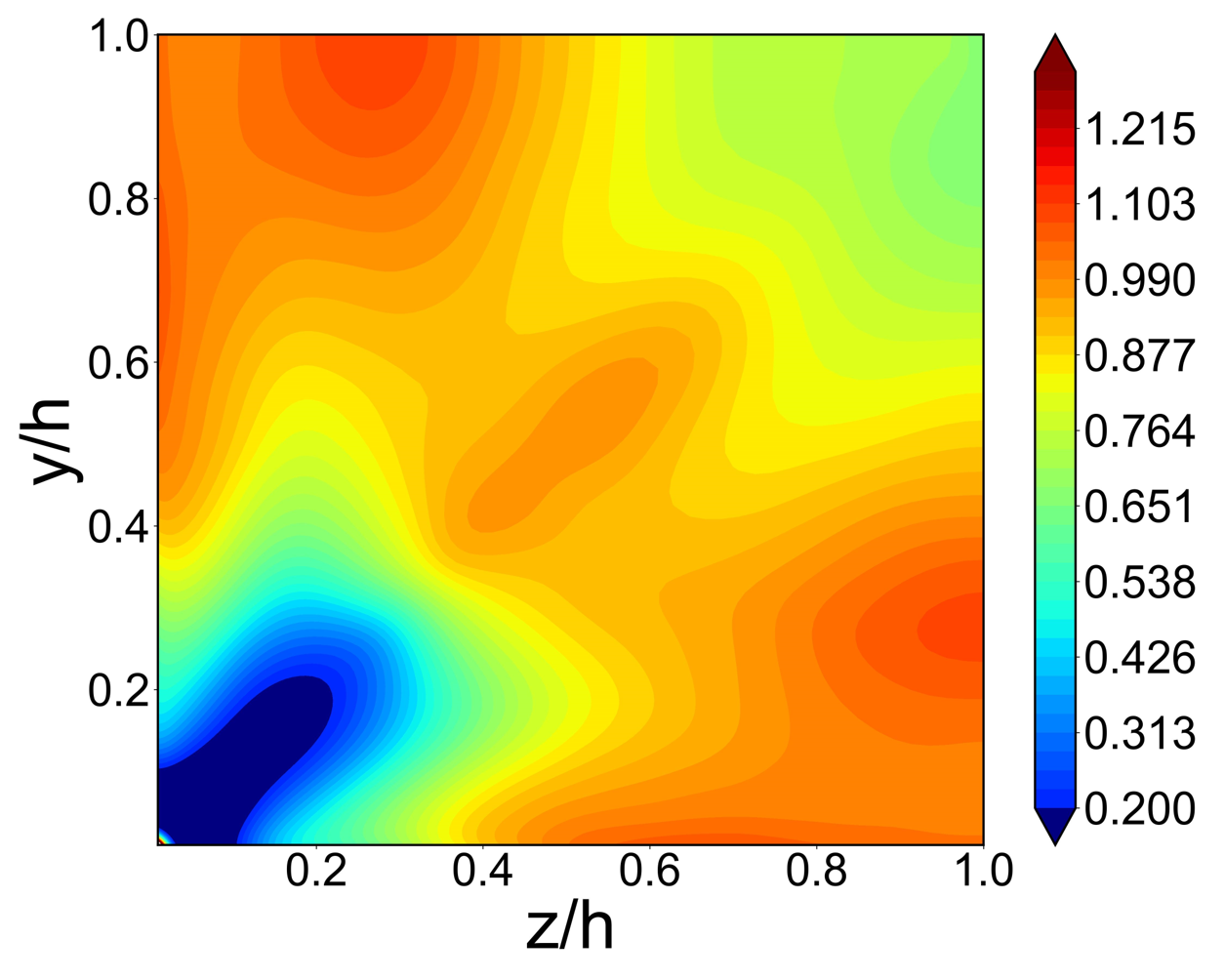}}\subfloat[Full parameters]{\begin{centering}
\includegraphics[width=3.6cm]{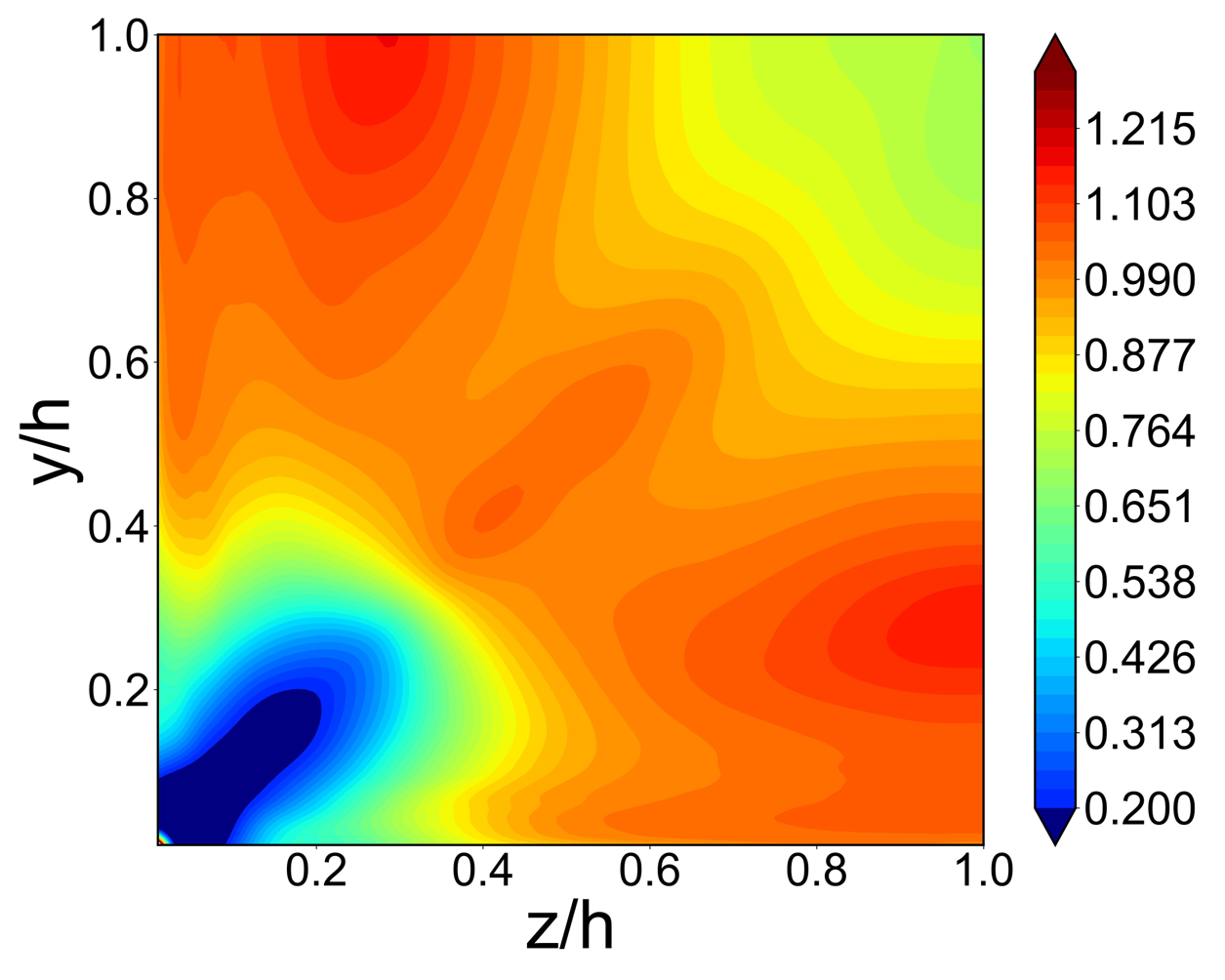}
\par\end{centering}
\centering{}}
\par\end{centering}
\begin{centering}
\subfloat[SFS - 10 parameters]{\begin{centering}
\includegraphics[width=3.6cm]{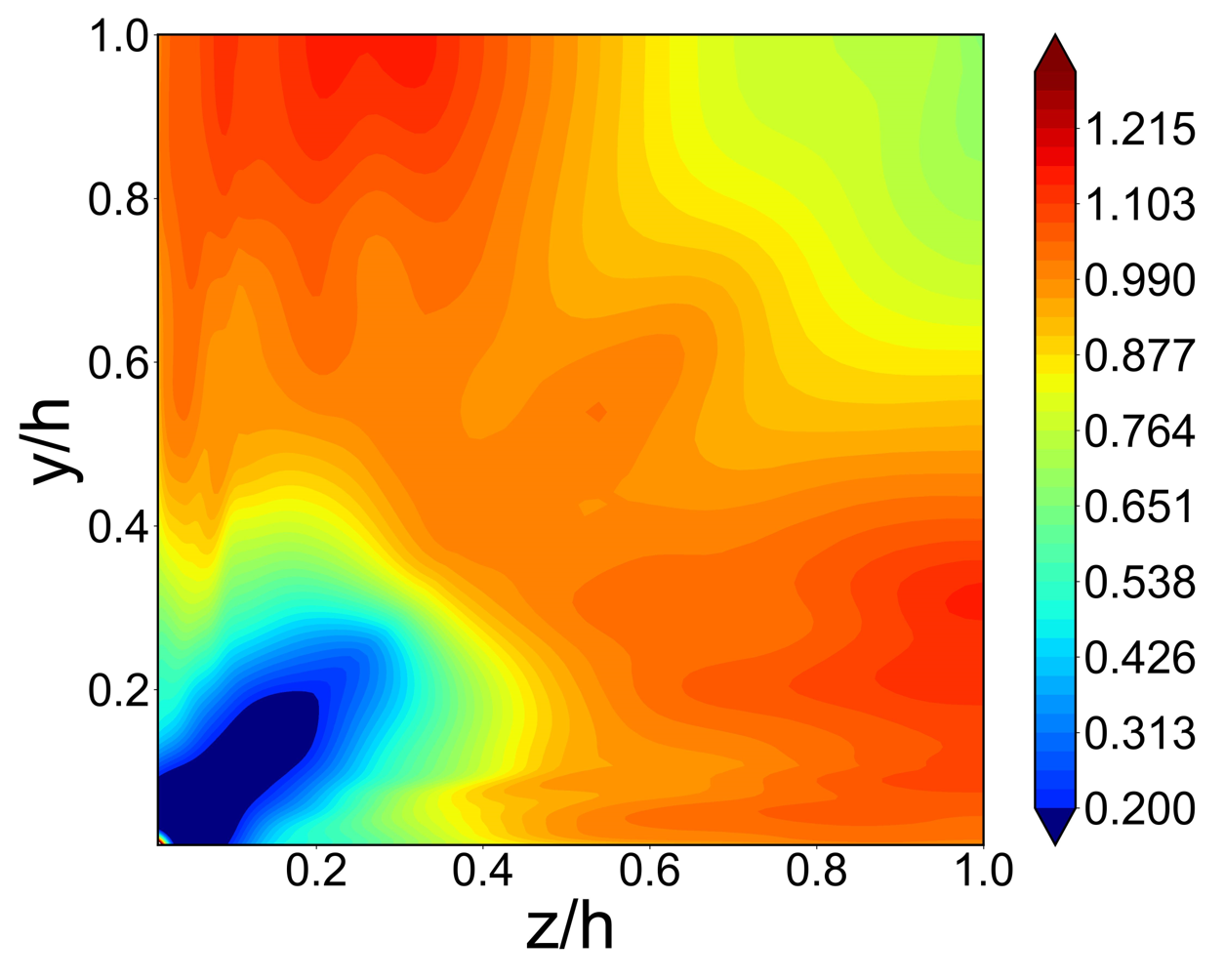}
\par\end{centering}
}\subfloat[SGE-TD - 10 parameters]{\begin{centering}
\includegraphics[width=3.6cm]{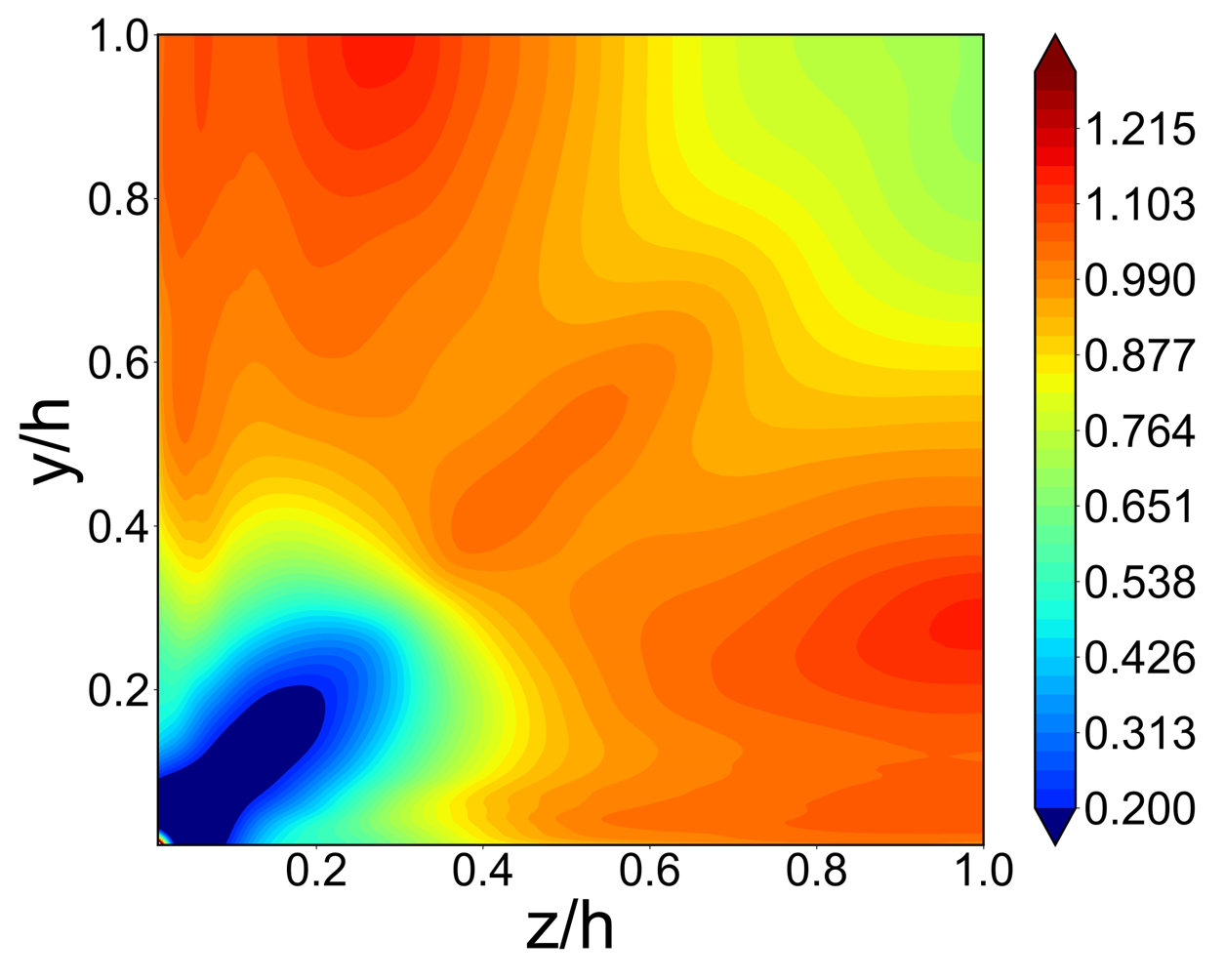}
\par\end{centering}
}
\par\end{centering}
\begin{centering}
\subfloat[SFS - 8 parameters]{\centering{}\includegraphics[width=3.6cm]{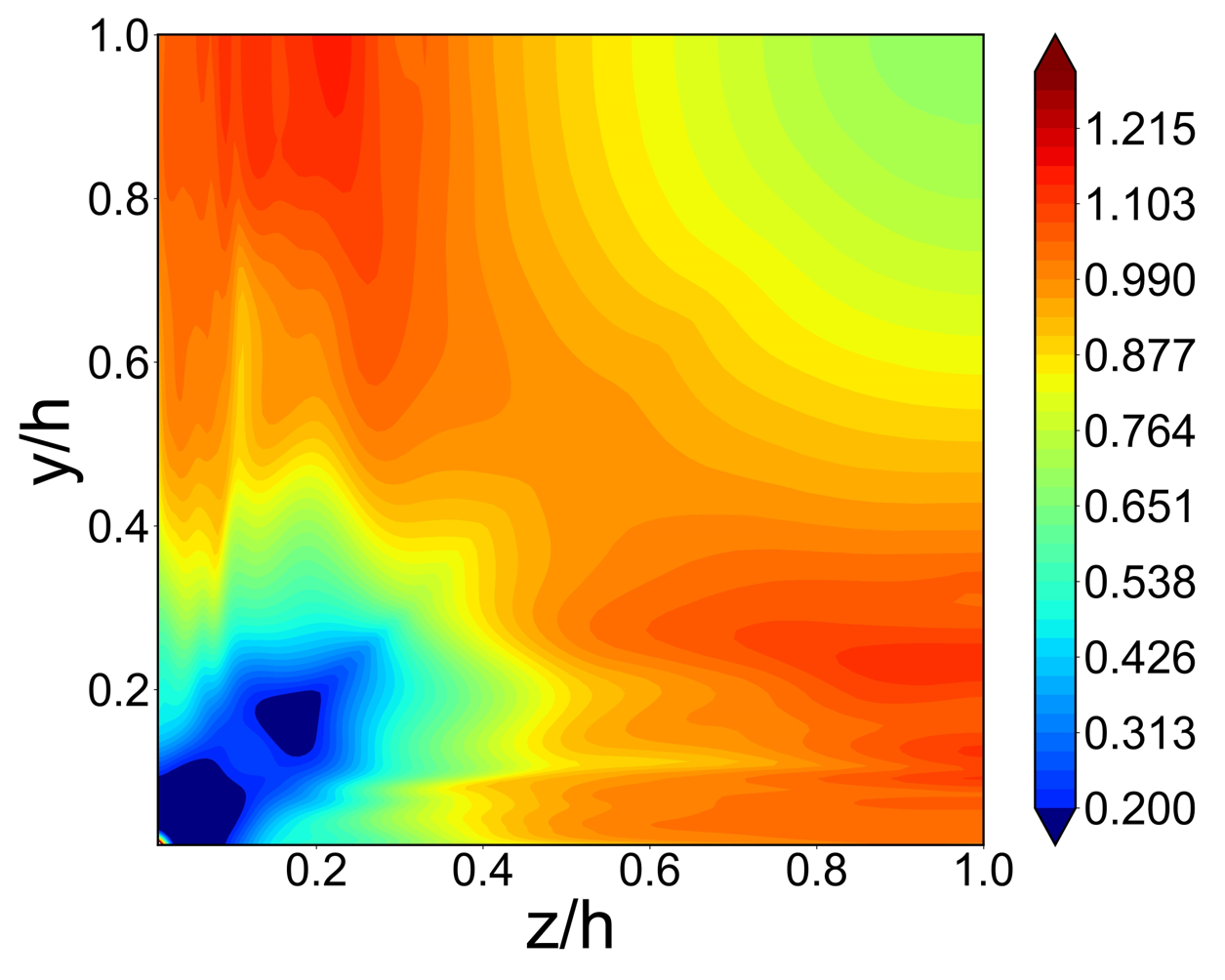}}\subfloat[SGE-TD - 8 parameters]{\centering{}\includegraphics[width=3.6cm]{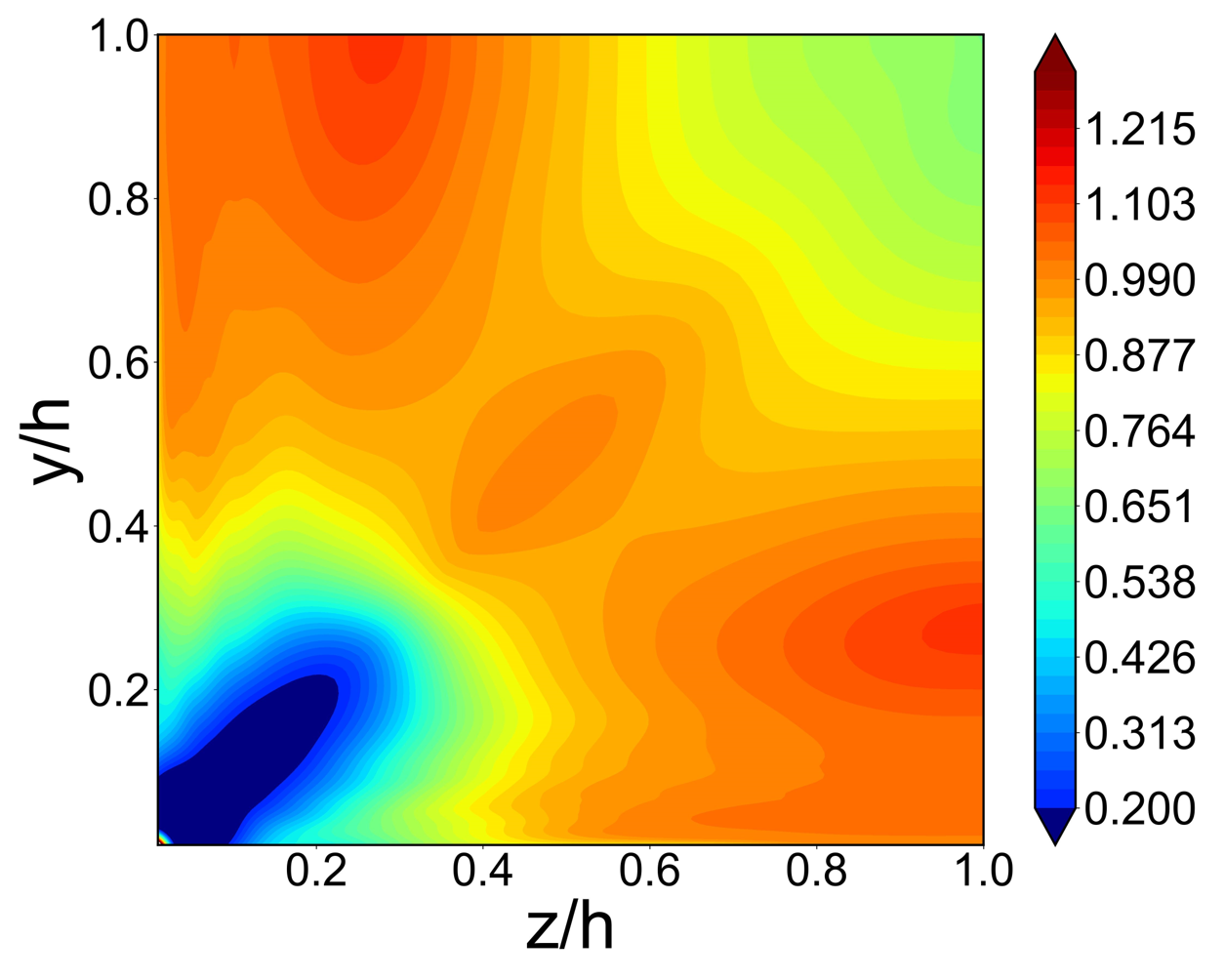}}
\par\end{centering}
\begin{centering}
\subfloat[SFS - 6 parameters]{\begin{centering}
\includegraphics[width=3.6cm]{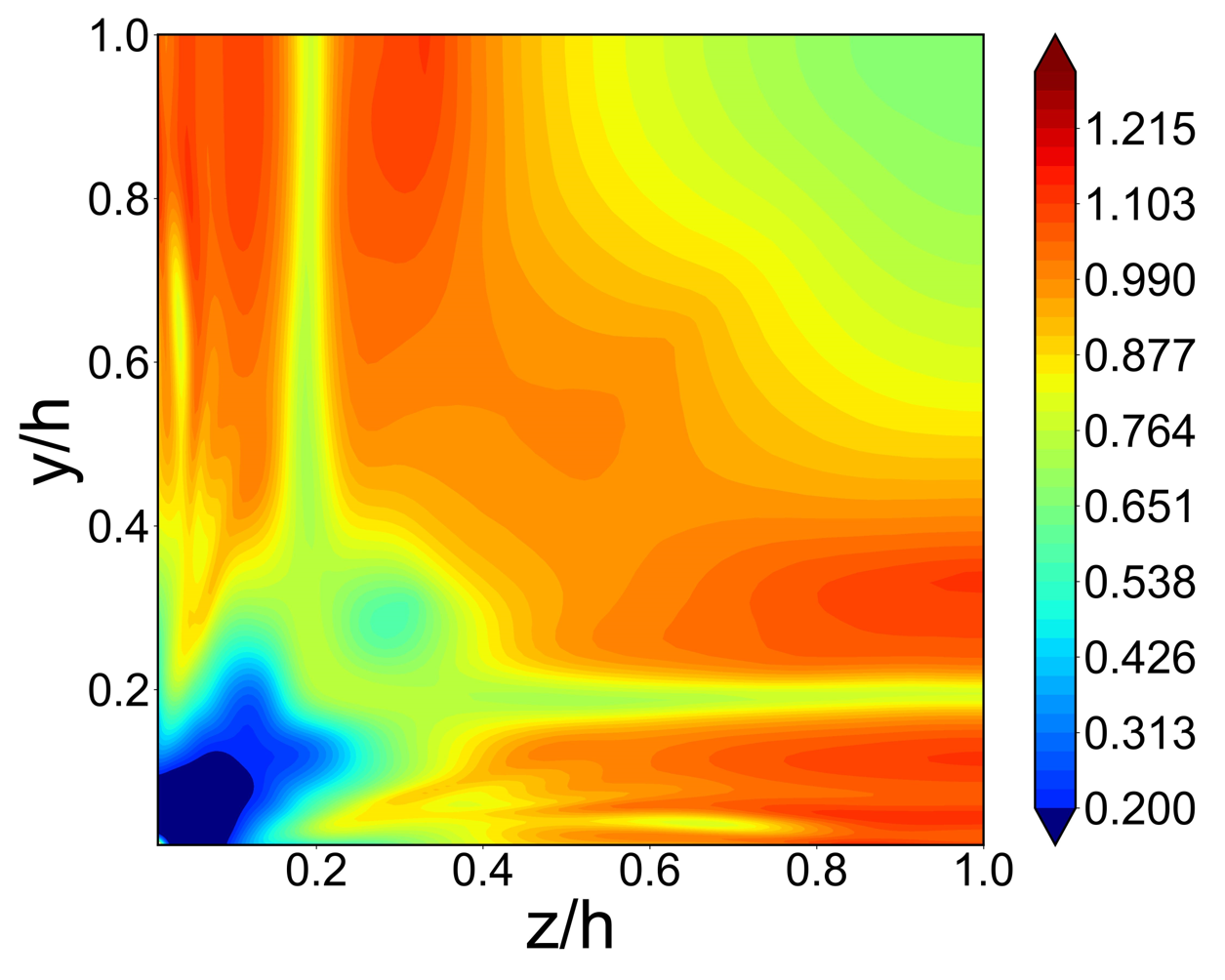}
\par\end{centering}
\centering{}}\subfloat[SGE-TD - 6 parameters]{\centering{}\includegraphics[width=3.6cm]{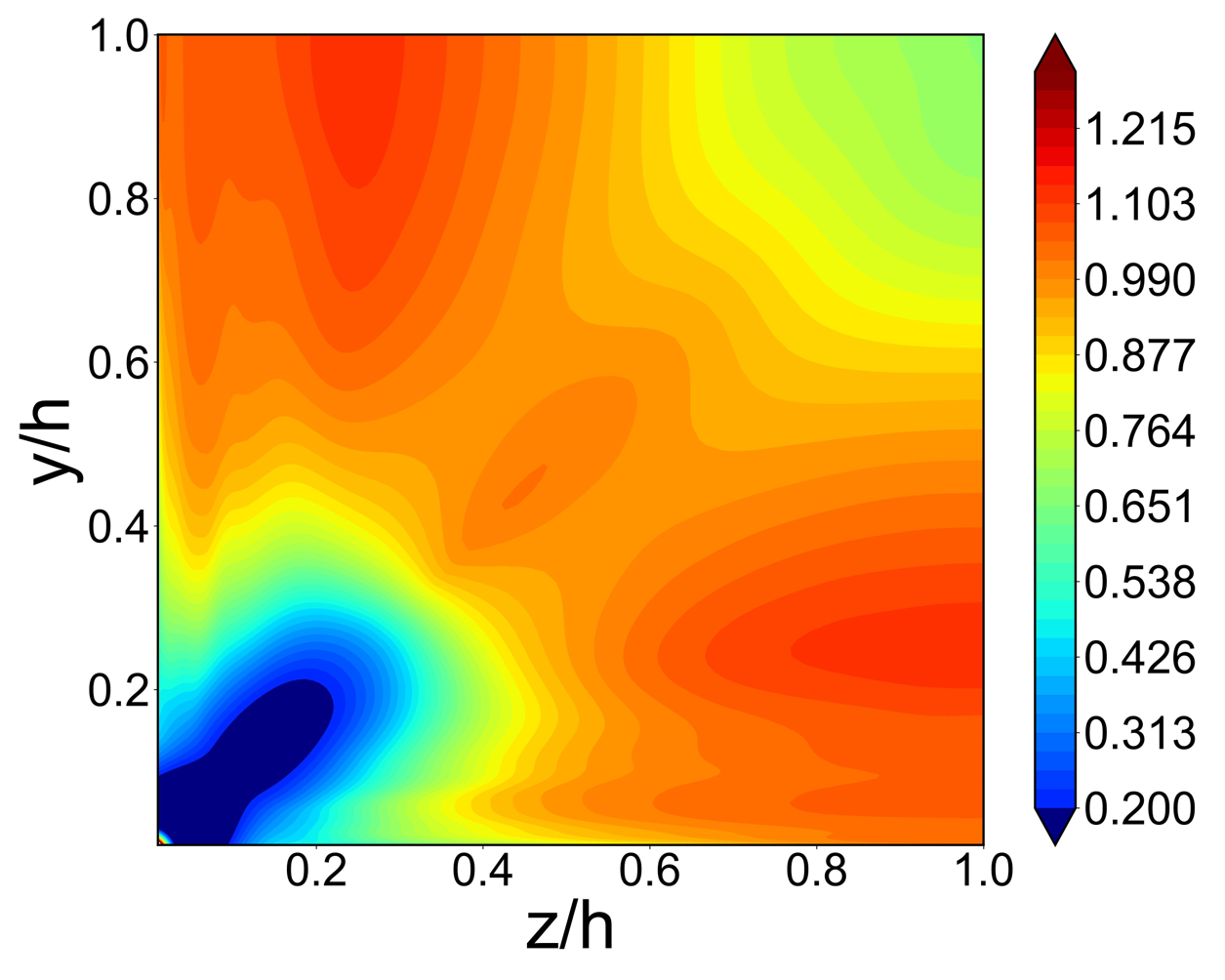}}
\par\end{centering}
\centering{}\subfloat[SFS - 3 parameters]{\centering{}\includegraphics[width=3.6cm]{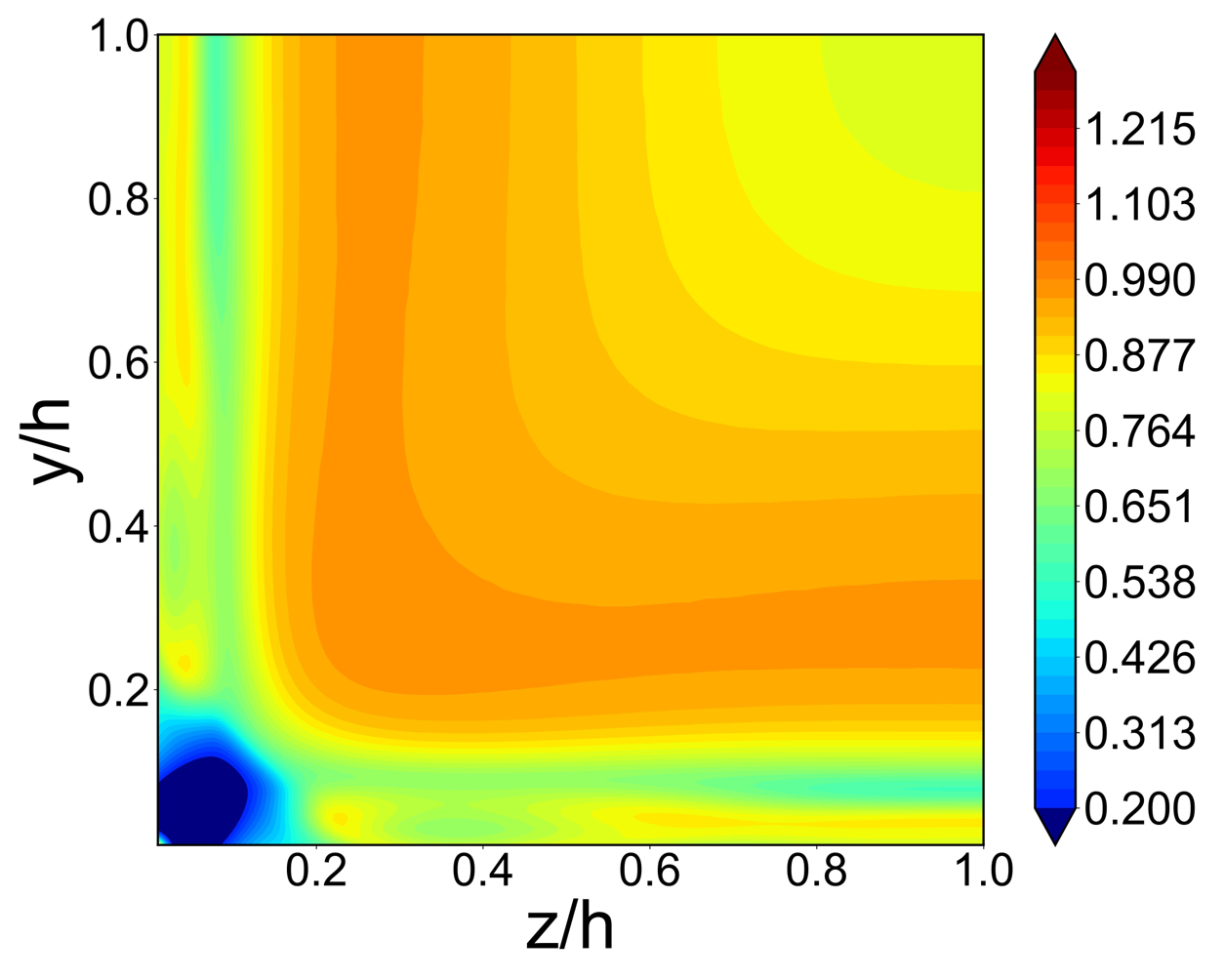}}\subfloat[SGE-TD - 3 parameters]{\centering{}\includegraphics[width=3.6cm]{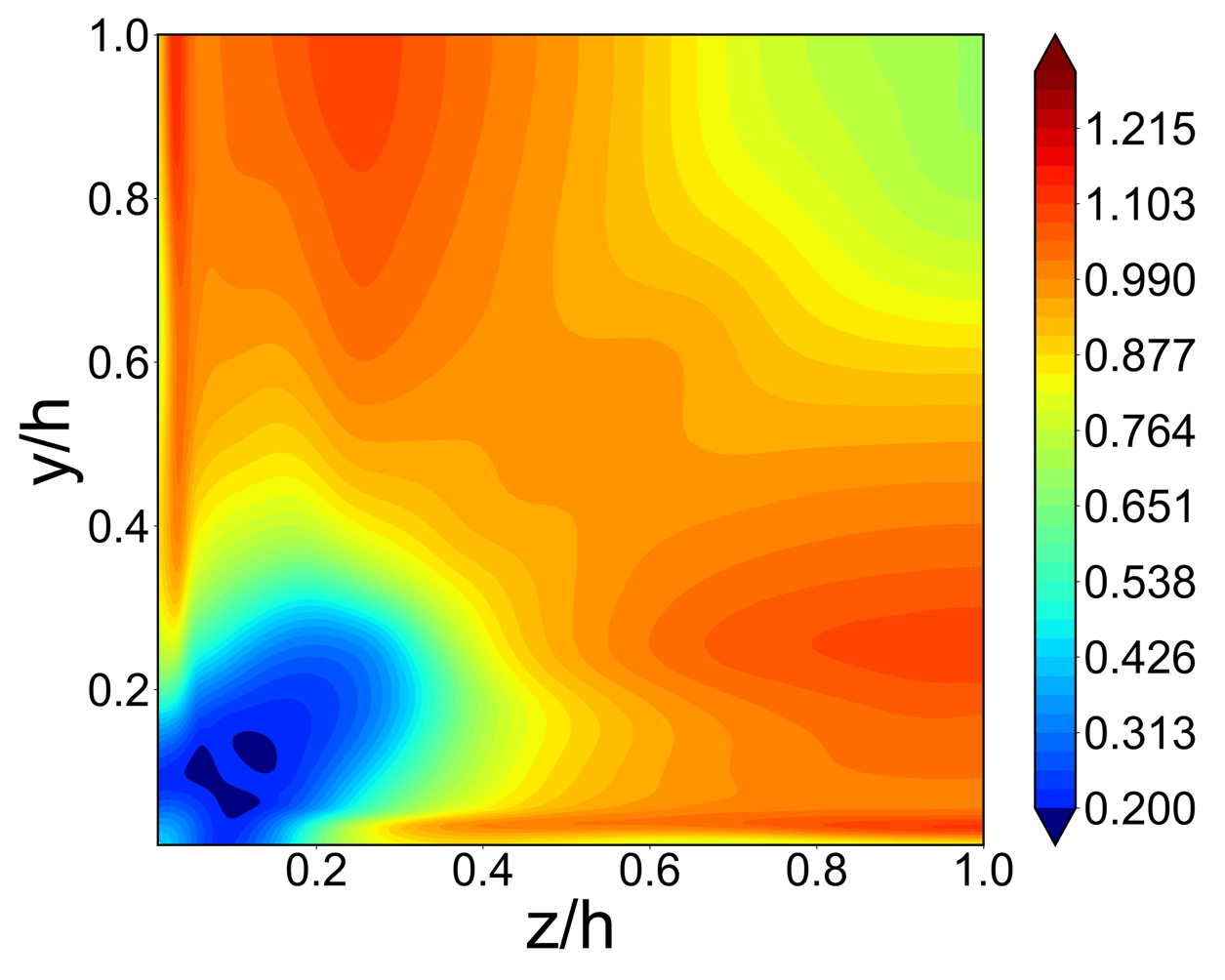}}\caption{Contours of $Pr_{t}$ from DNS, full parameter model, and reduced
models ($S_{3}$, $S_{6}$, $S_{8}$, and $S_{10}$) using SFS and
SGE-TD for turbulent heat transfer in a duct with $Pr=0.7$ (\textcolor{black}{2L-60N/L}).
Only the left-bottom quadrant of the domain is shown.\label{fig:Contours-of-differences Pr_t}}
\end{figure}

Fig.~\ref{fig:Contours-of-differences Pr_t} show the contours of
$Pr_{t}$ from the DNS, full parameter model, and reduced models ($S_{3}$,
$S_{6}$, $S_{8}$, and $S_{10}$) using SFS and SGE-TD. All models
were trained for 60 times using the database of $Pr$=0.2 and 2.0
without $Pr$=0.7. Then, the results for $Pr$=0.7 predicted using
the full parameter and reduced models are averaged over trained ANNs,
in order to estimate the practical prediction accuracy with reduced
training uncertainty. As expected, $Pr_{t}$ using the full parameter
model resembles the features of the DNS result most closely. The results
of each reduced model improve with the number of the parameters, but
those using SFS show distorted streaks near the walls. SGE-TD, on
the other hand, shows better prediction near the wall. SGE-TD shows
superior prediction in the center region of the plot as well compared
to SFS (particularly for $S_{3}$). Even for $S_{3}$, SGE-TD shows
higher resemblance with the DNS result while SFS completely fails
to do so. It may also be emphasized that the effect of $k$ is particularly
important. Despite the fact that $S_{10}$ of SFS and SGE-TD differ
only by a single parameter ($\varepsilon_{MTV}$ and $k$), $S_{10}$
with $k$ (SGE-TD) predicts $Pr_{t}$ better than the one without
it.

\begin{figure}[H]
\centering{}\includegraphics[width=6cm]{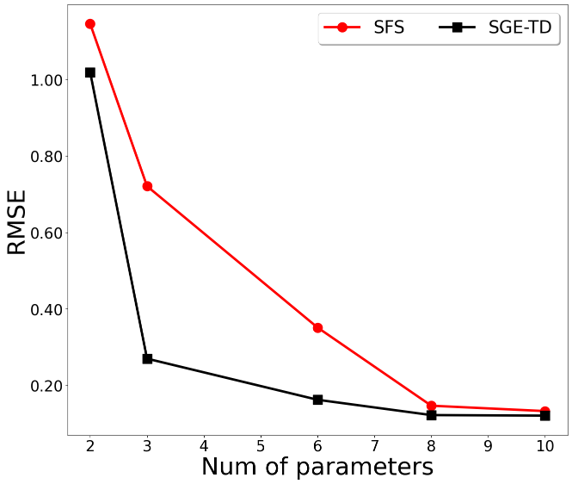}\caption{RMSE \textcolor{black}{(validation)} of predicted $Pr_{t}$ with 1L-30N/L
using the parameter subsets reduced with 2L-60N/L.\label{fig:RMSE-of-predicted}}
\end{figure}

To check the compatibility of the parameter subset between different
hyper-parameters, Fig.~\ref{fig:RMSE-of-predicted} shows the RMSE
of predicted $Pr_{t}$ with 1L-30N/L using the subsets reduced with
2L-60N/L. Even when the reduced models are trained under lower hyper-parameters,
SGE-TD is superior in the RMSE than SFS, although the difference decreases
as the parameter dimension increases.

\begin{figure}[H]
\centering{}\subfloat[1L-\textcolor{black}{15N/L}]{\centering{}\includegraphics[width=4cm]{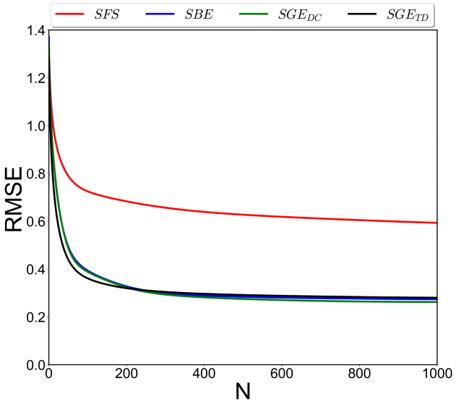}}\subfloat[1L-30\textcolor{black}{N/L}]{\centering{}\includegraphics[width=4cm]{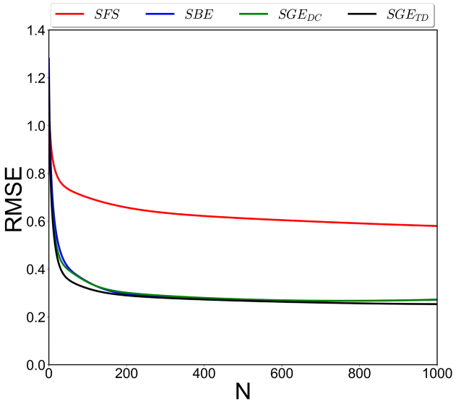}}\subfloat[2L-60\textcolor{black}{N/L}]{\centering{}\includegraphics[width=4cm]{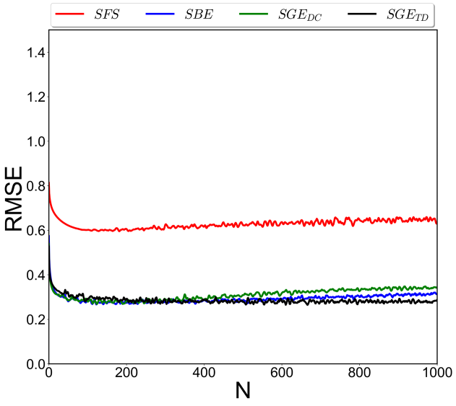}}\caption{RMSE \textcolor{black}{(validation)} of $S_{6}$ reduced with \textcolor{black}{1L-30N/L}
versus training epoch with different hyper-parameters for turbulent
heat transfer in a duct. RMSEs are averaged over trials. \label{fig:RMSE over hyperparameters}}
\end{figure}

To examine the effects of different subsets from the wrapper methods
on ANN training, Fig.~\ref{fig:RMSE over hyperparameters} shows
the RMSE with respect to epochs using $S_{6}$ reduced with \textcolor{black}{1L-30N/L}
for three different hyper-parameters. SGE-TD has the lowest RMSE out
of all methods. This shows relatively faster training speed, although
the differences from the other methods are not very large except for
SFS. SFS shows the highest RMSE value followed by SBE and SGE-DC.

\section{Conclusions\label{sec:Conclusions}}

Data-driven modeling is an emerging topic of interest for their capability
to reconstruct functional expressions of spatially distributed quantities
derived from high-fidelity simulations and experiments. These databases,
however, typically feature large number of related parameters and
high non-linearity which often lead to inefficient and inaccurate
training of the models. Thus, identification of primary parameters
and systematic dimensionality reduction are of utmost importance,
especially so in the field of turbulence modeling. In this study,
ANN-based wrapper methods are analyzed and a novel approach is proposed
to address a weakness of the existing methods. Due to the overfitting
issue and random nature of ANN training, consistency of the parameter
subsets over selection trials may be compromised especially in a high
parameter dimension.

In order to relax this issue, we defined two performance evaluation
metrics to evaluate consistency-over-trials (CoT) of the selected
subsets. Also, novel subset selection indices were devised based on
the gradient of the model function, under an idea that excessive weights
from the overfitting distort the partial derivatives while reducing
the RMSE. This leads to two sequential gradient-based elimination
(SGE) methods that minimize the loss in the total derivative of reduced
ANN models (SGE-TD) or the directional consistency of the gradient
(SGE-DC) at each elimination step. A large RMSE in the trained ANN
model leads to even larger error in its gradient, which can make the
SGE selection inconsistent. As a solution to this issue, it was found
that SGE using the trained ANNs with relatively smaller RMSEs (50\%
below the average) can successfully remove redundant and irrelevant
parameters. The training overhead could be minimized by using this
approach only for the final subset. When this approach was applied
to a manufactured subset selection problem, perfect success rates
were achieved. To further evaluate the SGE methods, these and other
existing ANN-wrapper methods (i.e., SFS and SBE) are tested for two
turbulence modeling applications: bubbly flow in a pipe and heat transfer
in a duct flow.

The ANN-wrapper methods were applied to the reduced modeling of the
bubble size ($D_{sm}$) in turbulent bubbly flows in a pipe. Using
the results of experimental studies, a database for modeling $D_{sm}$
was collected and the non-dimensional 5-parameter model of \citep{Jung2019}
is used as the full parameter model. Using SGE, the size of optimally
reduced subset was smaller compared to the SFS (and RF) method and
similar to the SBE method. Out of all the ANN-wrapper methods, only
SFS selected $J_{G}/J_{L}$ over $\alpha_{G}$, a well-known parameter
for modeling $D_{sm}$. Rest of the methods selected nearly identical
parameters. The CoT indices were also evaluated. SBE and SGE-TD showed
ideal CoT for all cases. SGE-DC showed slightly low CoT with all trained
ANNs but produced ideal CoT with well-trained ANNs. SFS showed relatively
lower CoT than the others.

To analyze the ANN-wrapper methods for a higher-dimensional database,
modeling $Pr_{t}$ in a turbulent duct flow was considered to reduce
a 12-parameter model composed of the physical parameters usually available
from RANS simulations. The database was obtained from a DNS study.
From the heat map of the correlation coefficients, complex relationships
and the overlapping effects among the parameters were observed. As
a result, SGE-TD showed the highest CoT indices followed by SBE, SGE-DC,
and SFS in order. Higher CoT of SGE-TD was more pronounced for a higher-dimensional
subset. Except for SFS showing relatively lower CoT and higher RMSE,
all methods preferred selecting $\nu_{t}$ and $k$ for 6-parameter
subsets. From the contours of $Pr_{t}$ with different $Pr$ from
the training data, SGE-TD showed far superior prediction of $Pr_{t}$
compared to SFS across all regions of the duct. Even for heavily reduced
(3-parameter) subset, SGE-TD was relatively effective in $Pr_{t}$
prediction. SFS, on the other hand, produced far inferior $Pr_{t}$
prediction, which may further verify the importance of $k$ and $\nu_{t}$.
From a comparative study on the hyper-parameters and training speed,
it was found that a reduced parameter subset is compatible among hyper-parameters,
and SGE-TD showed the fastest training speed although the differences
from the other methods are not very large.

In summary, the new ANN-wrapper method can successfully remove unnecessary
parameters using only the gradient information, leading to optimal
parameter subsets. Besides, more robust parameter selection with higher
CoT compared to the existing methods can be achieved, which is useful
for modeling physical problems with high-dimensional complexity using
ANN.
\begin{acknowledgments}
This work was supported by the National Research Foundation of Korea
(NRF) grants funded by the Korea government (MSIP) (No. 2017M2A8A4018482,
2017R1A2B3008273, and 2022R1F1A1074931).
\end{acknowledgments}

\section*{Appendix. The budget equations for kinetic and thermal energy in
turbulent flows\label{sec:Appendix.-The-budget}}

From the theories on turbulent flows, the budgets of the TKE ($k=\frac{1}{2}\overline{u_{j}^{\prime}u_{j}^{\prime}}$)
and temperature variance (TV, $\overline{T^{\prime}T^{\prime}}$)
have been considered for an analysis of the energy transfer by fluctuation
components. Similarly, the budgets for the mean kinetic energy (MKE)
and mean temperature variance (MTV) are considered. Their governing
equations are written as
\begin{equation}
\underbrace{\frac{1}{2}\frac{D\overline{u}_{j}\overline{u}_{j}}{Dt}}_{SD_{MKE}}=\underbrace{-\frac{\partial\overline{u}_{i}\overline{u_{i}^{\prime}u_{j}^{\prime}}}{\partial x_{j}}}_{C_{MKE}}\underbrace{+2\nu\frac{\partial\overline{u}_{i}\overline{S}_{ij}}{\partial x_{j}}}_{D_{MKE}}\underbrace{-\frac{1}{\rho}\frac{\partial\overline{u}_{j}\overline{p}}{\partial x_{j}}}_{\Pi_{MKE}}\underbrace{+\overline{u_{i}^{\prime}u_{j}^{\prime}}\frac{\partial\overline{u}_{i}}{\partial x_{j}}}_{-P_{TKE}}-\underbrace{2\nu\overline{S}_{ij}\overline{S}_{ij}}_{\varepsilon_{MKE}},\label{eq : mean vel budget}
\end{equation}
\begin{equation}
\underbrace{\frac{1}{2}\frac{D\overline{T}\,\overline{T}}{Dt}}_{SD_{MTV}}=\underbrace{-\frac{\partial\overline{T}\,\overline{u_{j}^{\prime}T^{\prime}}}{\partial x_{j}}}_{A_{MTV}}\underbrace{+\frac{\alpha}{2}\frac{\partial^{2}\overline{T}\,\overline{T}}{\partial x_{j}\partial x_{j}}}_{D_{MTV}}\underbrace{+\overline{T^{\prime}u_{j}^{\prime}}\frac{\partial\overline{T}}{\partial x_{j}}}_{-P_{TV}}-\underbrace{\alpha\frac{\partial\overline{T}}{\partial x_{j}}\frac{\partial\overline{T}}{\partial x_{j}}}_{\varepsilon_{MTV}}.\label{eq : mean temperature budget}
\end{equation}
where $S_{ij}$ denotes the strain rate tensor. The overbar ($^{-}$)
and prime ($'$) symbols denote averaged and fluctuation components
of a variable, respectively. The term $\varepsilon_{MKE}$ and $\varepsilon_{MTV}$
denote the MKE dissipation and MTV dissipation, respectively.

\bibliographystyle{elsarticle-num-names}
\addcontentsline{toc}{section}{\refname}\bibliography{references}

\end{document}